\documentclass[preprint2]{aastex61} 
 
\begin{document} 
 
\title{Extreme radio flares and associated X-ray variability from
young stellar objects in the Orion Nebula Cluster}                      
 

\author[0000-0001-8694-4966]{Jan Forbrich}
\affiliation{Centre for Astrophysics Research, 
School of Physics, Astronomy and Mathematics, University of Hertfordshire, 
College Lane, Hatfield AL10 9AB, UK}
\affiliation{Harvard-Smithsonian Center for Astrophysics, 60 Garden St, 
Cambridge MA 02138, USA}

\author{Mark J. Reid}
\affiliation{Harvard-Smithsonian Center for Astrophysics, 60 Garden St, 
Cambridge MA 02138, USA}

\author{Karl M. Menten}
\affiliation{Max-Planck-Institut f\"ur Radioastronomie, Auf dem
H\"ugel 69, 53121 Bonn, Germany}

\author{Victor M. Rivilla}
\affiliation{Osservatorio Astrofisico di Arcetri, Largo Enrico
Fermi 5, 50125, Firenze, Italy}

\author{Scott J. Wolk}
\affiliation{Harvard-Smithsonian Center for Astrophysics, 60 Garden St, 
Cambridge MA 02138, USA}

\author{Urvashi Rau}
\affiliation{National Radio Astronomy Observatory, P.O. Box O,
Socorro, NM 87801, USA}

\author{Claire J. Chandler}
\affiliation{National Radio Astronomy Observatory, P.O. Box O,
Socorro, NM 87801, USA}                                                 
 
\begin{abstract} 
Young stellar objects are known to exhibit strong radio variability
on timescales of weeks to months, and a few reports have documented
extreme radio flares, with at least an order of magnitude change in 
flux density on timescales of hours to days.  However, there have 
been few constraints on the occurrence rate of such radio flares or 
on the correlation with pre-main sequence X-ray flares,
although such correlations are known for the Sun and nearby
active stars.  Here we report simultaneous deep VLA radio and {\it
Chandra} X-ray observations of the Orion Nebula Cluster, targeting
hundreds of sources to look for the occurrence rate of extreme radio
variability and potential correlation with the most extreme X-ray 
variability. We identify 13 radio sources with extreme radio variability,
with some showing an order of magnitude change
in flux density in less than 30~minutes. All of these sources 
show X-ray emission and variability, but only on timescales $<$1h
do we find clear correlations with extreme radio flaring.
Strong X-ray variability does not predict the extreme radio sources 
and vice versa. Radio flares thus provide us with a new perspective 
on high-energy processes in YSOs and the irradiation of their protoplanetary
disks. Finally, our results highlight implications for interferometric
imaging of sources violating the constant-sky assumption.               
\end{abstract} 
 
 
 
 
\section{Introduction} 
 
Only a few examples of extreme radio flaring from young stellar objects 
(YSOs) -- defined as flux density changing by more than an order of magnitude 
in less than a few hours -- have been reported (e.g., \citealp{bow03,for08}). Other reported YSO radio variability is either less pronounced or becoming apparent on longer timescales of weeks to months (e.g., \citealp{fel93b,zap04,riv15}).
However, the true occurrence rate is not known due to sensitivity limitations 
on short timescales. The spectral characteristics of such events
are also poorly explored and, thus, it has been difficult to ascertain 
the relevant emission processes. Nonthermal radio emission at centimeter
wavelengths in this context is usually interpreted as gyrosynchrotron radiation,
created by mildly relativistic electrons gyrating in magnetic fields. The
millimeter wavelength range provides access to synchrotron radiation from 
relativistic electron populations (e.g., \citealp{gue02}). While both X-ray and 
nonthermal radio emission in YSOs is thought to be produced in 
coronal-type activity (e.g., \citealp{fem99}), X-ray flares are much better
studied than radio flares (e.g., \citealp{wol05,get08a,get08b}), and one outstanding question
is whether or not radio and X-ray flares are physically related.         
 
An empirical relation between the time-averaged radio and X-ray luminosities, 
the  G\"udel-Benz (GB) relation \citep{gue93,ben94}, has been established
for a wide range of active stars.  For the Sun one also observes
correlated of X-ray and radio variability, the Neupert effect \citep{neu68}, 
where one observes magnetic energy release, from particle
acceleration and injection to energy transformation and heating.
In this relation, seen in some but not all flares, the time-integrated
radio emission, which traces the energy injection, is proportional to
the thermal X-ray emission.  This phenomenon has been observed for a 
nearby active star \citep{gue02n} and for one YSO \citep{get11}.
 
In contrast to well studied X-ray flares from YSOs, only a handful of 
radio flares have been documented.  \citet{bow03} reported an increase in
flux density at a wavelength of 3.5~mm of a factor of 4--8 over hours for
the evolved weak-line T~Tauri star GMR~A.  Similar flares were reported from
the weak-line T~Tauri binary V773~Tau \citep{mas06} and the classical
T~Tauri binary DQ~Tau \citep{sal08}, most likely due to synchrotron
radiation from inter-binary magnetic interactions.  At a longer
wavelength of 1.3~cm, \citet{for08} reported an even stronger flare,
with an increase in flux density of $>10$ in a few hours, toward
the deeply embedded protostar termed the Orion Radio Burst Source
(ORBS).                                                                 
 
Both X-ray and radio flares provide clues related to the high-energy
irradiation of the vicinities of YSOs and their protoplanetary
disks.  This irradiation may impact planet formation and planetary habitability.
However, the connection between radio and X-ray emission remains unclear. 
This is partly due to observational limitations, since comparisons of flares
on short timescales require simultaneous observations of phenomena that are
not predictable and often weak.  The radio flare of GMR~A is one of
only a few cases with simultaneous radio and X-ray coverage \citep{bow03}. 
Clearly, a larger sample of YSO extreme radio variability is needed to provide
constraints on the association with X-ray flares.  

We here continue the analysis of the first simultaneous radio and X-ray
observations of YSOs following the sensitivity upgrade of the National Radio
Astronomy Observatory's Karl G. Jansky Very Large Array (VLA). The upgraded 
VLA is an interferometer with broad bandwidth and excellent instantaneous 
$(u,v)$-coverage to enable high-sensitivity studies of YSOs. Using the VLA
and {\it Chandra}, we have targeted the Orion Nebula Cluster (ONC), maximizing
the chances of detecting YSO flares, since we can observe hundreds of YSOs in 
a single pointing. We have already published a catalog of 556 compact sources
in the cluster, based on the entire concatenated radio dataset \citep{for16}. 
The increased sensitivity of the VLA is now enabling access to shorter 
timescales than in previous such experiments, which have focused on 
time-averaged properties \citep{gag04,for07,ost09,for11,fow13}.  

In order to constrain the  occurrence of extreme radio flares in these YSOs, 
akin to the event reported by \citet{for08}, we measured the radio light-curves
of the all of these sources,  looking for extreme changes in flux density, 
where ``extreme'' is defined as changes of more than an order of magnitude on 
timescales $<$2~days. Subsequently, we assessed the simultaneous X-ray 
variability of this sample of radio flares. The goal of this paper is to begin 
the exploration of YSOs in the radio--X-ray time domain by first identifying the
most extreme variability.                                               
 
\section{Observations and data analysis} 
 
While details of our Orion observations and the underlying source catalog can 
be found in \citet{for16}, we here briefly recount the main features of this
experiment. We observed a single pointing position in the ONC for a total of 
30~h over five days in the VLA's highest-resolution  A-configuration, using two 
spectral bands of 1~GHz width within the C-band (4-8~GHz) receivers, centered at
4.736~GHz and 7.336~GHz. During the first two epochs, the array was being re-
configured to the A configuration from the BnA configuration.
The VLA observations were accompanied by 24~h of nearly simultaneous 
{\it Chandra} observations. The observing log is listed in 
Table~\ref{tab_obslog}. In the following, all source numbers
refer to the source catalog in \citep{for16}, with the prefix [FRM2016]. 
 
\begin{deluxetable}{rrr} 
\tablecaption{Observing log\label{tab_obslog}} 
\tablehead{ 
\colhead{Epoch}    & \colhead{Observatory} & \colhead{Time range}\\ 
} 
\startdata 
1 & VLA           & 2012-Sep-30 07:33 -- 15:01\\ 
  & {\it Chandra} & -- \\ 
2 & VLA           & 2012-Oct-02 07:25 -- 14:53\\ 
  & {\it Chandra} & 2012-Oct-02 07:02 -- 15:16\\ 
3 & VLA           & 2012-Oct-03 13:20 -- 16:19\\ 
  & {\it Chandra} & 2012-Oct-03 12:53 -- 15:56\\ 
4 & VLA           & 2012-Oct-04 10:49 -- 16:00\\ 
  & {\it Chandra} & 2012-Oct-04 08:21 -- 16:13\\ 
5 & VLA           & 2012-Oct-05 08:28 -- 16:11\\ 
  & {\it Chandra} & 2012-Oct-05 10:42 -- 16:17\\ 
\enddata 
\end{deluxetable}

We have sliced the entire dataset into short time intervals and
imaged independently. Without precise constraints
on the most relevant timescale, and given the trade-off between high
time resolution and sensitivity, we produced three sets of images
over timescales of 6 hours, 30 minutes, and 6 minutes. We here focus
on imaging in Stokes-$I$.  
The latter two sets of images are limited somewhat by problems associated
with varying $(u,v)$-coverage for some resolved sources, varying
source strength during the integration period, and their distribution
over a wide field comparable to the individual antenna primary beamwidth.

Imaging was done in the Common Astronomy Software Applications package 
(CASA, \citealp{mcm07})
as had been done for the concatenated dataset \citep{for16}, including the same
approximate correction for the primary beam attenuation.
For the two sets of maps with higher time resolution, with different
synthesized beam shapes and sizes per time bin, we chose a smaller
pixel size of 0$\farcs$04, instead of 0$\farcs$10.   Since we left the
image size at $8192\times8192$ pixels (for computational reasons), this
resulted in a field of view of $5\farcm4\times5\farcm4$, which excluded 
some sources in the outermost portion of the primary beam. However,
the images still covered 507 of the 556 sources in our catalog. 
Additionally, prior to imaging, we time averaged 10~sec, in order to 
speed up the computations, at a modest cost of a few percent in flux 
density loss for sources near the edges of the maps (e.g., KM Ori 
\citealp{for16}).  Only a few of the mapped time intervals were discarded
based on poor image quality (see below).                                               
 
As for our original catalog, we fitted all sources using ``jmfit'' 
of the Astronomical Image Processing System
(AIPS\footnote{\url{http://www.aips.nrao.edu}}).
We assumed a model of a single Gaussian component to determine
the peak flux density, provided the peak was within a few pixels of
the catalog position and had a signal-to-noise ratio greater than 4.  
Such fits work best for point sources, since otherwise the
changing synthesized beam can interact with the complex source structure
to introduce spurious variability. Many sources detected in the
concatenated image were not detected with the lower sensitivity afforded
by the higher time resolution. 
To assess the inter-epoch calibration, we selected the 223 sources
detected at all five epochs.  Removing the 10\% most variable ones, 
left 201 sources.  Averaging their peak flux densities yield identical values
to within a few percent, 0.520$\pm$0.015~mJy/beam, indicating
very good inter-epoch calibration.                                      

This paper describes only the most extreme varying sources that comprise only 
2--3\% of radio sources.  Many sources exhibit significant variability below
our extreme threshold, but they are beyond the scope of this paper.
At low levels of variability, systematic effects like 
the impact of the changing $(u,v)$ coverage on resolved sources and the wideband
primary beam are difficult to quantify and remove.

\subsection{X-ray observations} 
 
Our radio observations for epochs 2--5 had simultaneous
X-ray observations by \textit{Chandra} (see Table~\ref{tab_obslog}).
The {\it Chandra} observations were reduced using acis\_extract \citep{bro10},
and the catalog of the {\it Chandra} Orion Ultra-deep Project (COUP) 
\citep{get05} was used as an input catalog for point
source extraction. The code acis\_extract extracts information from multiple
observing epochs, taking into account different spacecraft parameters
and, in particular, differing point-spread functions at different epochs, 
and it combines this information into global metrics across all epochs. 
Of particular interest here are the extracted X-ray lightcurves and 
variations of median photon energy as a sensitive tracer of
spectral changes. While the code also extracts spectra and concatenates
these across observing epochs, our focus here is on these time
series metrics.                                                         
 
\section{Radio analysis} 
 
\begin{figure} 
\includegraphics[width=1.1\linewidth]{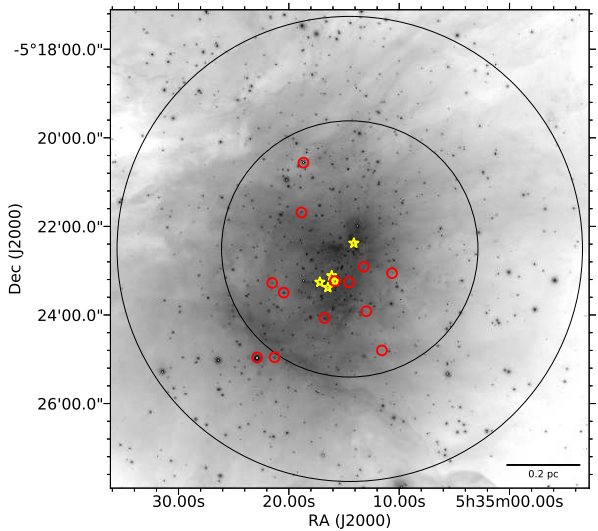}
\caption{Distribution of the 13 radio sources showing extreme radio
variability (by a factor $>10$), at any of the three time-scales discussed 
in the text, overlaid on the VISION-$K_S$ image from \citet{mei16}.
For orientation, the locations of the Trapezium (${\theta }^{1}$
Ori A) and the BN object are marked with star symbols. The circles
indicate the smallest (i.e., highest-frequency) and largest C-band
HPBW primary beams, and the scale bar indicates the physical scale
at a distance of 414 pc.\label{fig_view}}                               
\end{figure}

A total of 13 sources showed extreme radio variability (greater than a factor 
of $\Delta S$=10) on timescales of less than 2 days.  Their positions are shown in 
Figure~\ref{fig_view} and details are listed in Table~\ref{tab_exvar}, including 
i) the maximum variability factor,
ii) the minimum timescale where variability by a factor of at least 10, and 
iii) the peak flux densities from the concatenated data, as derived by \citet{for16}. 
Radio lightcurves for these sources at all imaging timescales are shown in
Figure~\ref{fig_var_all}.                                               

In \citet{for08} we used the BN object, a non-variable thermally emitting
source \citet{for16}, to assess the magnitude of systematic effects affecting 
measured variability.   This source is number 162 in our catalog; it is bright 
and marginally resolved, and it could show effects both due to the wideband 
primary beam and to changing $(u,v)$-coverage.  In our time series data, the 
standard deviation of its peak flux density is $<11\%$, and we conclude that 
systematics have not significantly affected our findings presented here.

\subsection{Individual epochs} 

We find only 244 out of 556 previously catalogued sources are detected
in all five epochs, partly due to the lower sensitivity compared to the 
concatenated data and partly due to variability. The rms noise levels 
range between 12 and 19~$\mu$Jy\,bm$^{-1}$ due to the different durations 
of the individual epochs.  Seven sources meet our criteria for extreme
variability (see Table~\ref{tab_exvar}) at these imaging integration times.  
Two sources are outside of the primary beam half-power radius at the higher 
frequency band and, thus, are not included in the followong 
discussion\footnote{Sources 358 and 459 are varying by factors of $>18$ and 
$>20$, respectively, on a timescale of two days.}.  
Source 515 displays the greatest change in flux density, a factor of 57.  
Some sources were detected in just one epoch, but strong enough to produce 
a detection in the concatenated dataset.         

We note that the averages of the peak flux densities of the individual epochs 
do not always agree with the values from imaging with the concatonated dataset
\citep{for16}.  This discrepancy is a result of strong variability during
the imaging integration time, which can scatter power away from the source
in a complicated manner. This finding is a warning that flux densities and
spectral indices of strongly variable sources can be erroneous when the
integration time is longer than the variability time scale.

\subsection{30 min intervals} 
 
\begin{figure*} 
\includegraphics[width=\linewidth]{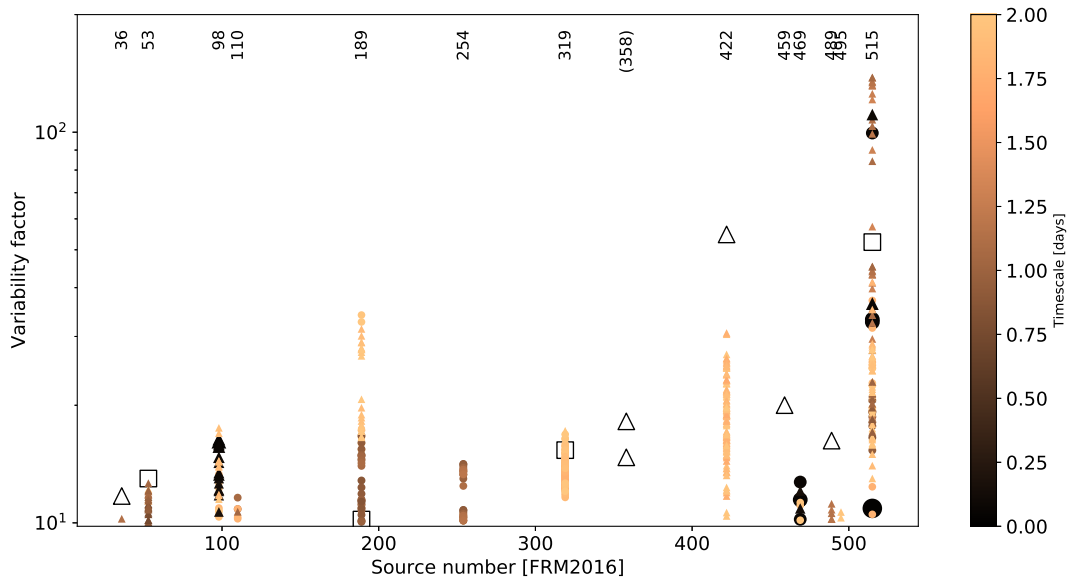} 
\caption{Selection of extremely variable sources at a resolution
of 30~min. The variability factors $>10$ from all two-point comparisons
are plotted, on a logarithmic scale and by source number. The color
encodes the associated timescale for the respective point pair, which
is also highlighted by symbol size. Triangles denote lower limits.
Finally, the large open symbols show the selection for data from
the individual epochs. Labels additionally designate the source numbers,
with source 358 in brackets because no data at high time resolution
is available.\label{fig_varhist1}}                                      
\end{figure*}

At 30~min time resolution we made 57 images, with median rms noise 
of 34~$\mu$Jy\,bm$^{-1}$.  Owing to reduced sensitivity from short
integrations, 76 (out of 507) sources were not detected in any image,
and only 48 sources have $>3\sigma$ detections in all 30~min bins.
All seven sources identified from individual epoch images are confirmed 
to be extreme variables in the 30-min images.  In addition, five other sources 
(98, 110, 254, 469, and 495) now show extreme variability, with sources 
110 and 469 displaying particularly extreme variations. 
In a few cases, the variability amplitude appears to decrease when
compared to the by-epoch assessment, but only when flux density upper
limits are involved, such that the resulting variability amplitudes
are statistically compatible with each other. A visualization of the selection
process is shown in Figure~\ref{fig_varhist1}.                          
 
\subsection{6 min intervals} 
 
Finally, we made 332 images from 6-min integrations.  (At this
sampling rate, nine integrations produced anomalous noise levels and
were discarded.)  Naturally, the median rms noise of 56~$\mu$Jy\,bm$^{-1}$
per image increased compared to the 30-min integrations, and 83 catalog
sources were not detected in any 6-min image.  
Only source number 400 (GMR~F) is detected in every single 6~min bin, 
but 70 sources have detections in at least half of the bins. 
Imaging artifacts due to poor $(u,v)$-coverage increase and produce 
occasional strong outliers in the lightcurves.                          
 
\begin{figure*} 
\includegraphics[width=\linewidth]{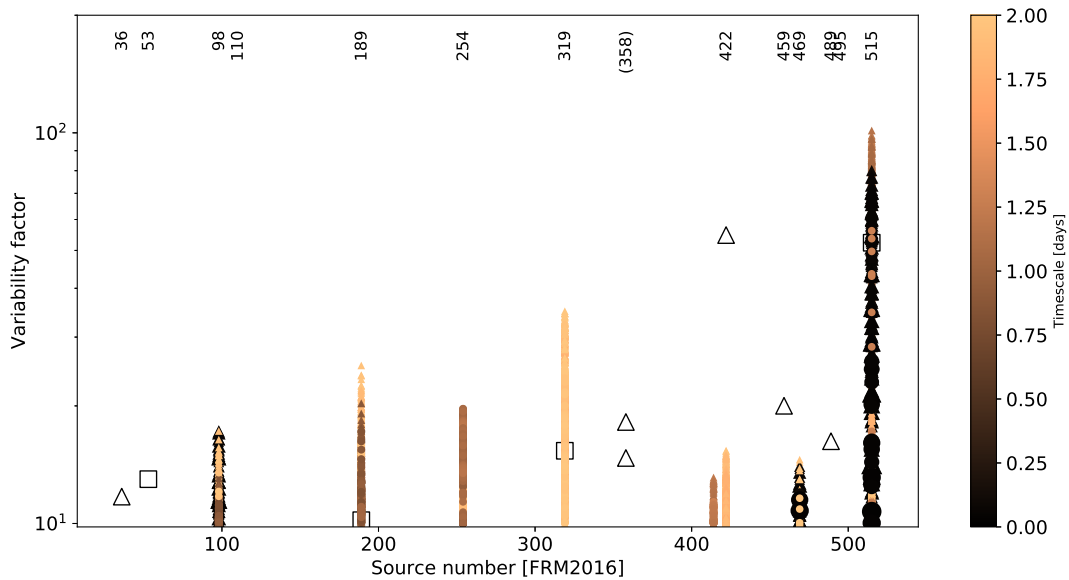} 
\caption{Selection of extremely variable sources at a resolution
of 6~min; see Figure~\ref{fig_varhist1} for details. \label{fig_varhist2}}
\end{figure*}

\begin{figure*} 
\begin{minipage}{\linewidth} 
\includegraphics*[bb= 60 219 511 560, width=0.2620\linewidth]{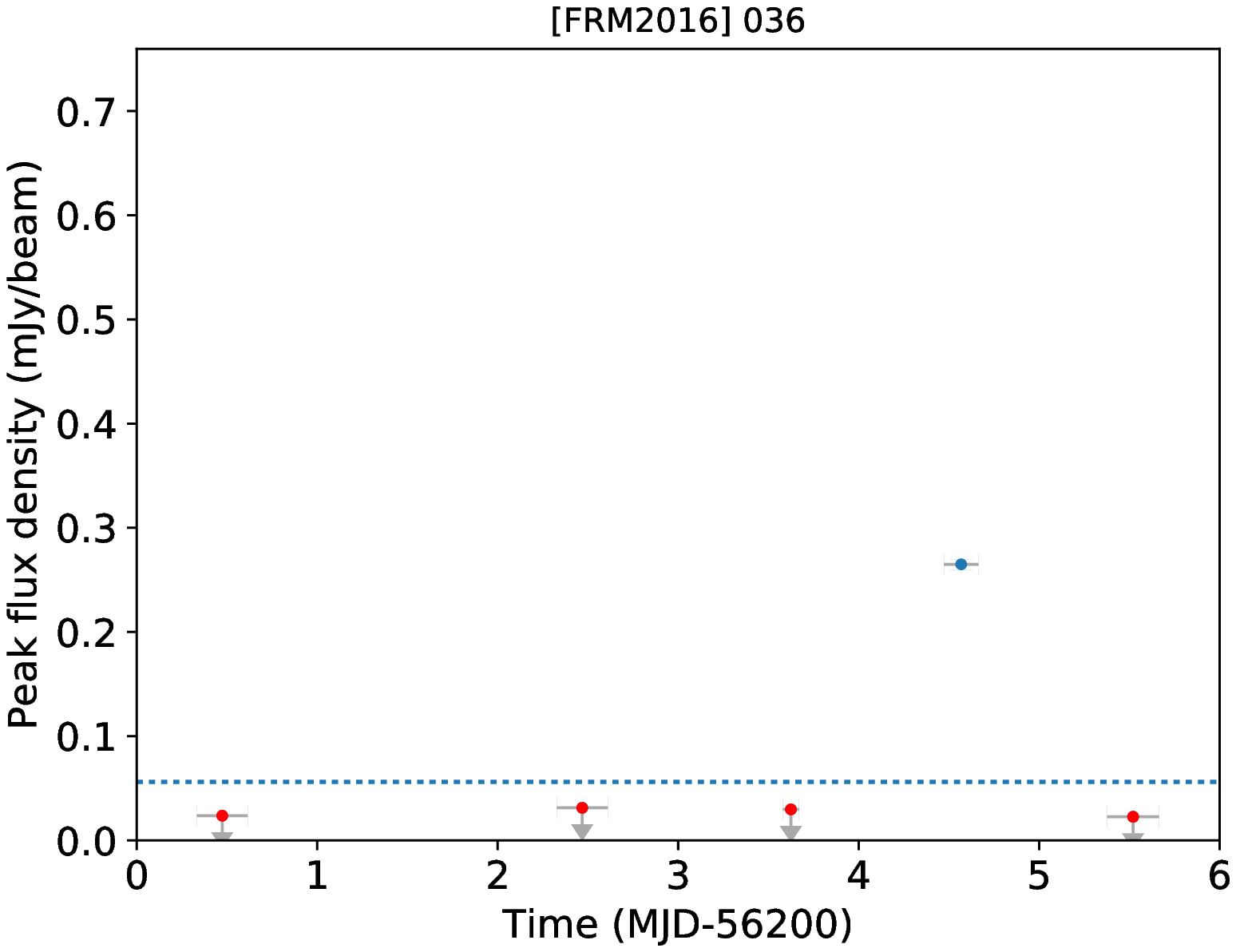}
\includegraphics*[bb=110 219 511 560, width=0.2330\linewidth]{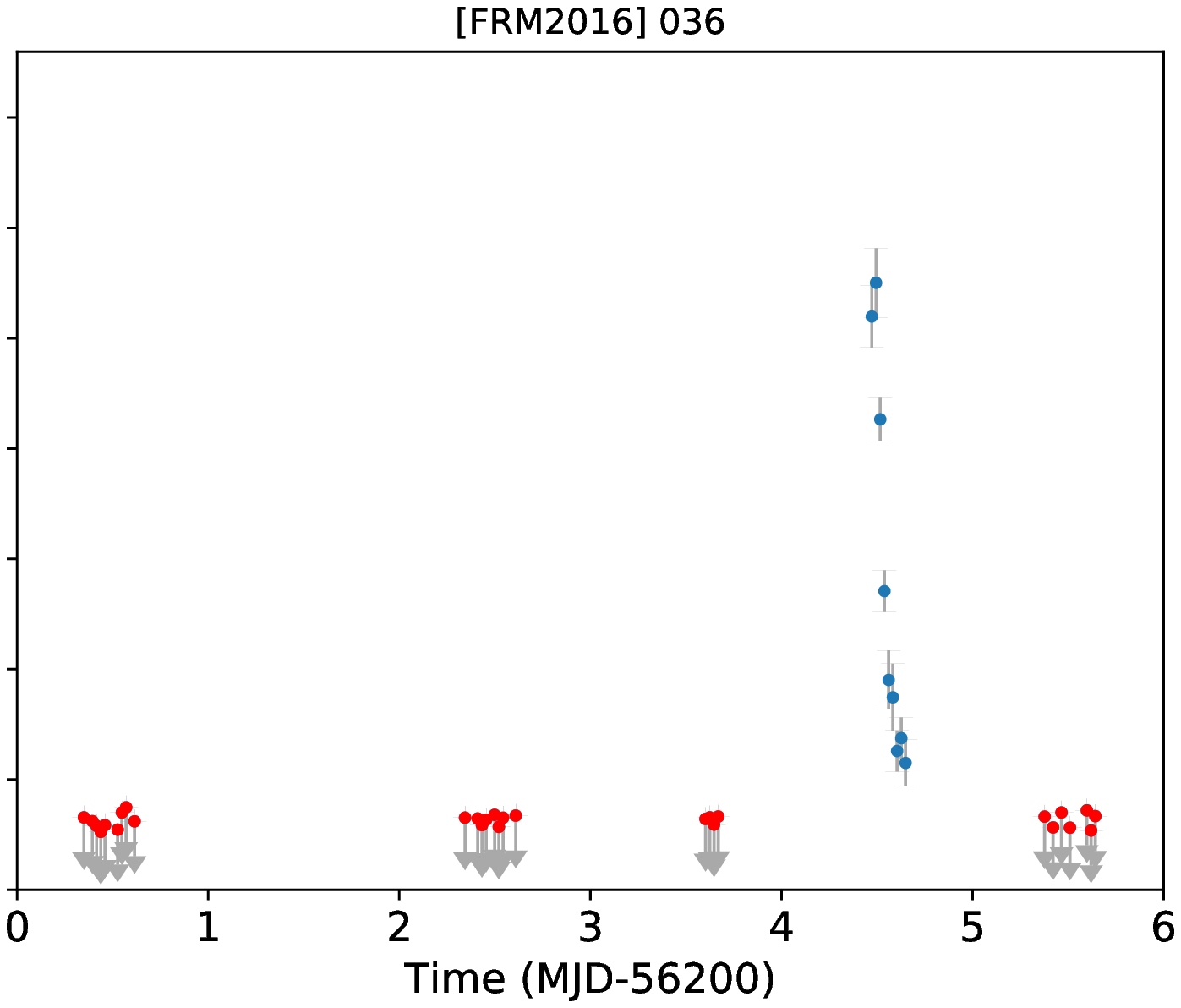}
\includegraphics*[bb=110 219 511 560, width=0.2330\linewidth]{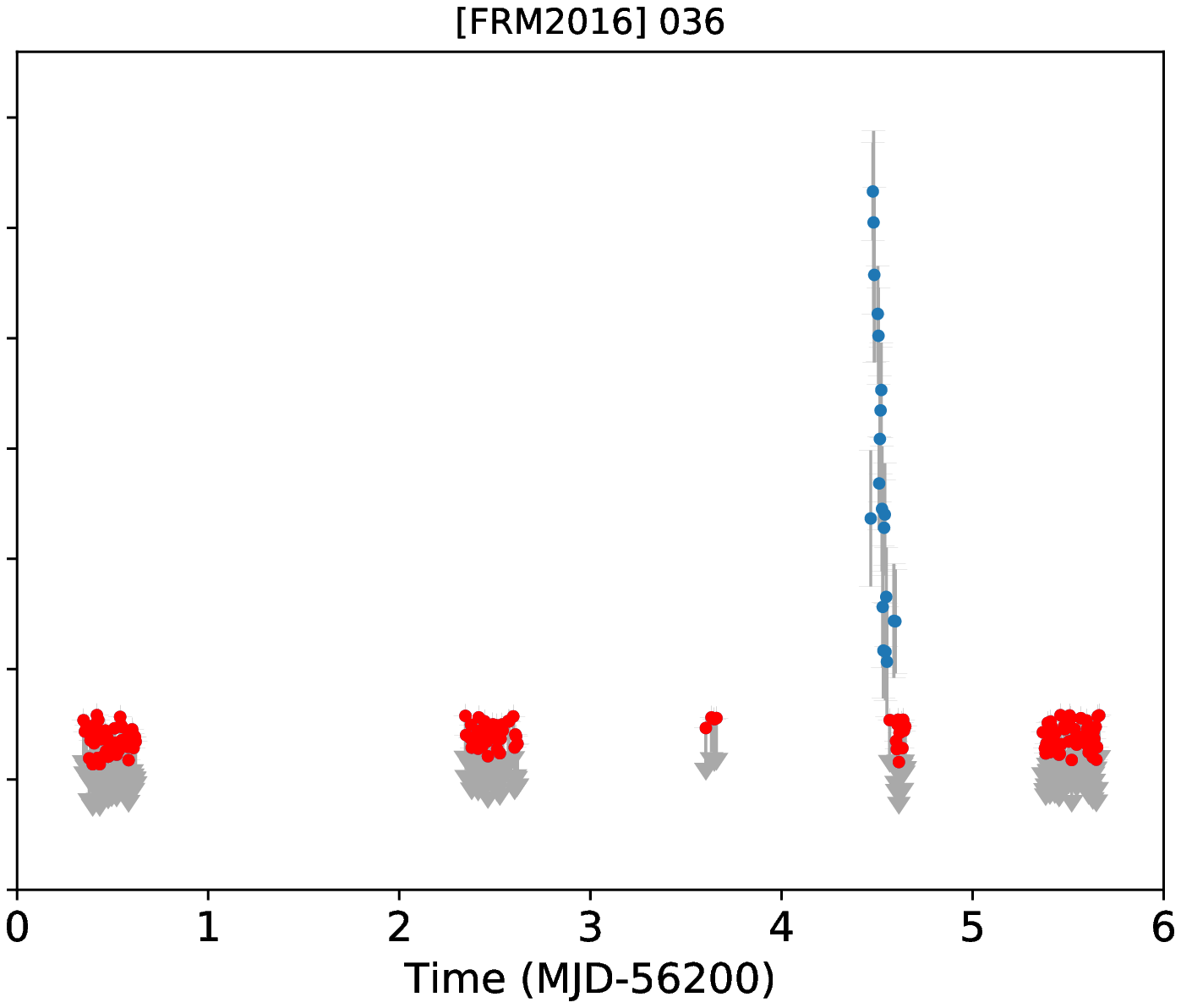}
\includegraphics*[bb=110 219 511 560, width=0.2330\linewidth]{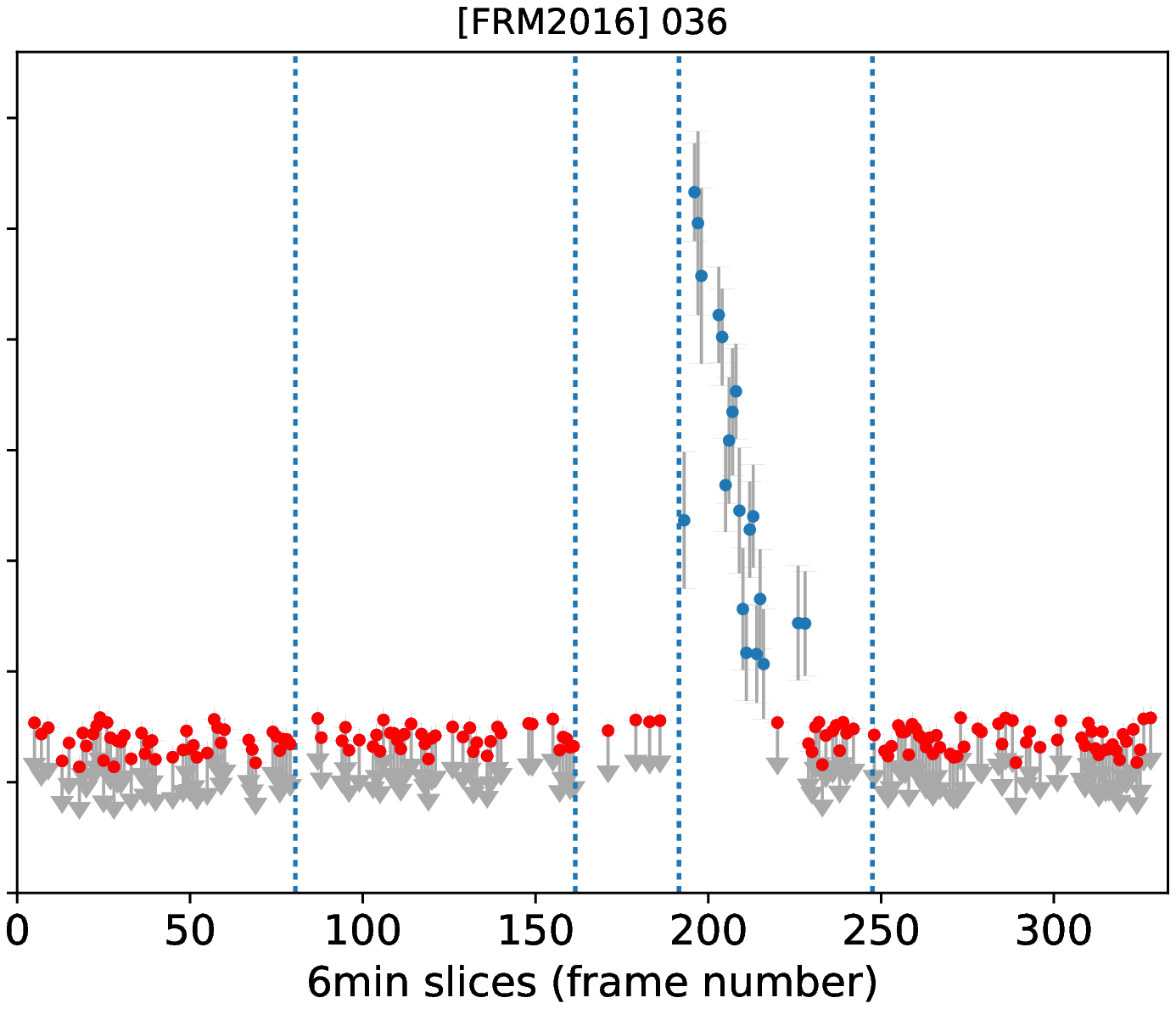}
\end{minipage} 
\vspace*{1mm} 
\begin{minipage}{\linewidth} 
\includegraphics*[bb= 60 219 511 560, width=0.2620\linewidth]{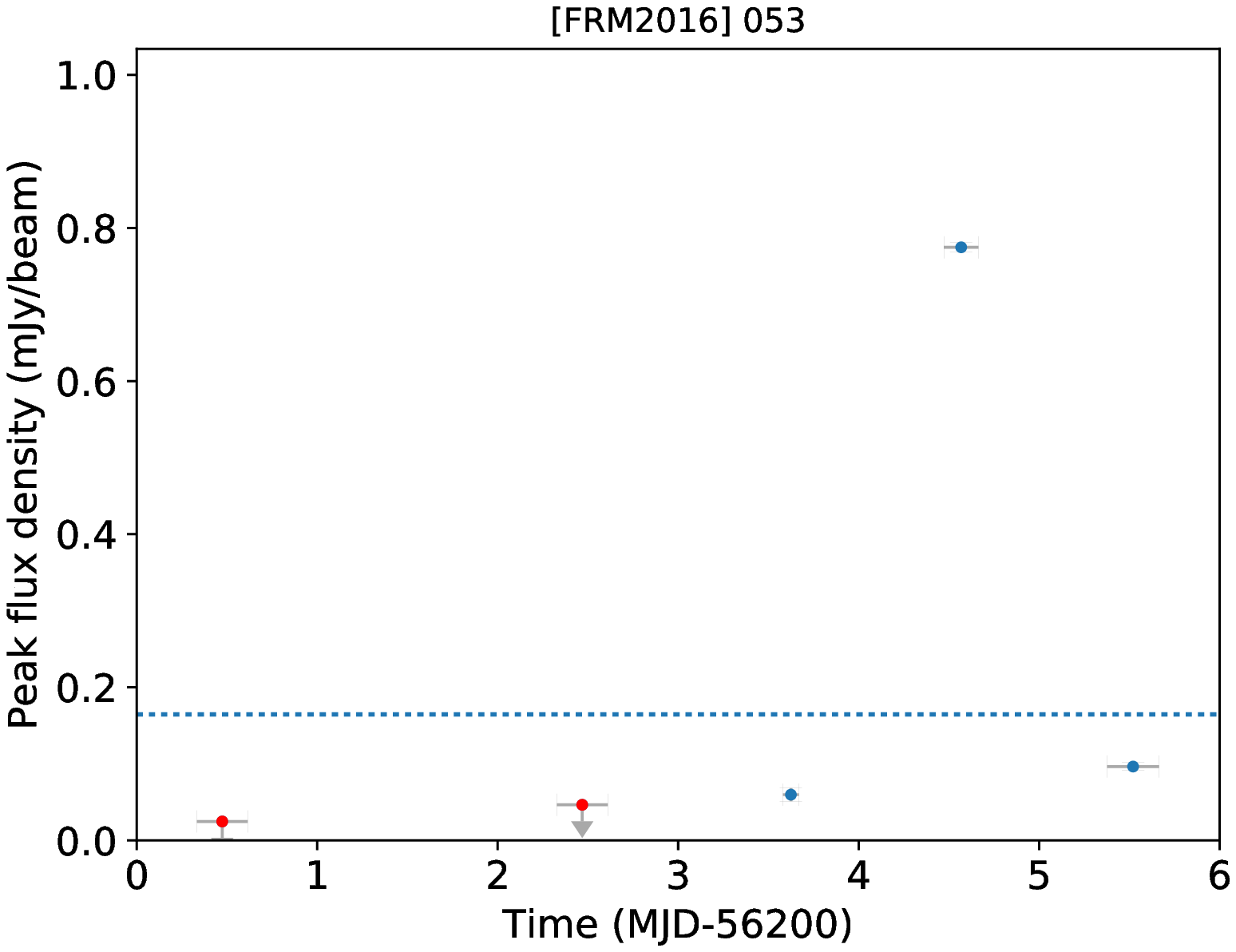}
\includegraphics*[bb=110 219 511 560, width=0.2330\linewidth]{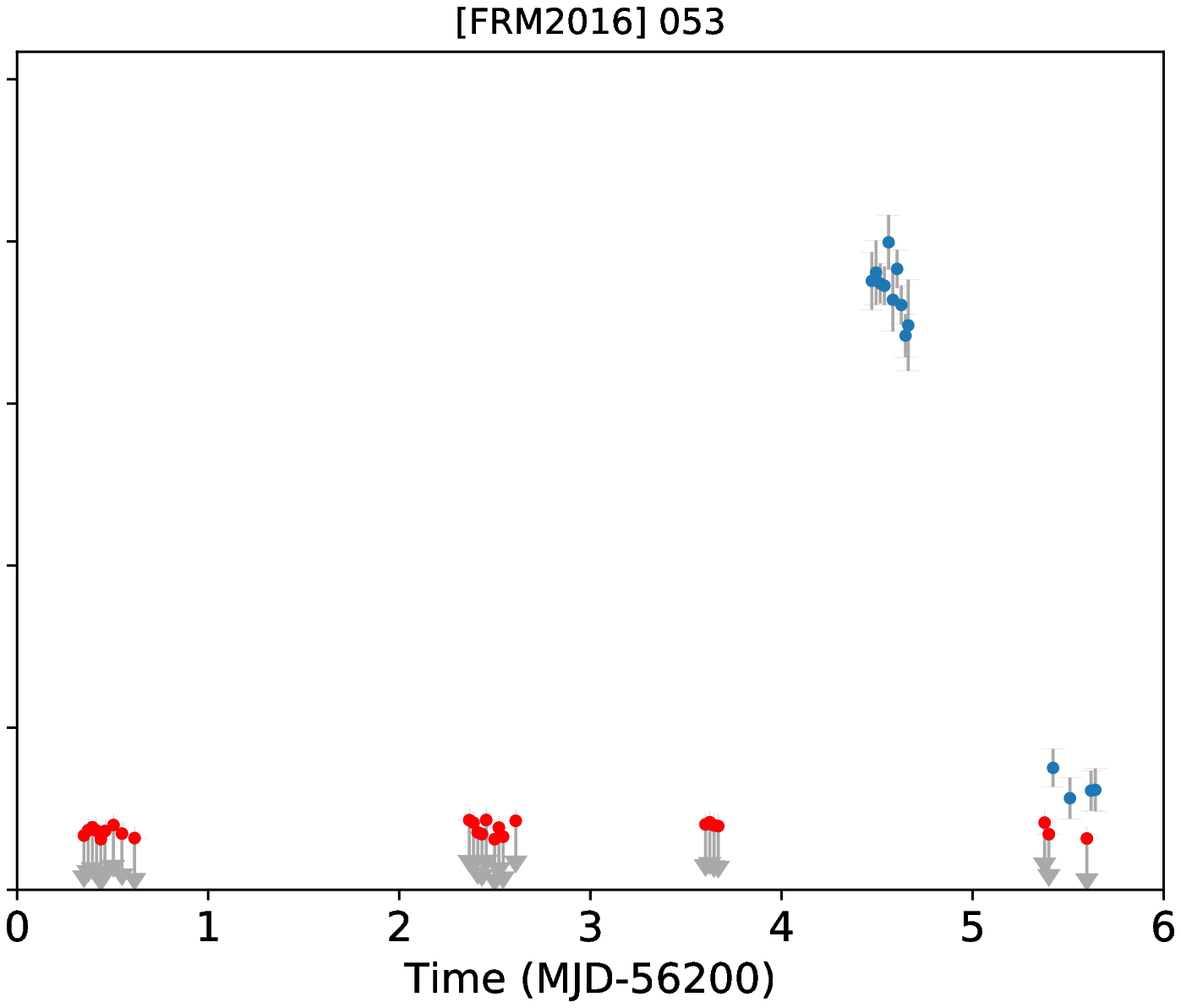}
\includegraphics*[bb=110 219 511 560, width=0.2330\linewidth]{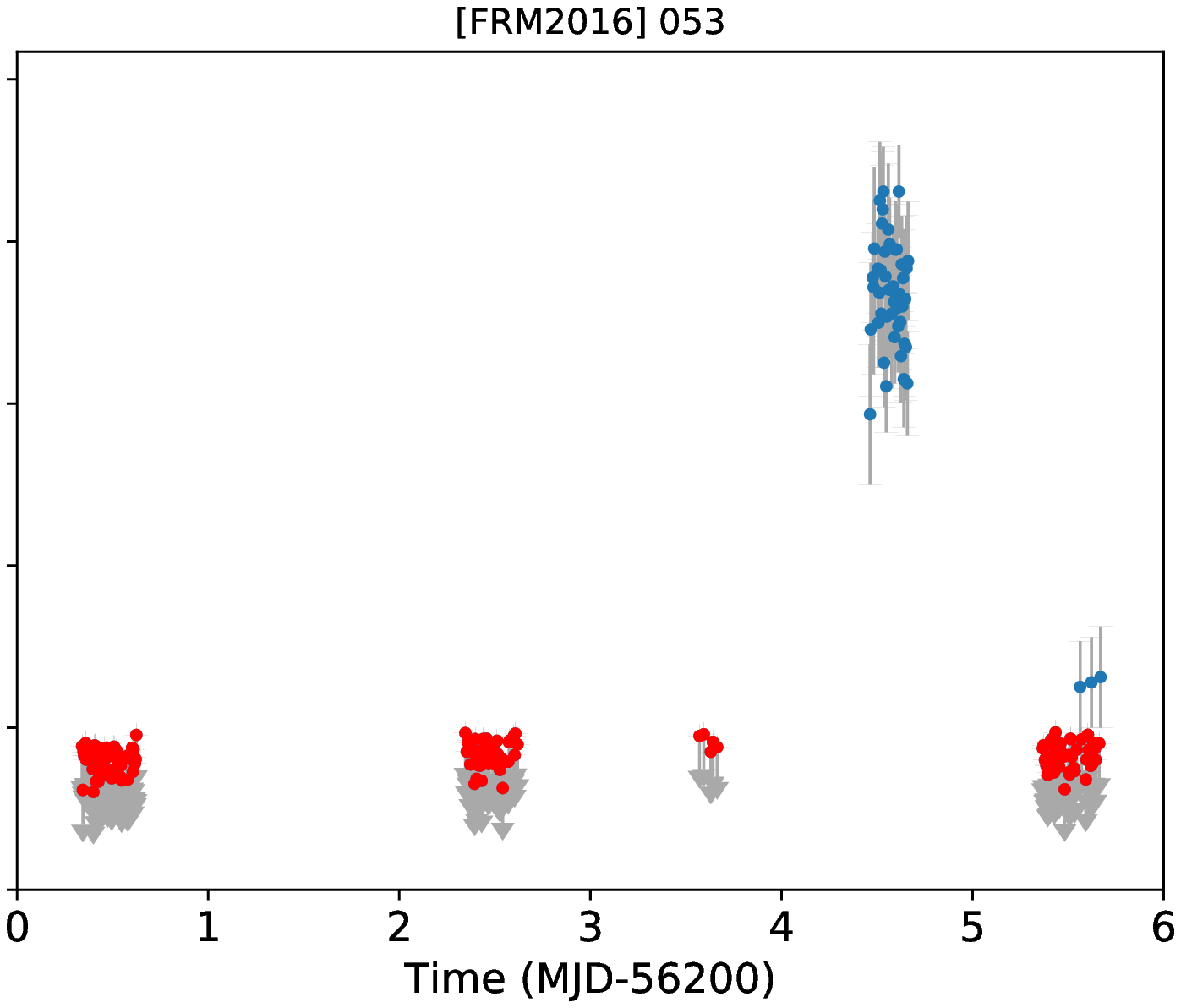}
\includegraphics*[bb=110 219 511 560, width=0.2330\linewidth]{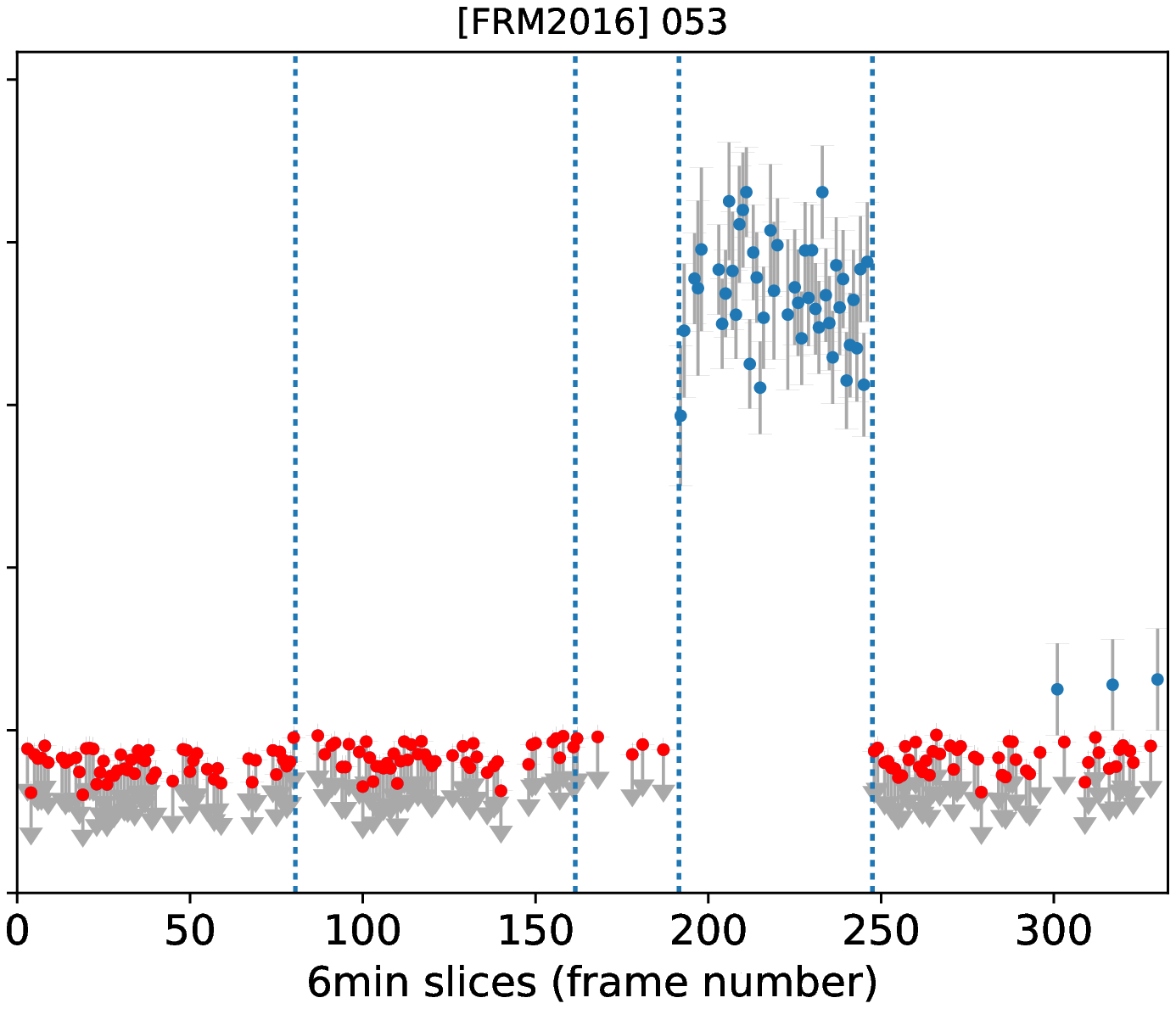}
\end{minipage} 
\vspace*{1mm} 
\begin{minipage}{\linewidth} 
\includegraphics*[bb= 60 219 511 560, width=0.2620\linewidth]{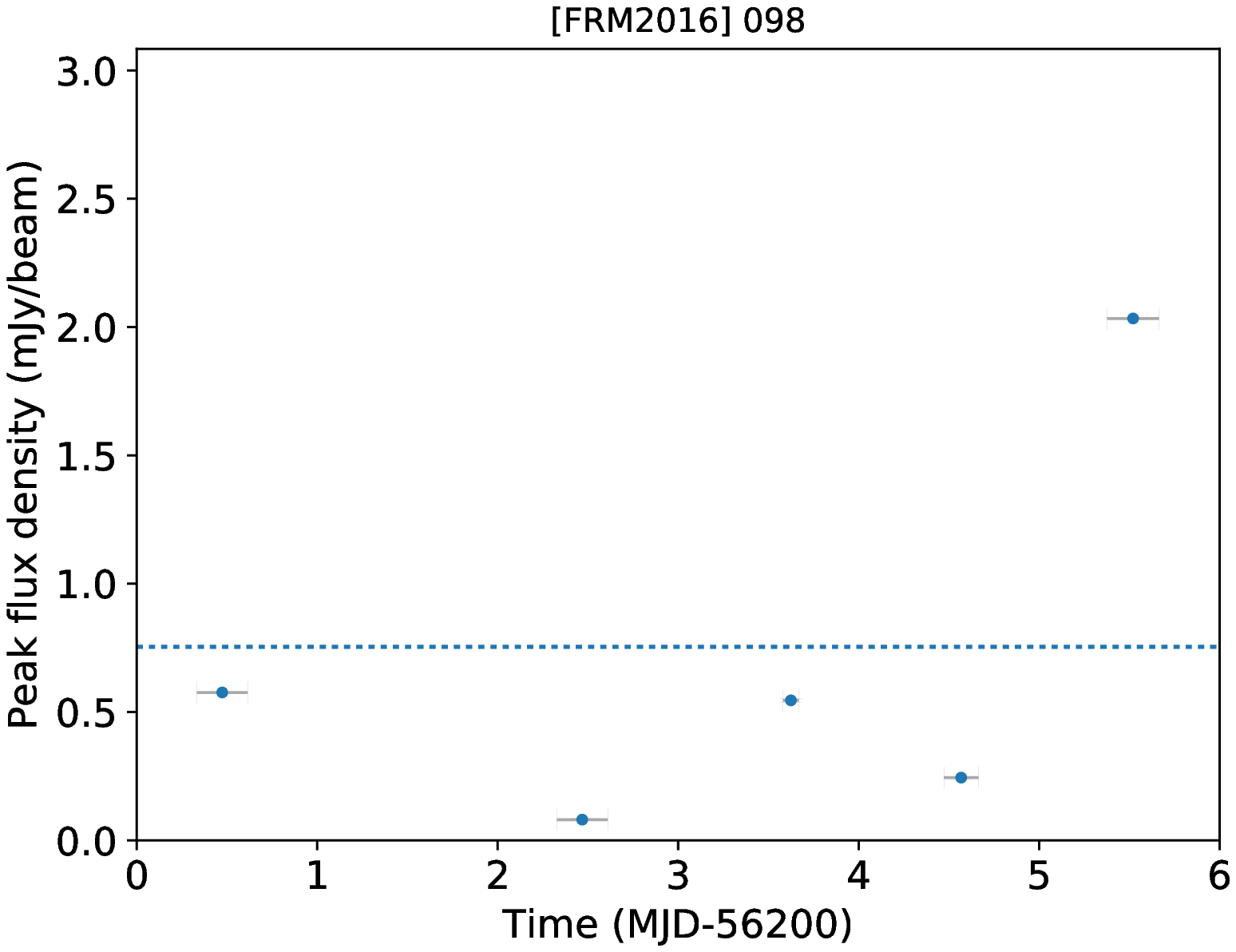}
\includegraphics*[bb=110 219 511 560, width=0.2330\linewidth]{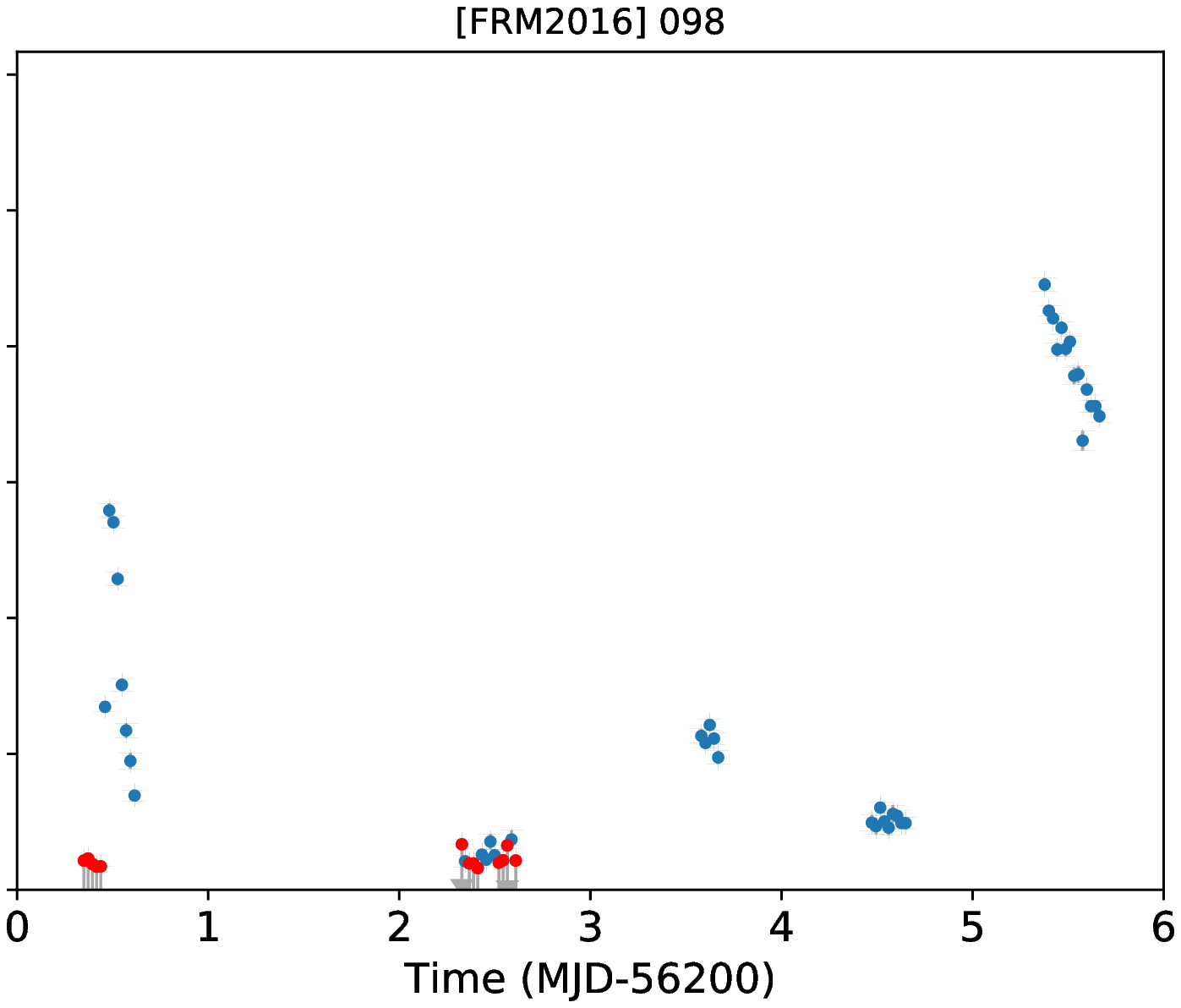}
\includegraphics*[bb=110 219 511 560, width=0.2330\linewidth]{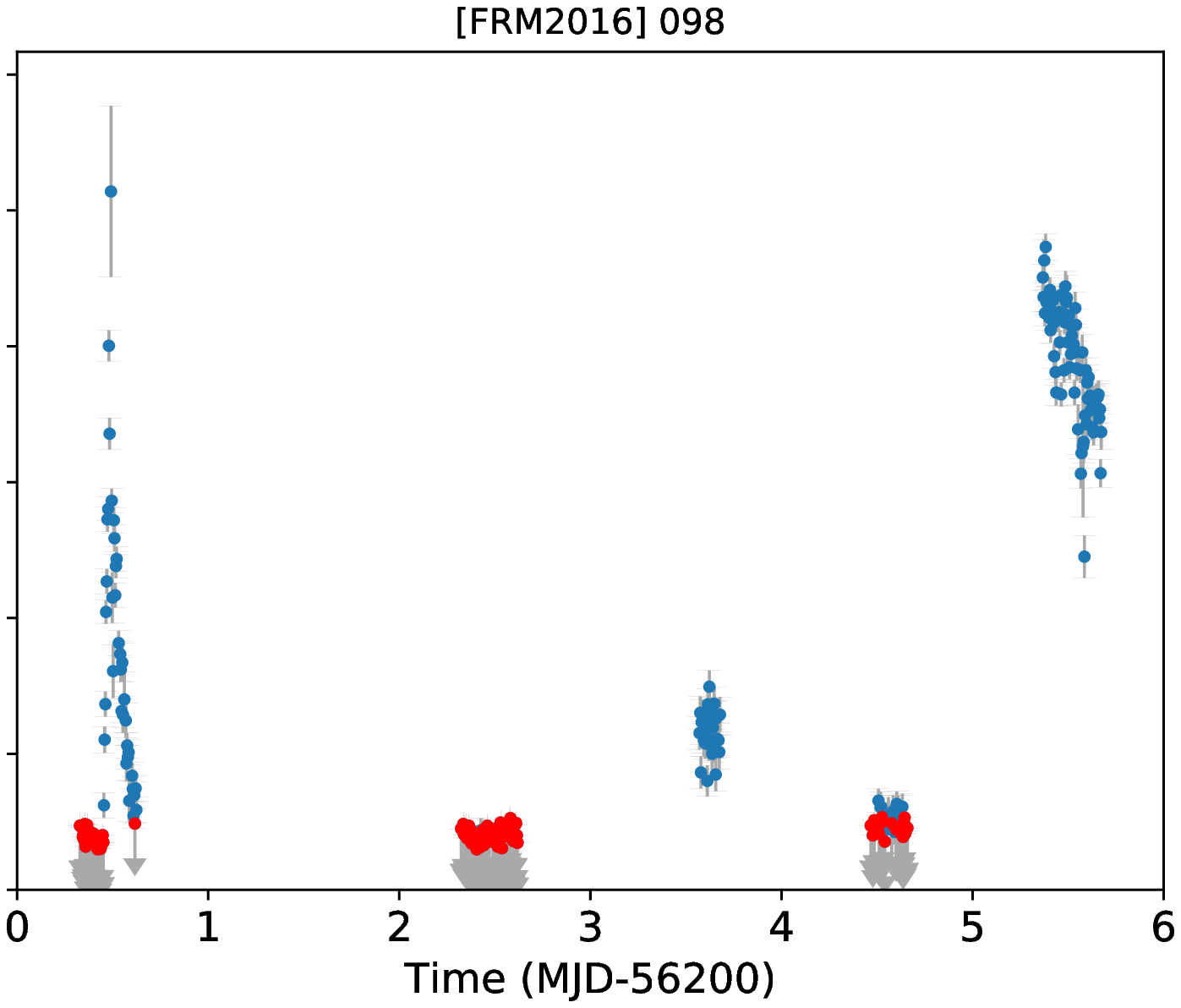}
\includegraphics*[bb=110 219 511 560, width=0.2330\linewidth]{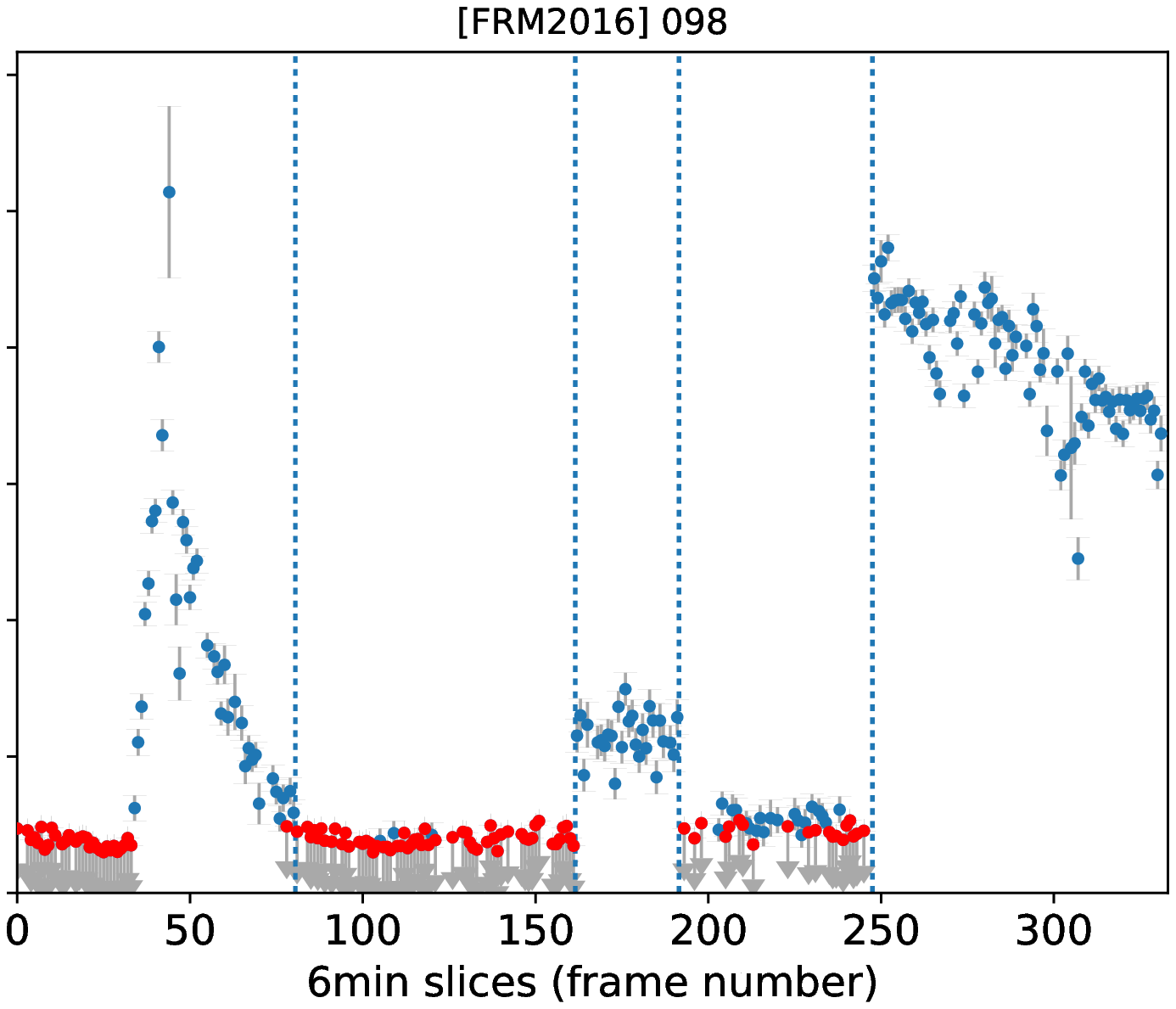}
\end{minipage} 
\vspace*{1mm} 
\begin{minipage}{\linewidth} 
\includegraphics*[bb= 60 219 511 560, width=0.2620\linewidth]{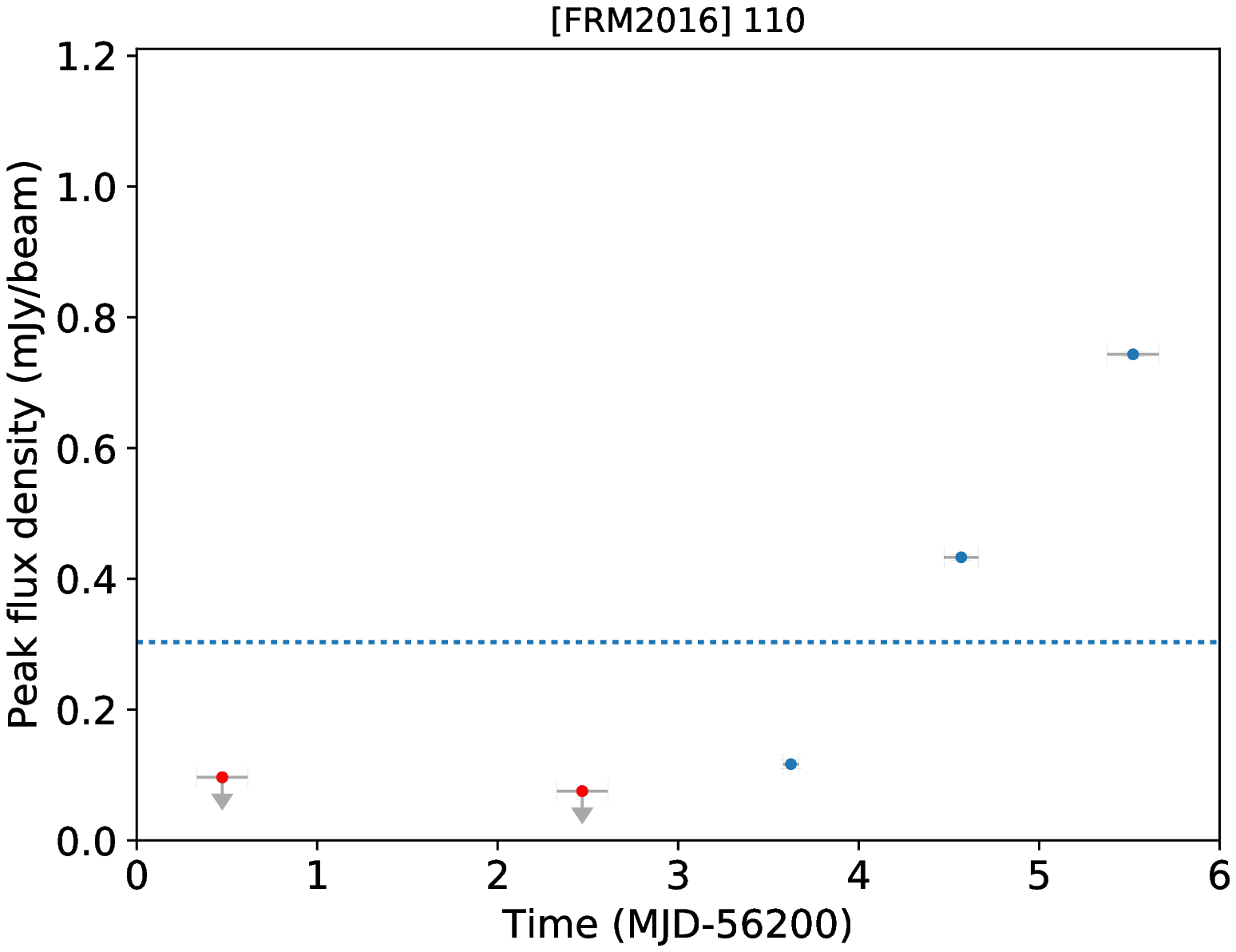}
\includegraphics*[bb=110 219 511 560, width=0.2330\linewidth]{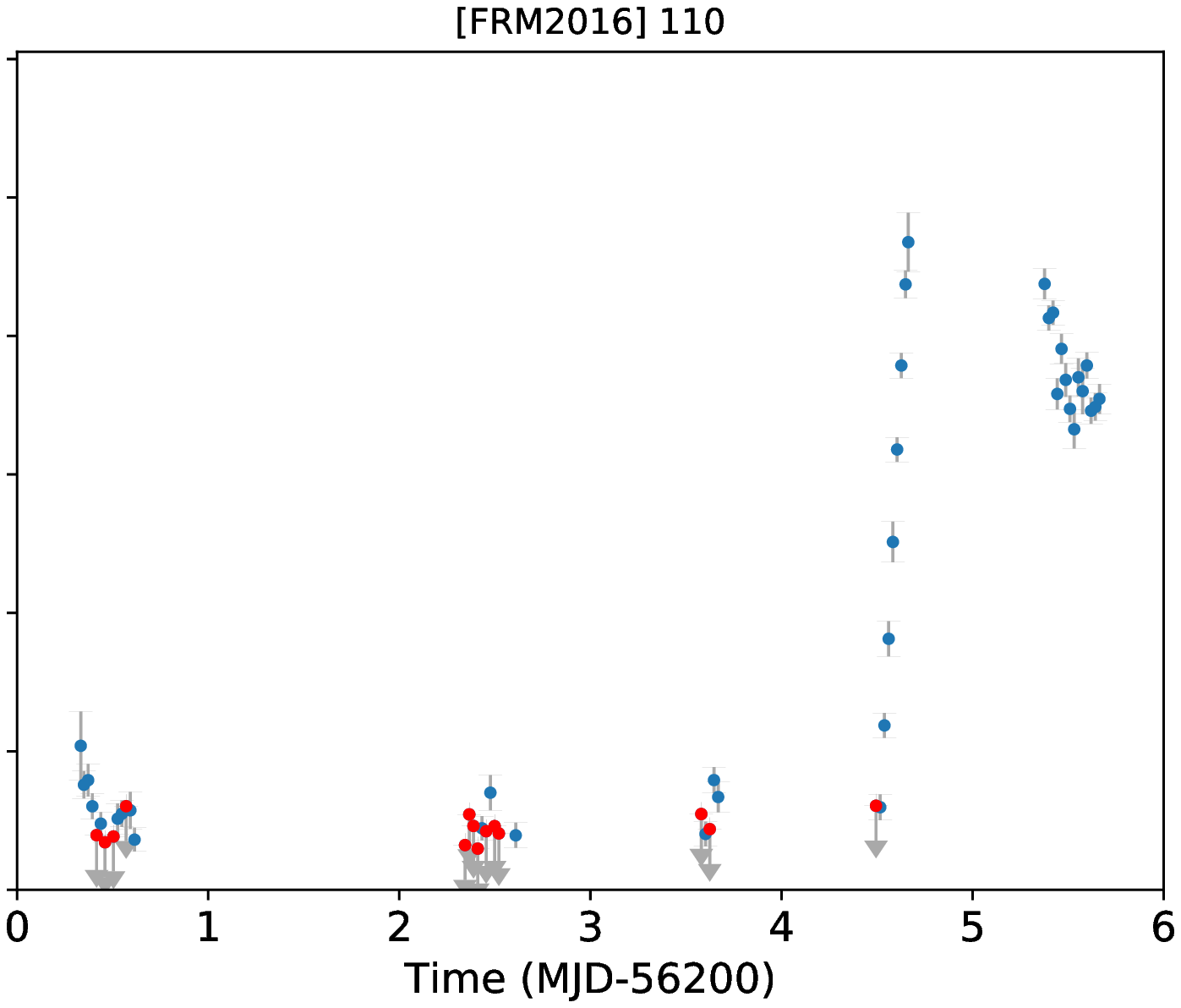}
\includegraphics*[bb=110 219 511 560, width=0.2330\linewidth]{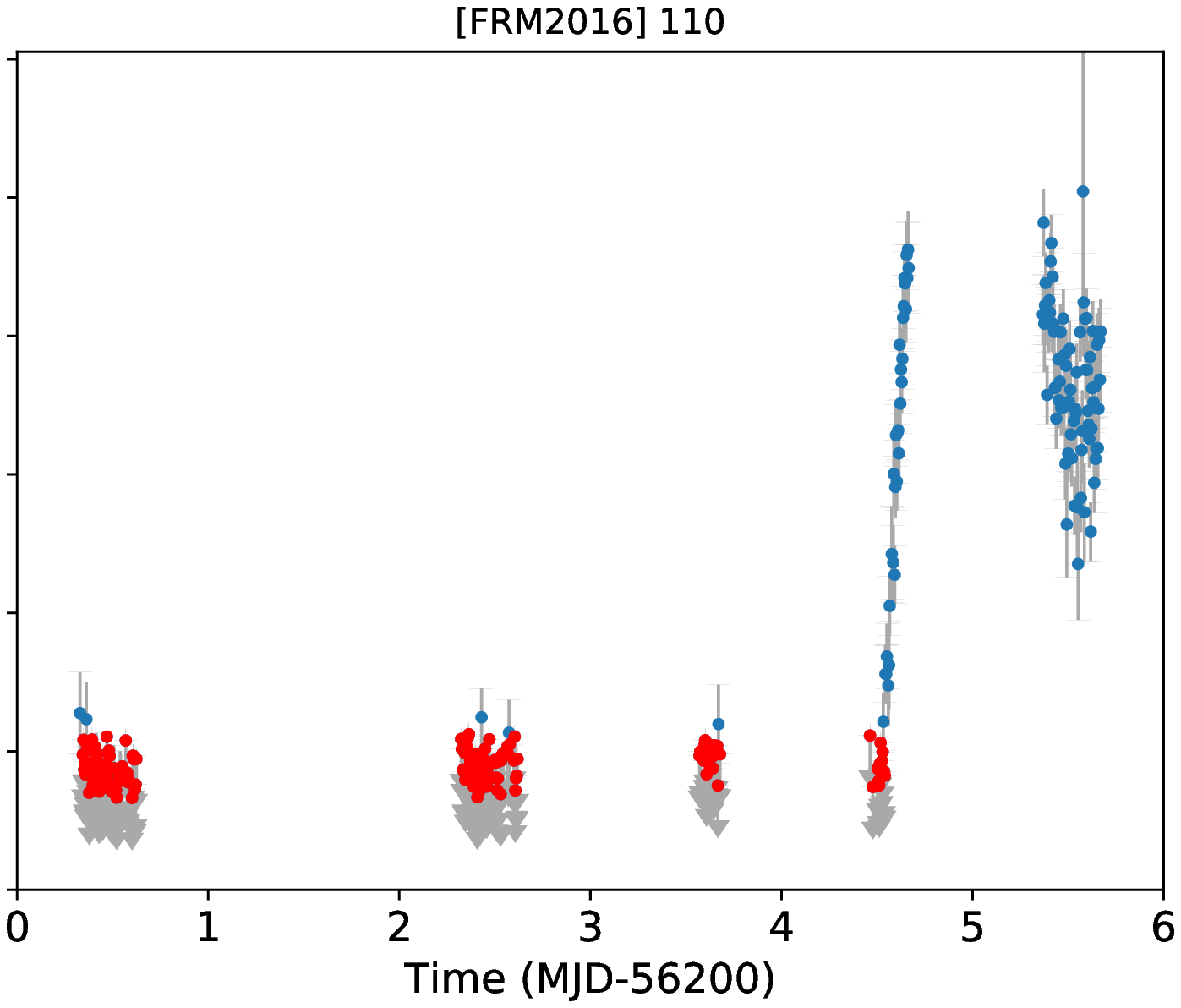}
\includegraphics*[bb=110 219 511 560, width=0.2330\linewidth]{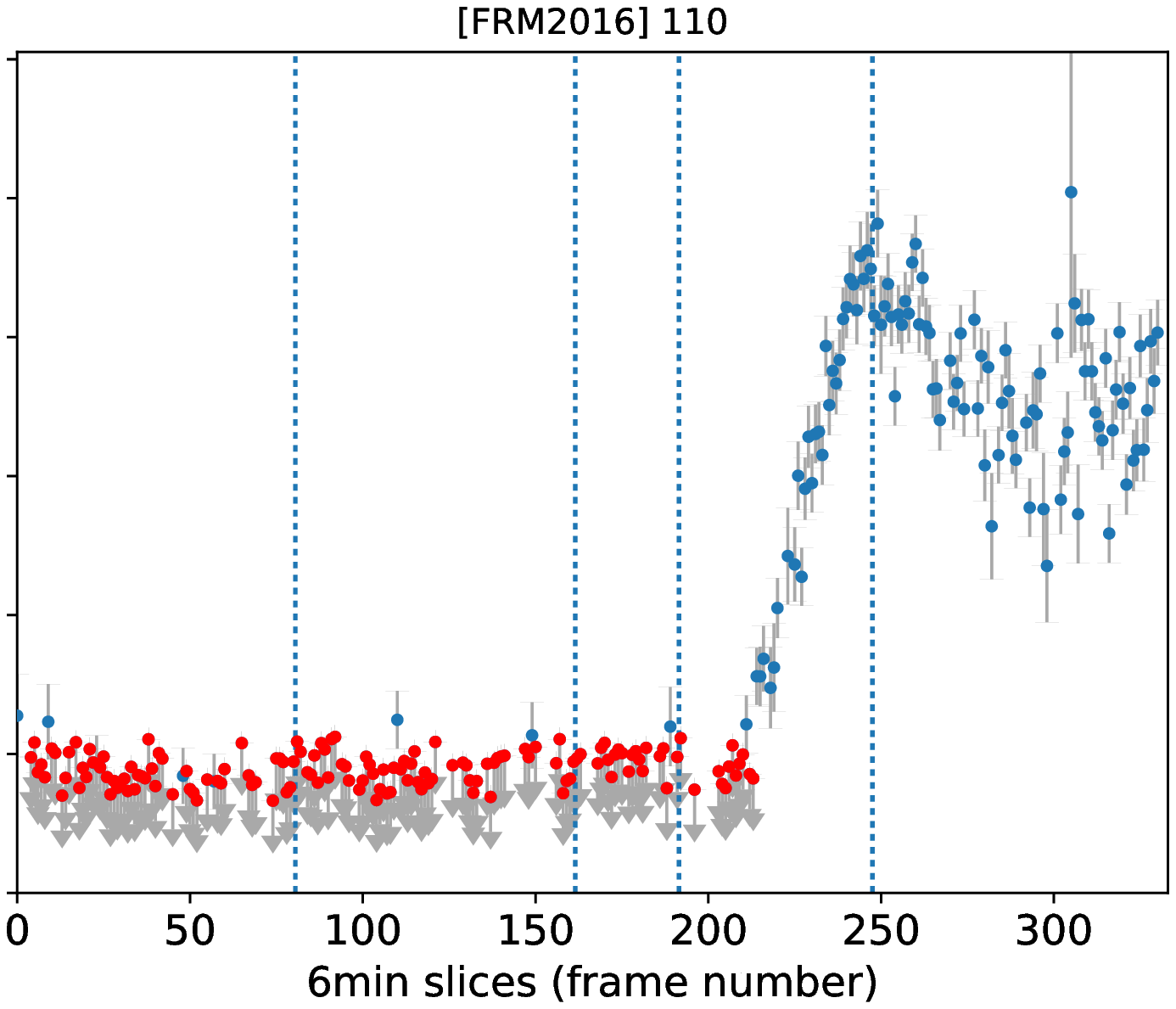}
\end{minipage} 
\vspace*{1mm} 
\begin{minipage}{\linewidth} 
\includegraphics*[bb= 60 219 511 560, width=0.2620\linewidth]{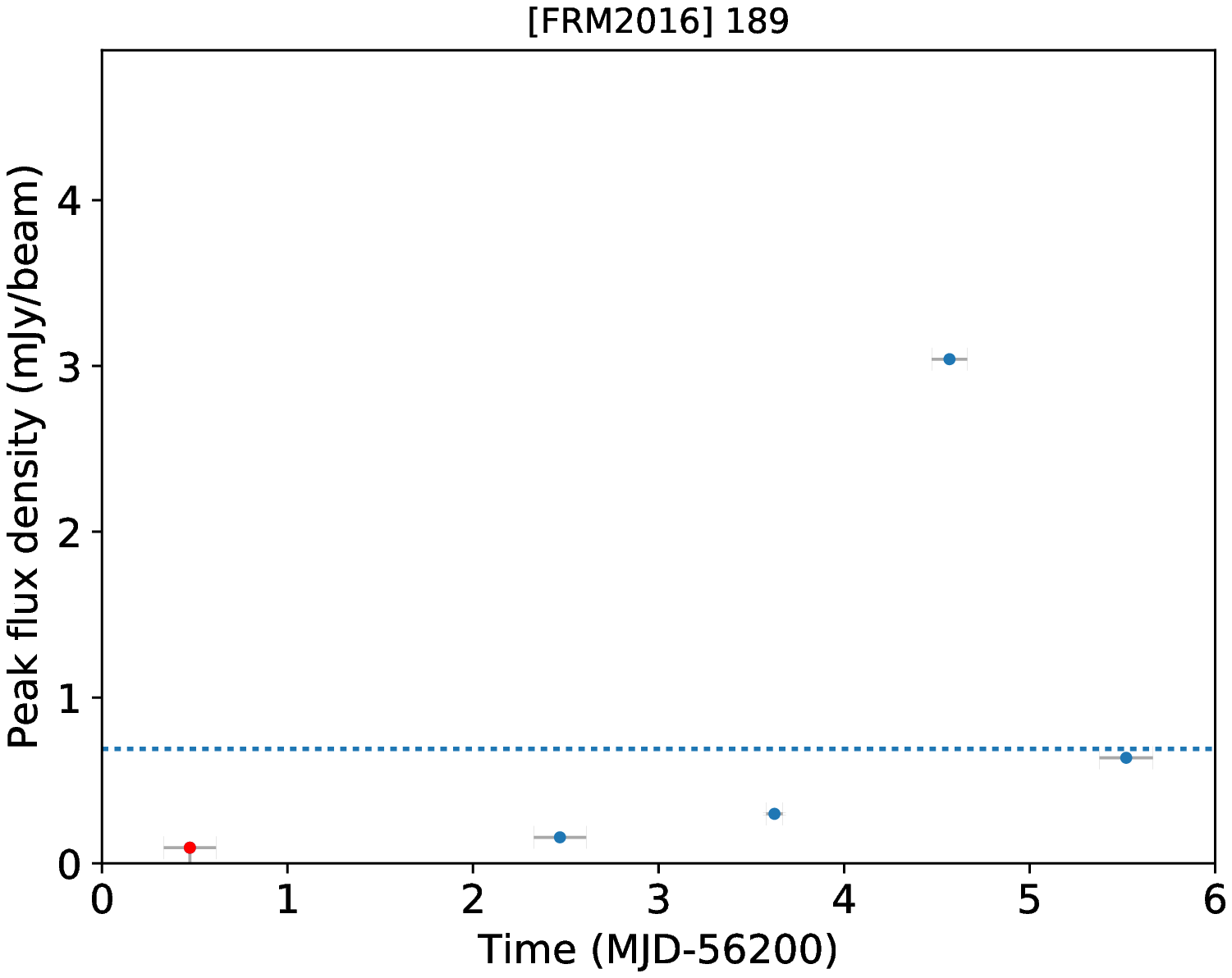}
\includegraphics*[bb=110 219 511 560, width=0.2330\linewidth]{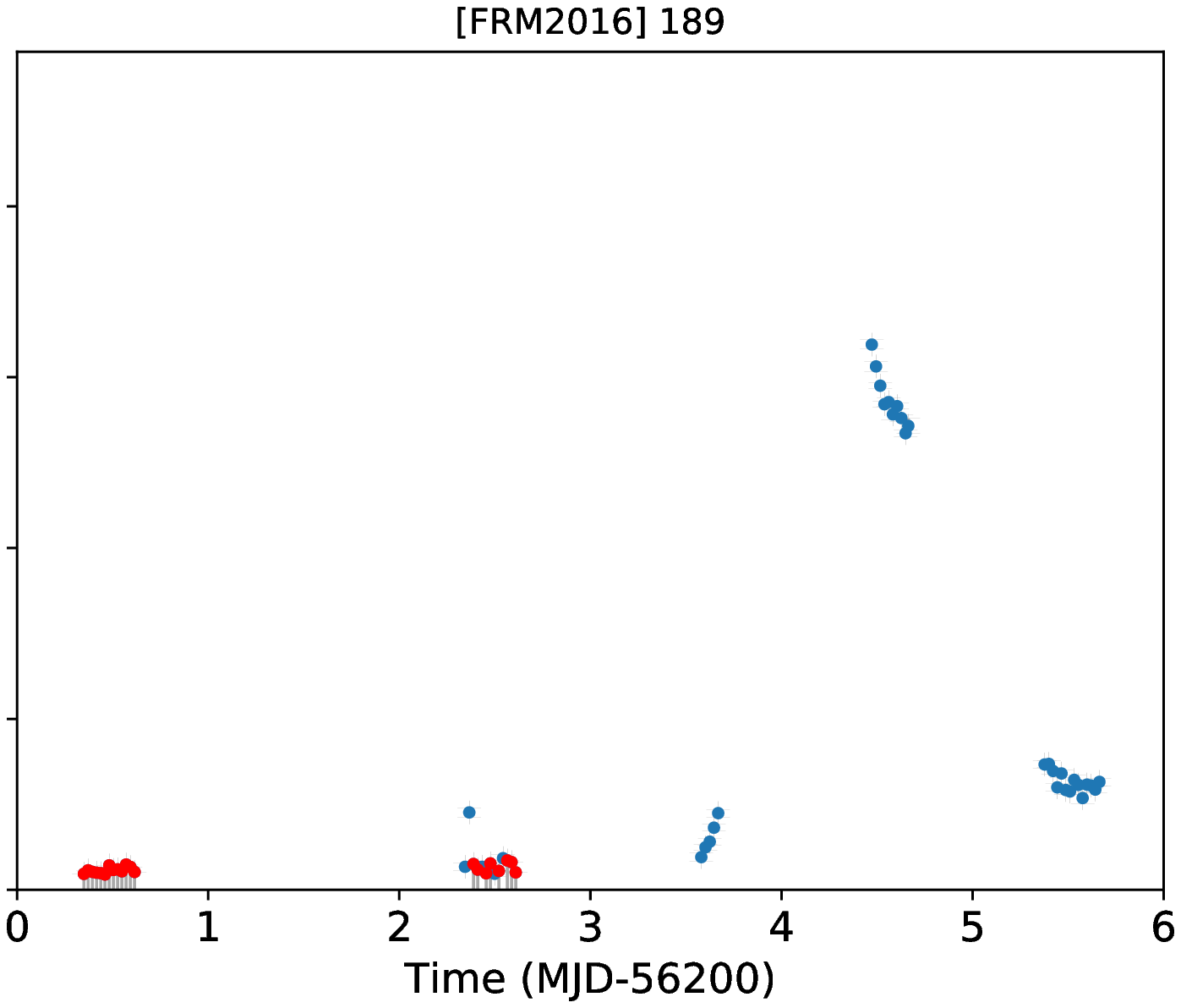}
\includegraphics*[bb=110 219 511 560, width=0.2330\linewidth]{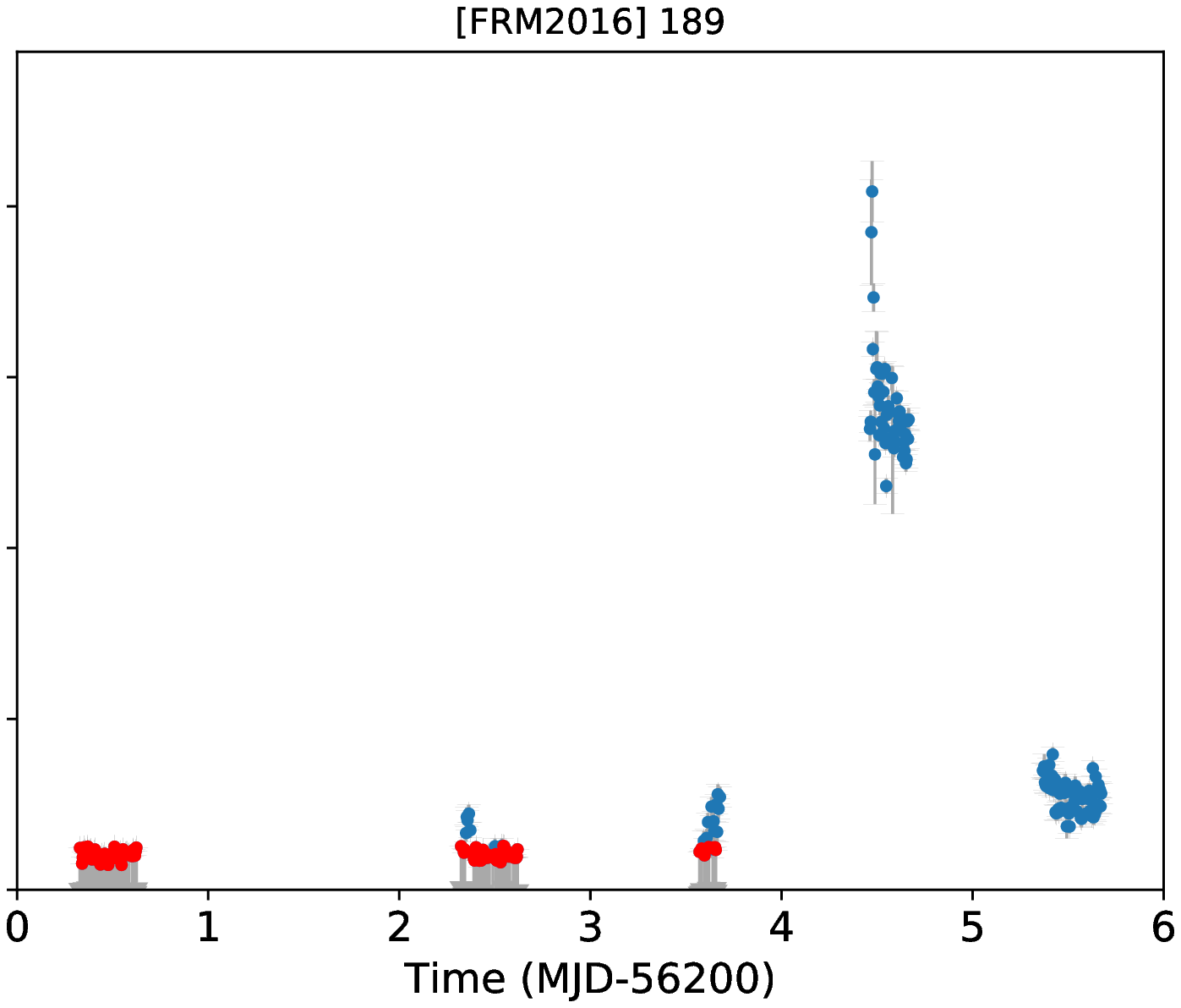}
\includegraphics*[bb=110 219 511 560, width=0.2330\linewidth]{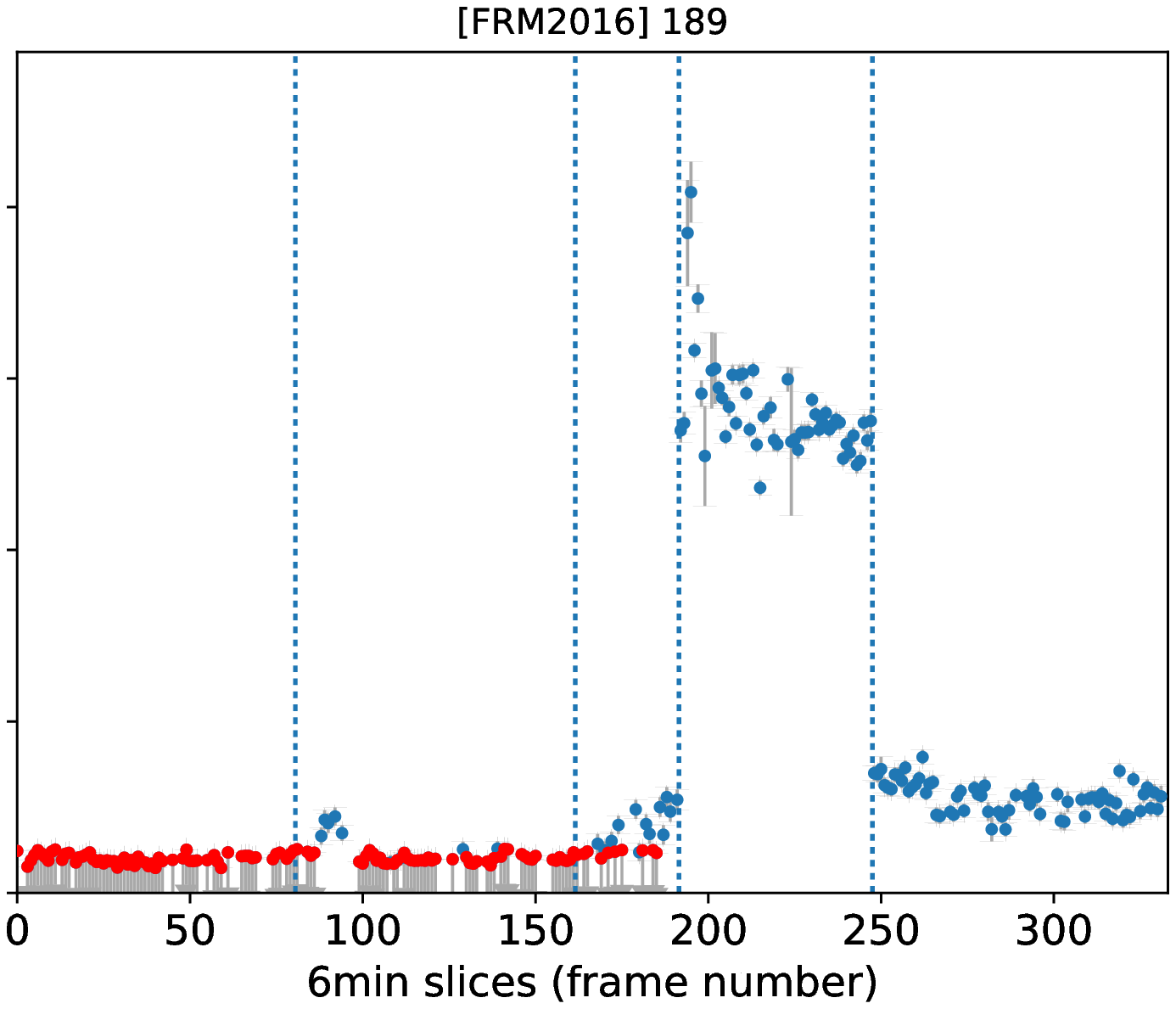}
\end{minipage} 
 
\caption{Radio lightcurve synopsis with peak flux densities corrected
for the primary beam response. Detections are marked in blue, errors
in grey, and upper limits in red. The left-hand panel compares the
peak flux density from the concatenated data (dotted line) with
the numbers from the individual epochs. The other three panels show
the data at 30~min resolution and at 6~min resolution, the latter
both with a time axis and sequential frames to highlight details.\label{fig_var_all}}
\end{figure*}

In total, we detect eight extreme variable sources with 6-min integrations.
Only one of these (number 414) had not been found in the longer time 
integrations.  This brings the total count of extreme variable sources to 13.  
For three sources (254, 319, 469), the amplitude of variability is greater 
than determined from the longer integrations, and more such cases may be 
hidden in cases where flux density upper limits are involved. 
The lightcurves of some sources show spectacular detail; sources 98, 469, and 
515 show strong flares  (Figure~\ref{fig_var_all}), with flux density changes 
by more than a factor of 10 over 0.4 to 0.7~h (see Table~\ref{tab_exvar}). 
Source 515 shows the strongest flare, with a peak shortly before the end of 
the observations.  The remaining sources show irregular variability, and many 
remain undetected over entire multi-hour epochs at this timescale.
A visualization of the selection process is shown in Figure~\ref{fig_varhist2}.
 
\subsection{Individual 90~sec scans} 
 
\begin{figure} 
\includegraphics[width=\linewidth]{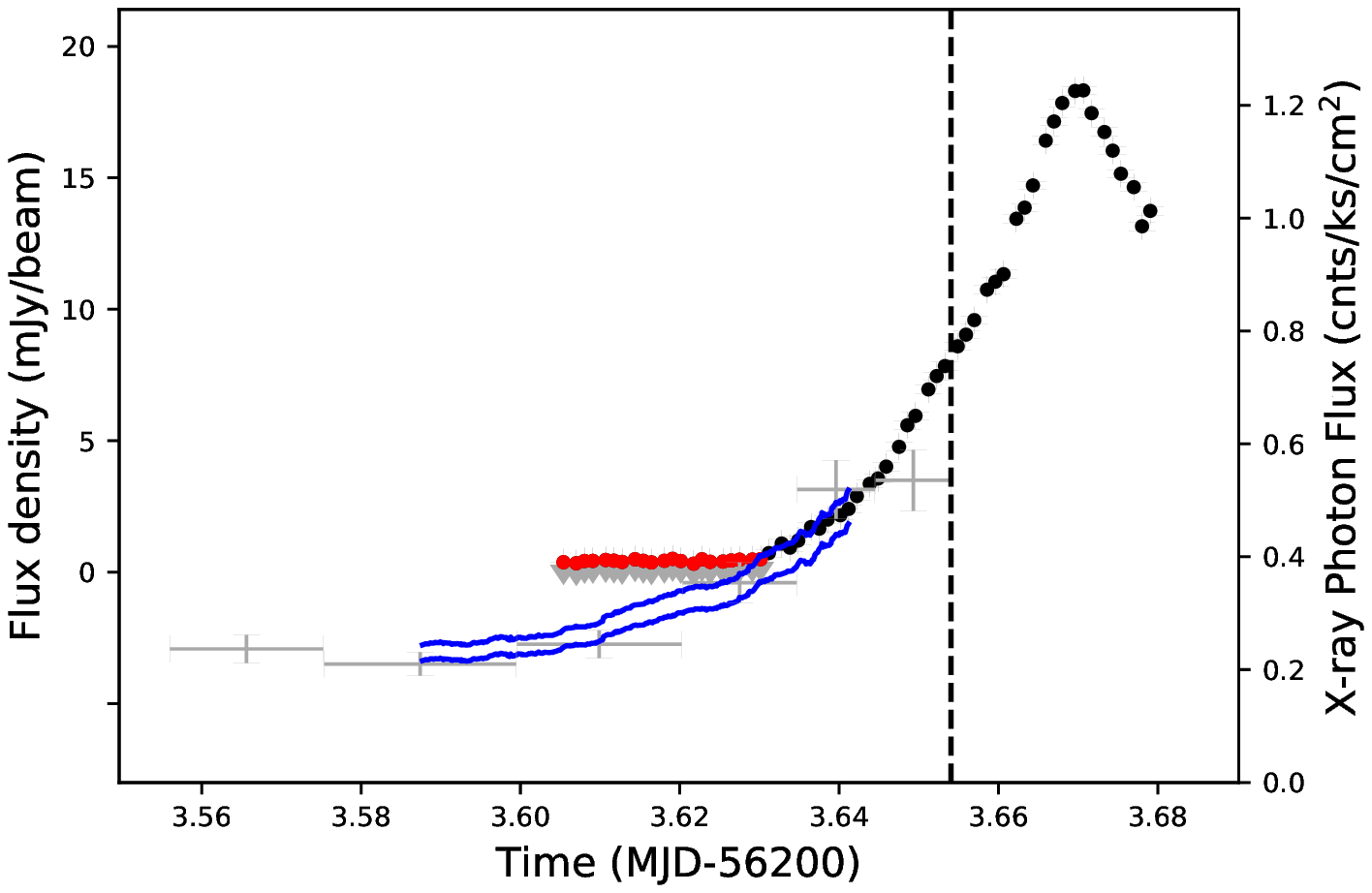}
\caption{Radio lightcurve of source 515, at a time resolution of
90~sec, showing only the flare in epoch 3, with data points in black
and upper limits in red (left axis). Shown additionally, in blue,
are the confidence intervals of the adaptively smoothed X-ray light
curve from acis\_extract (right axis; same as in Figure~\ref{fig_var_all}),
and a binned version of the same lightcurve with 100 counts per
bin (90 in the first bin). To show that both lightcurves likely
trace the same event, both scales have been linearly shifted with
respect to each other so that the lightcurves match up. The vertical
dashed line indicates the end of the {\it Chandra} observation. \label{fig_src515_1min}}
\end{figure}

To study the bright flare of source 515 in the third epoch in more
detail, we have imaged 60 individual 90~sec scans, and the light
curve is shown in Figure~\ref{fig_src515_1min}.
The flux densities are marginally higher than with a resolution of
6~min , and a turnover before the end of the observation now is 
visible more clearly than at a resolution of 6~minutes.

\subsection{Occurrence rate of extreme radio variability} 

We now estimate the occurrence rates of extreme radio variability
of Orion YSOs on different timescales. While our search has targeted
all 507 sources catalogued in \citet{for16} that lie within a more 
limited field of view, that full source count is not the best 
reference sample since, as we have discussed in that paper, several
of the radio sources are likely non-stellar. Assuming that the most
complete sample of YSOs in the inner ONC is the COUP X-ray sample
\citep{get05b} and noting that all 13 radio flare sources identified
here indeed have COUP X-ray counterparts, our reference sample thus
consists of those 222 radio sources among the 507 studied here that 
have X-ray counterparts.

From the cumulative radio observing time of these sources 
of about $222\times30=6,660$~hours, we find only three flares with a 
flux density change of more than a factor of 10 on timescales 
less than hour. Assuming a detectable flare duration of 2~h, 
we thus estimate that each of these sources displays on average such 
a flare only $\approx0.09\pm0.05$\% of the time, 
assuming a Poisson error in the flare count of $\pm\sqrt{3}$. 
Thus, the mean time between such flares is $2220\pm1280$~h, or 
$8.0\pm4.6$~Msec, or roughly every three months.  This is, of course, 
only a crude estimate, since it assumes a homogeneous underlying sample.

On time scales longer than $\approx1$~h, we have identified 13 sources with
order-of-magnitude changes in flux density on timescales of several hours to
days (although this number is uncertain because of unavoidable gaps in the 
observing schedule).  Thus, extreme variability on these longer timescales
seems to occur roughly once every three weeks. 
In the context of variability on longer timescales, we note finally that 
of the 13 variable sources that we identify here, a total of nine sources 
have been detected in previous VLA observations (see Table~\ref{tab_exvar}),
albeit with different sensitivity limits.

\begin{figure*} 
\begin{minipage}{\linewidth} 
\includegraphics*[bb= 60 219 511 560, width=0.2620\linewidth]{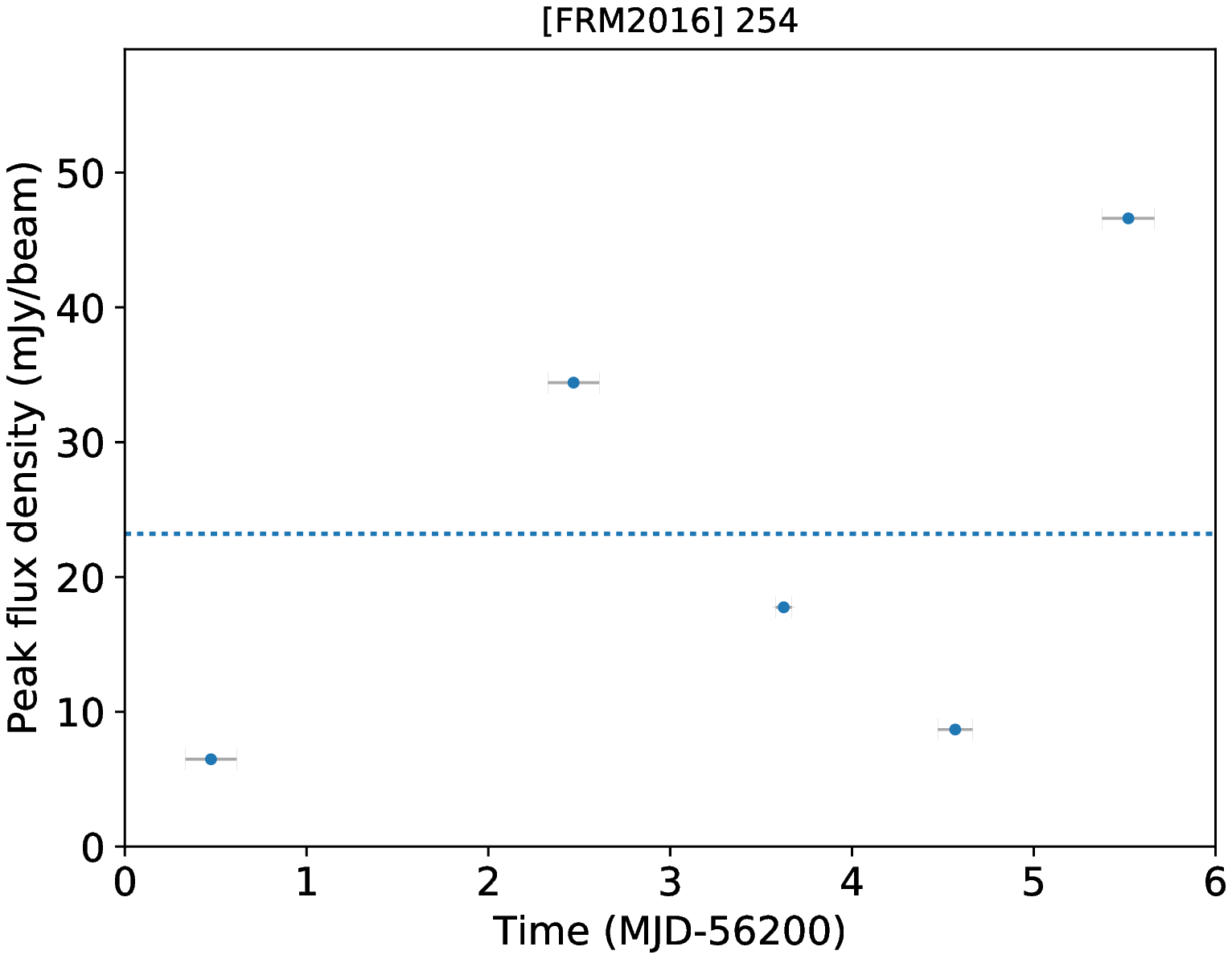}
\includegraphics*[bb=110 219 511 560, width=0.2330\linewidth]{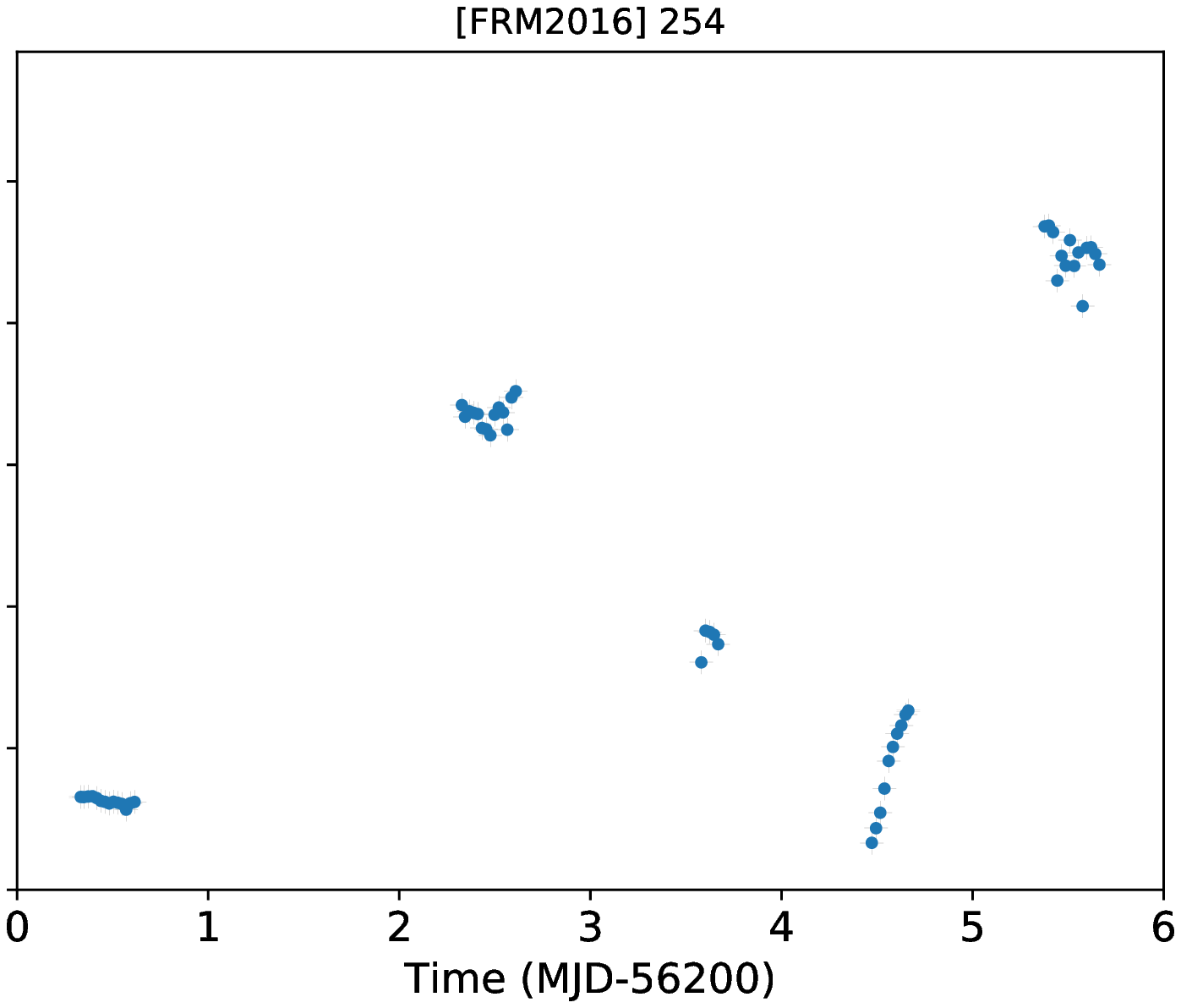}
\includegraphics*[bb=110 219 511 560, width=0.2330\linewidth]{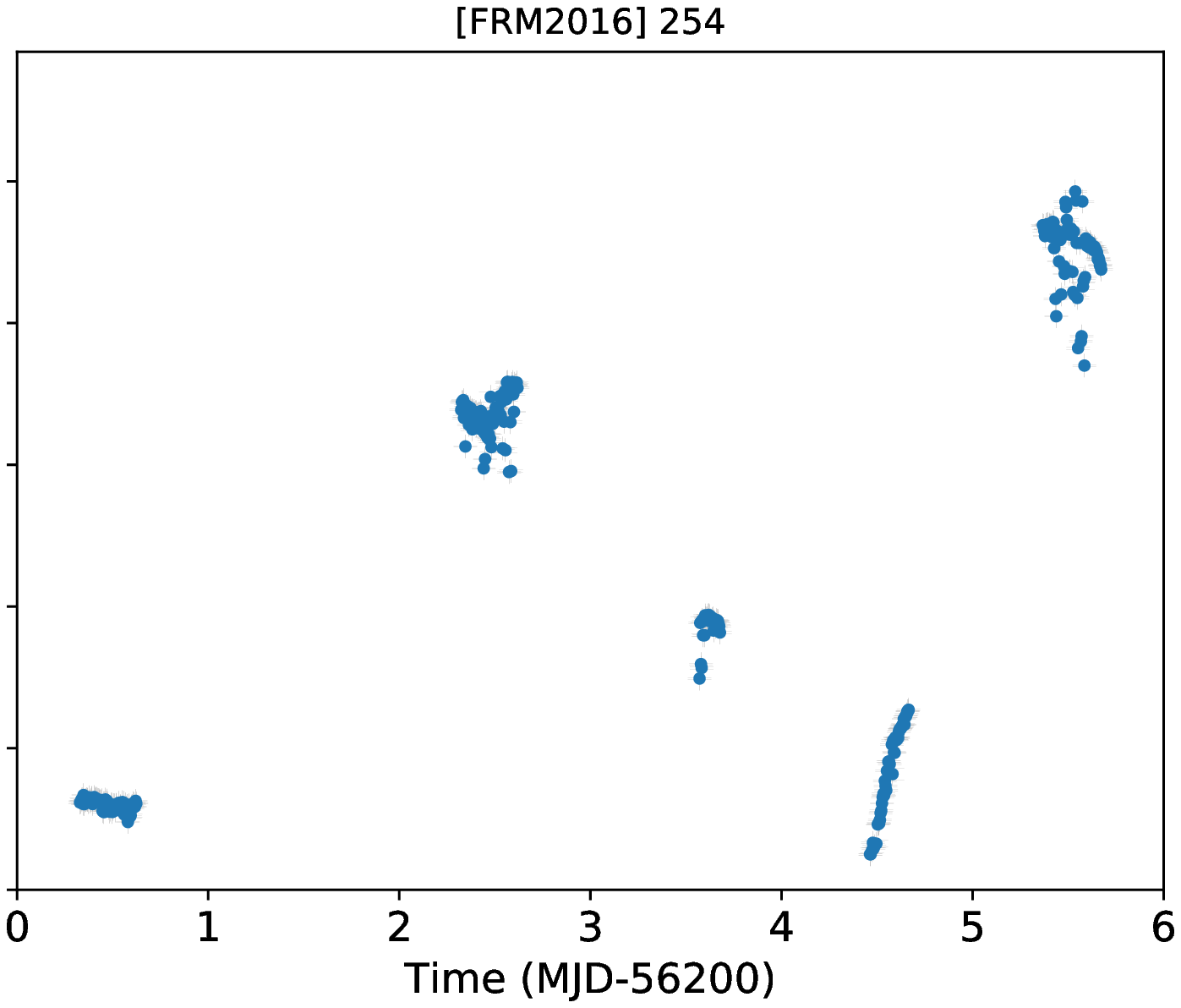}
\includegraphics*[bb=110 219 511 560, width=0.2330\linewidth]{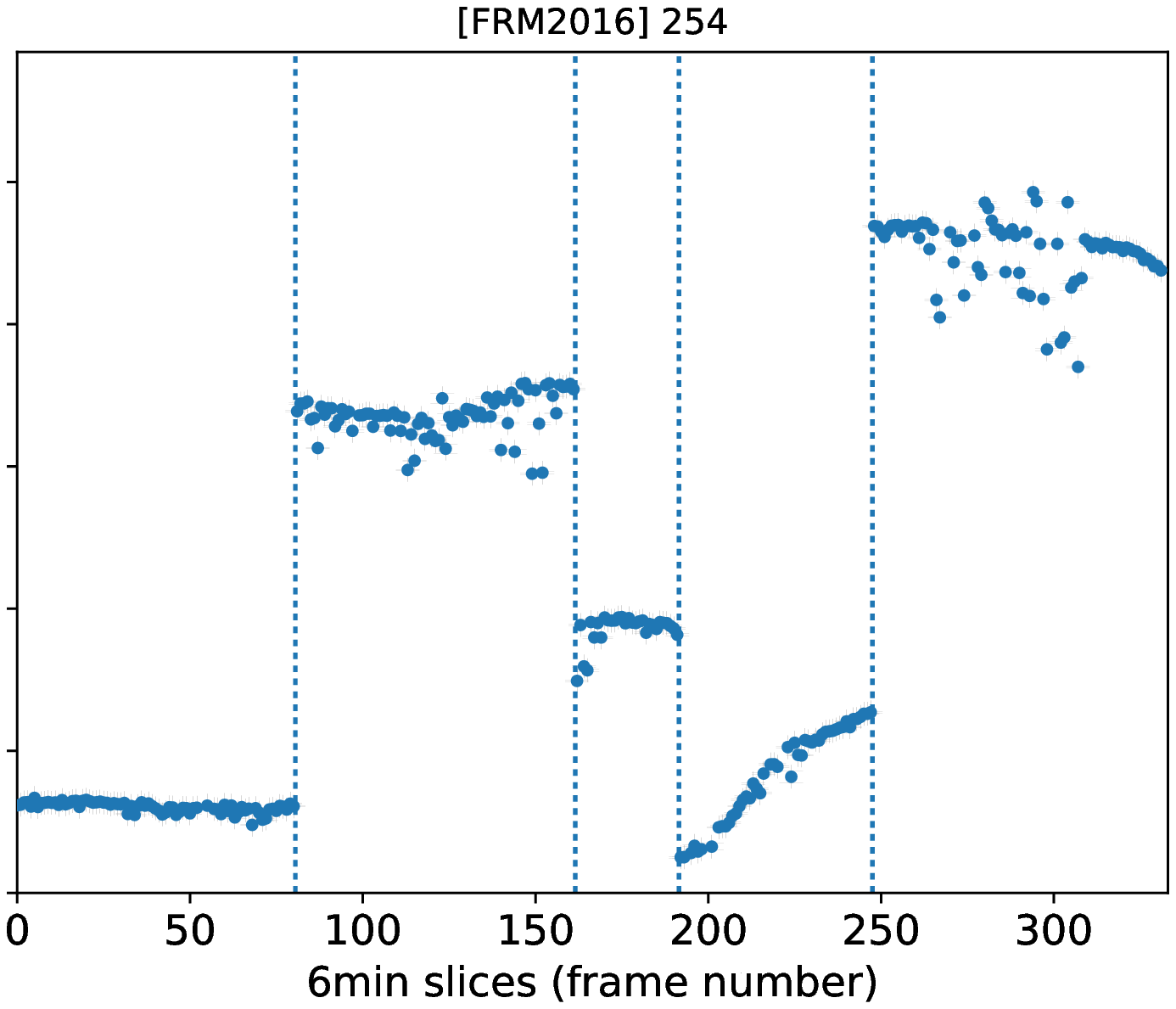}
\end{minipage} 
\vspace*{1mm} 
\begin{minipage}{\linewidth} 
\includegraphics*[bb= 60 219 511 560, width=0.2620\linewidth]{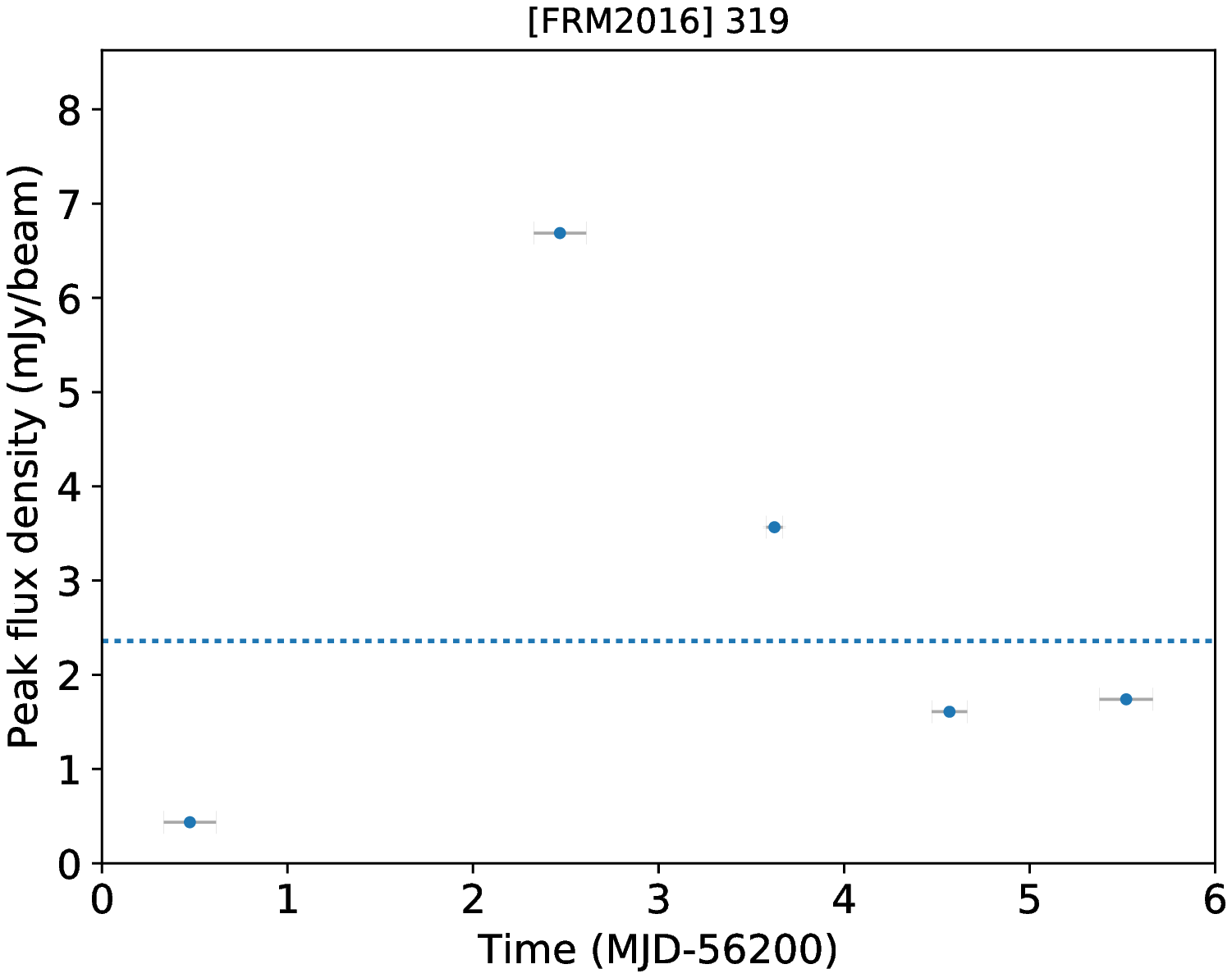}
\includegraphics*[bb=110 219 511 560, width=0.2330\linewidth]{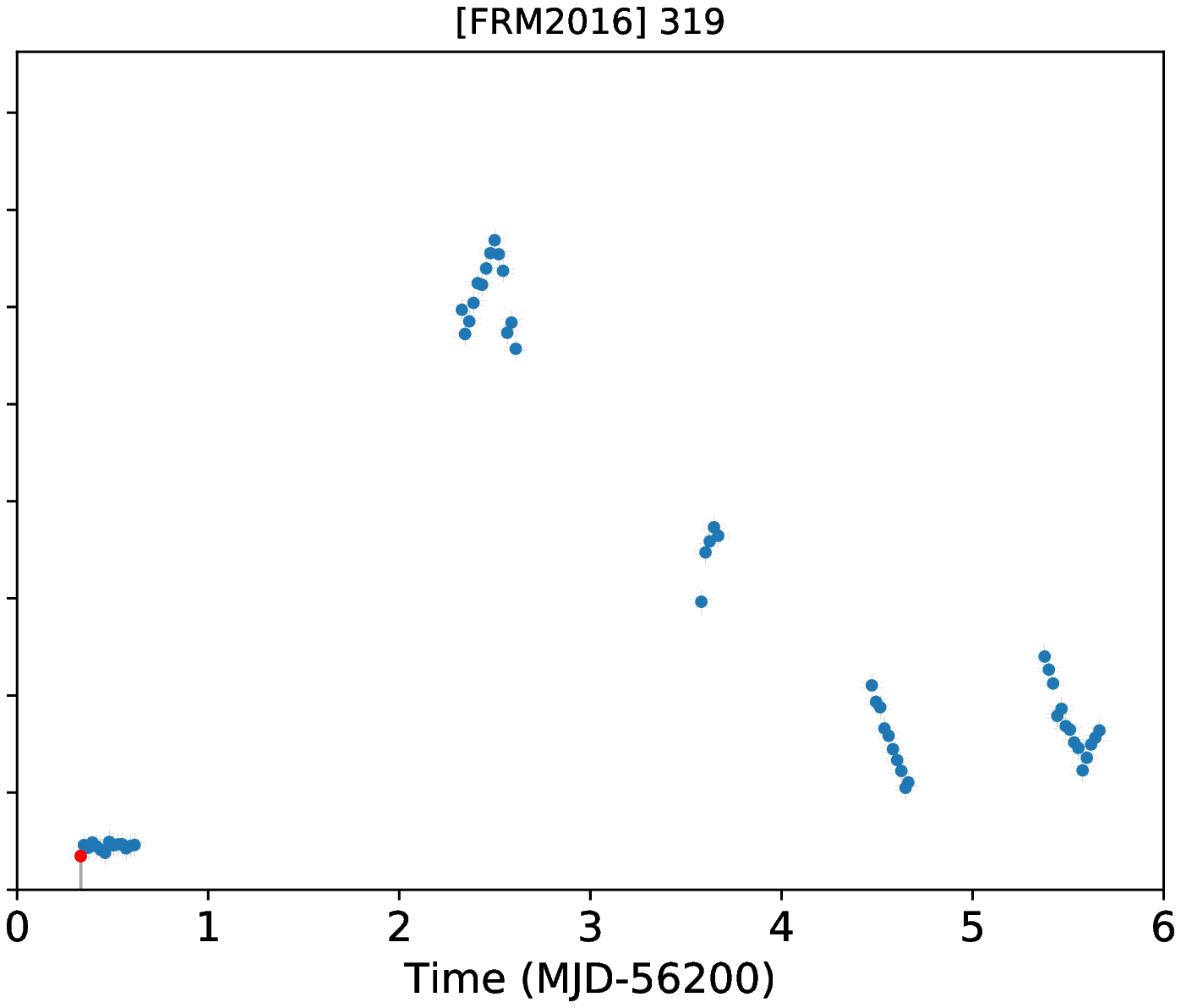}
\includegraphics*[bb=110 219 511 560, width=0.2330\linewidth]{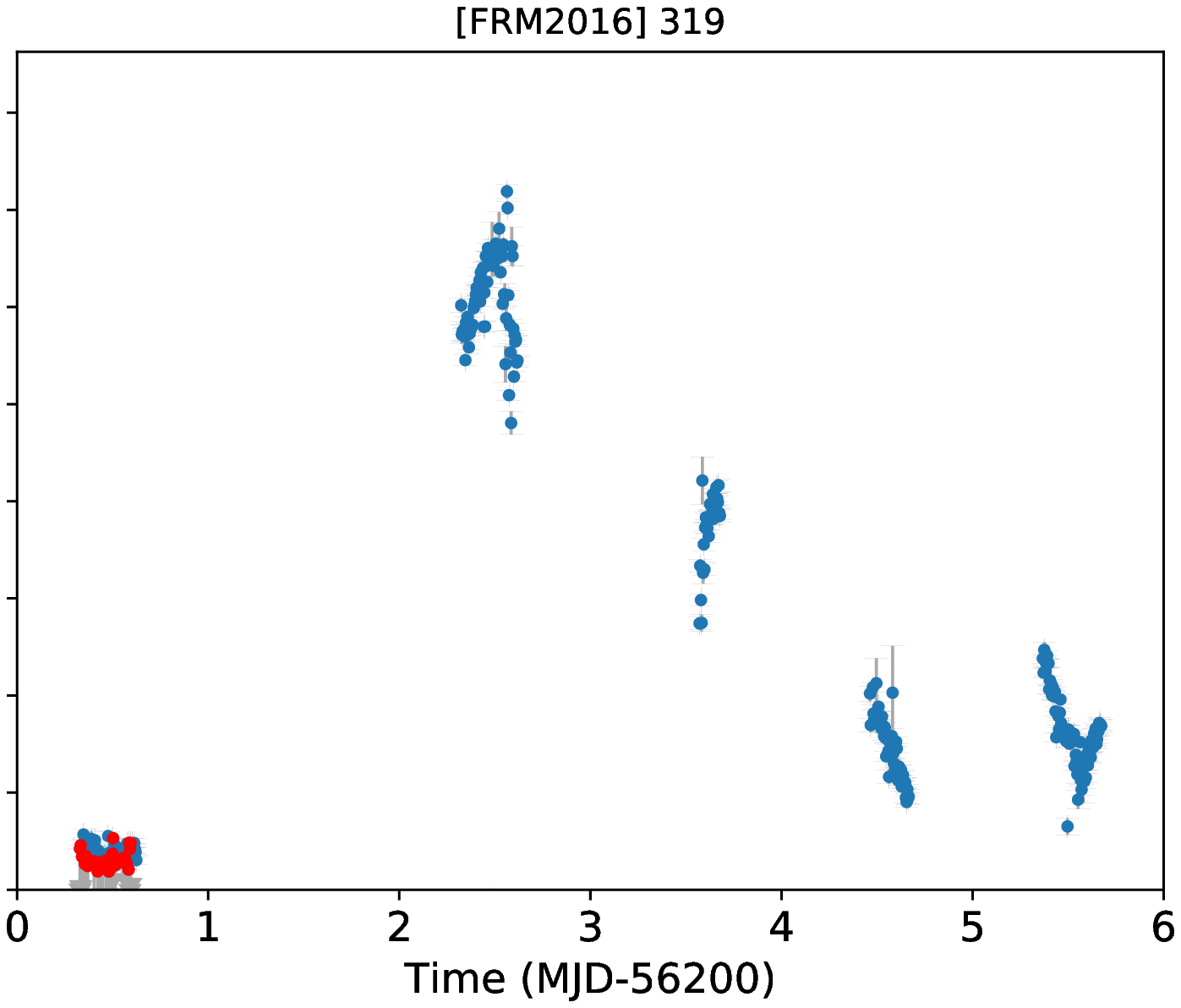}
\includegraphics*[bb=110 219 511 560, width=0.2330\linewidth]{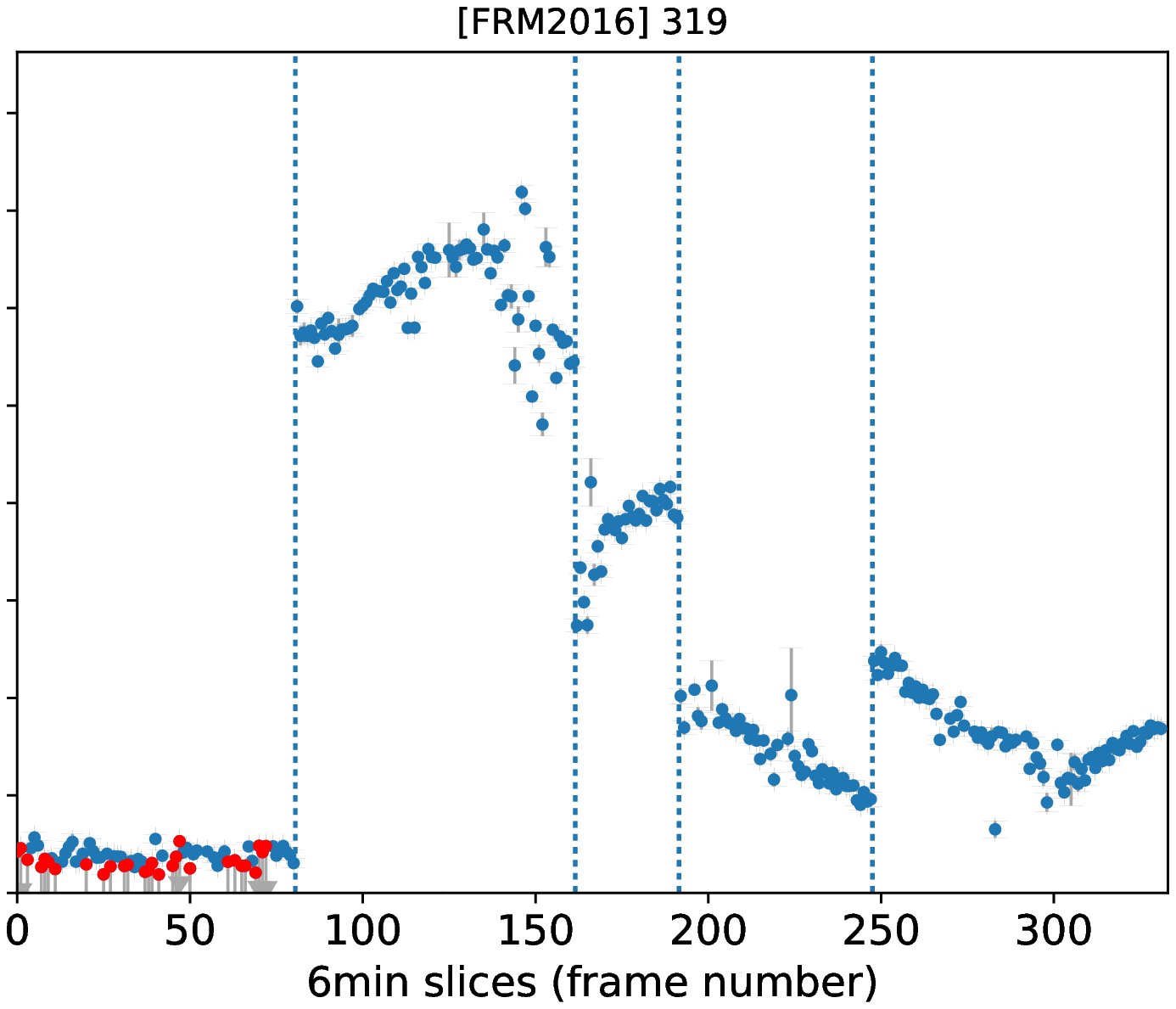}
\end{minipage} 
\vspace*{1mm} 
\begin{minipage}{\linewidth} 
\includegraphics*[bb= 60 219 511 560, width=0.2620\linewidth]{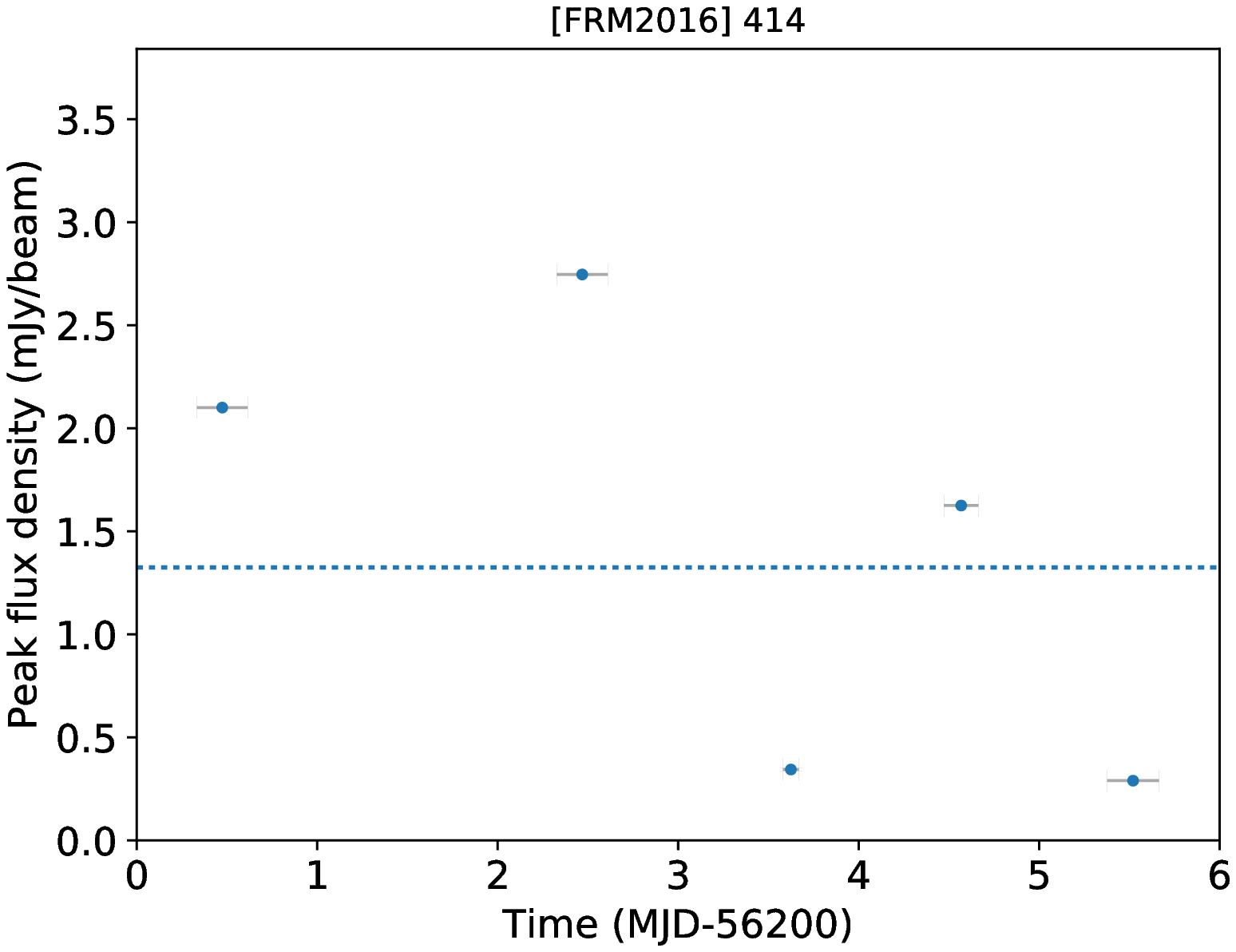}
\includegraphics*[bb=110 219 511 560, width=0.2330\linewidth]{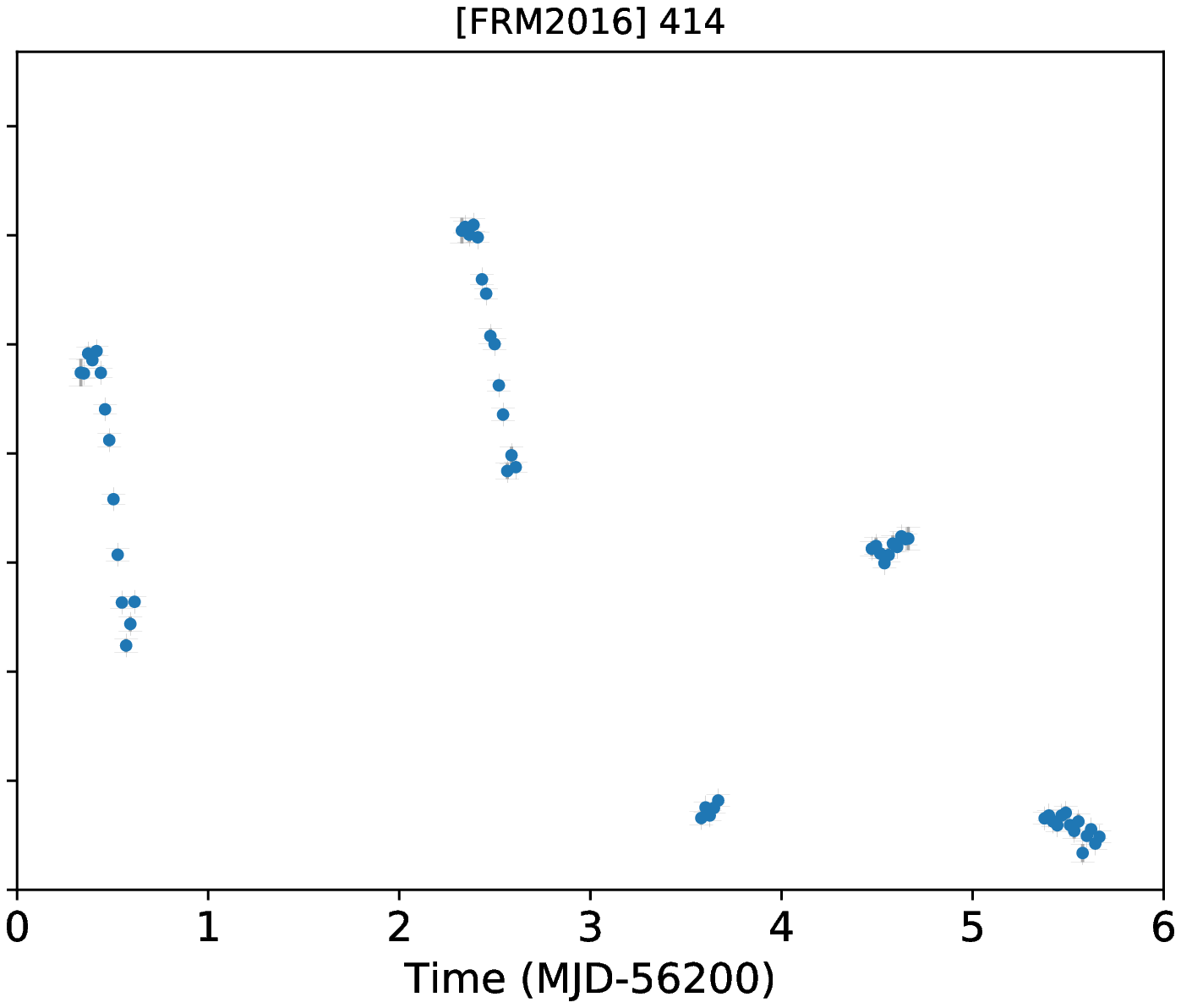}
\includegraphics*[bb=110 219 511 560, width=0.2330\linewidth]{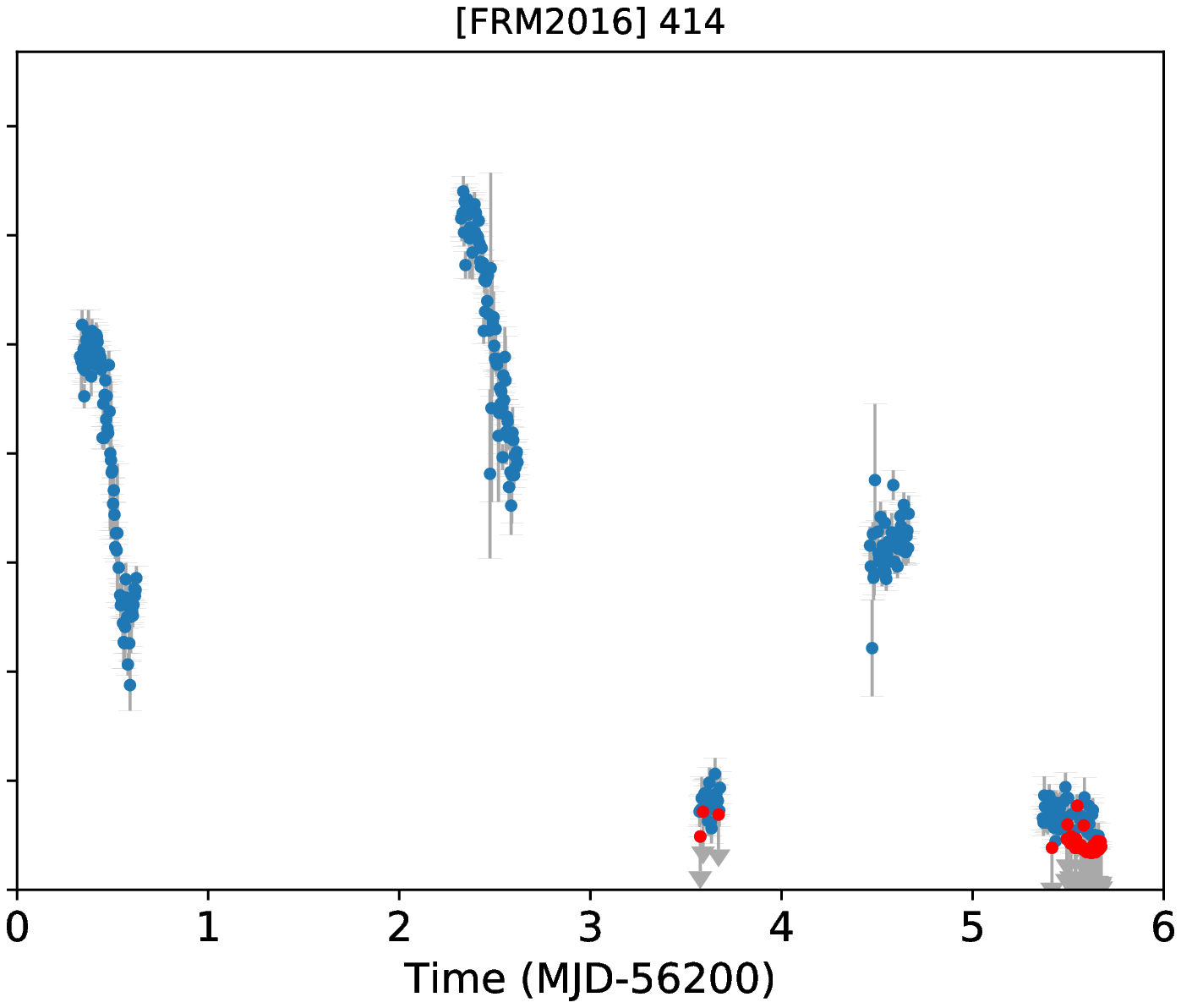}
\includegraphics*[bb=110 219 511 560, width=0.2330\linewidth]{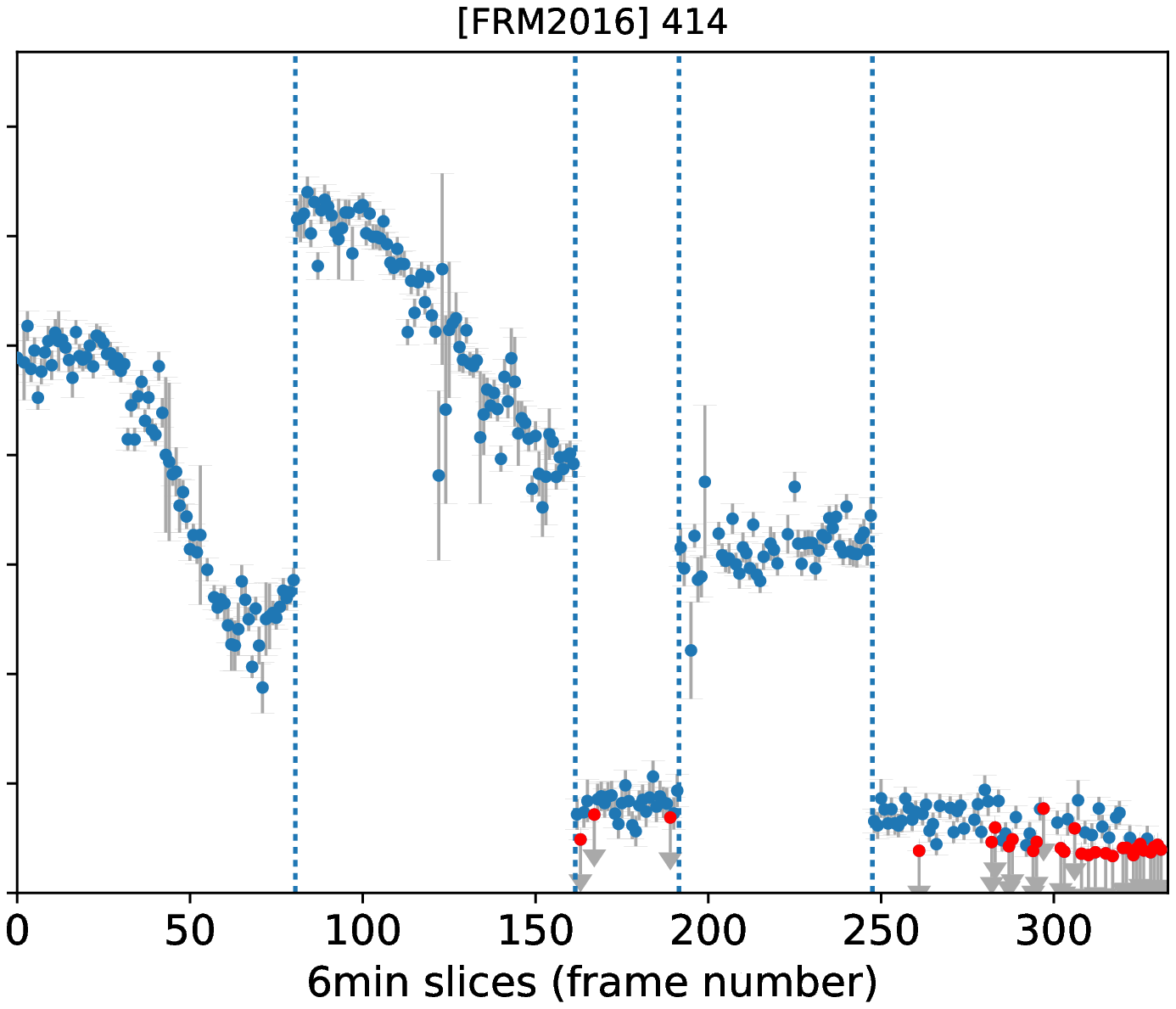}
\end{minipage} 
\vspace*{1mm} 
\begin{minipage}{\linewidth} 
\includegraphics*[bb= 60 219 511 560, width=0.2620\linewidth]{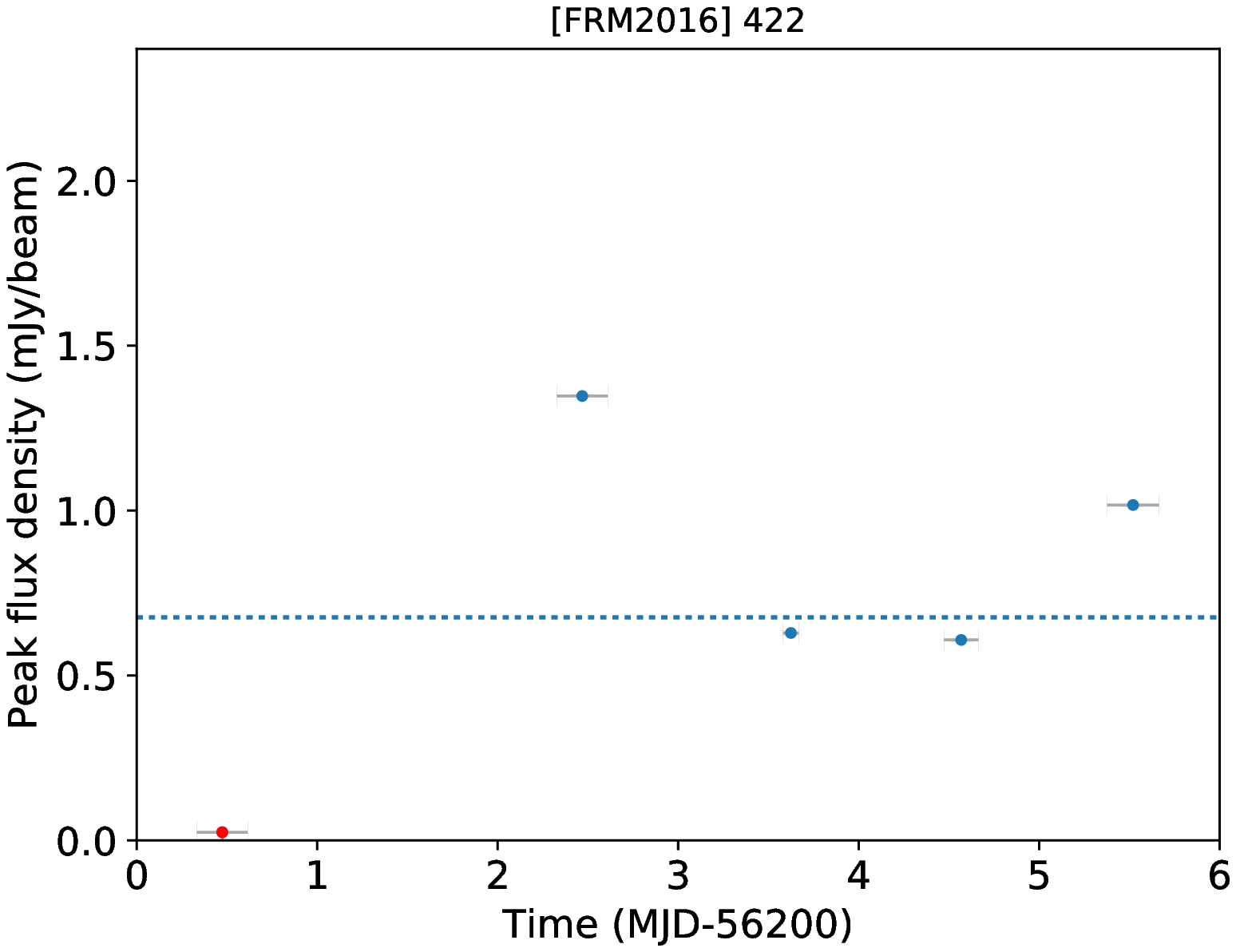}
\includegraphics*[bb=110 219 511 560, width=0.2330\linewidth]{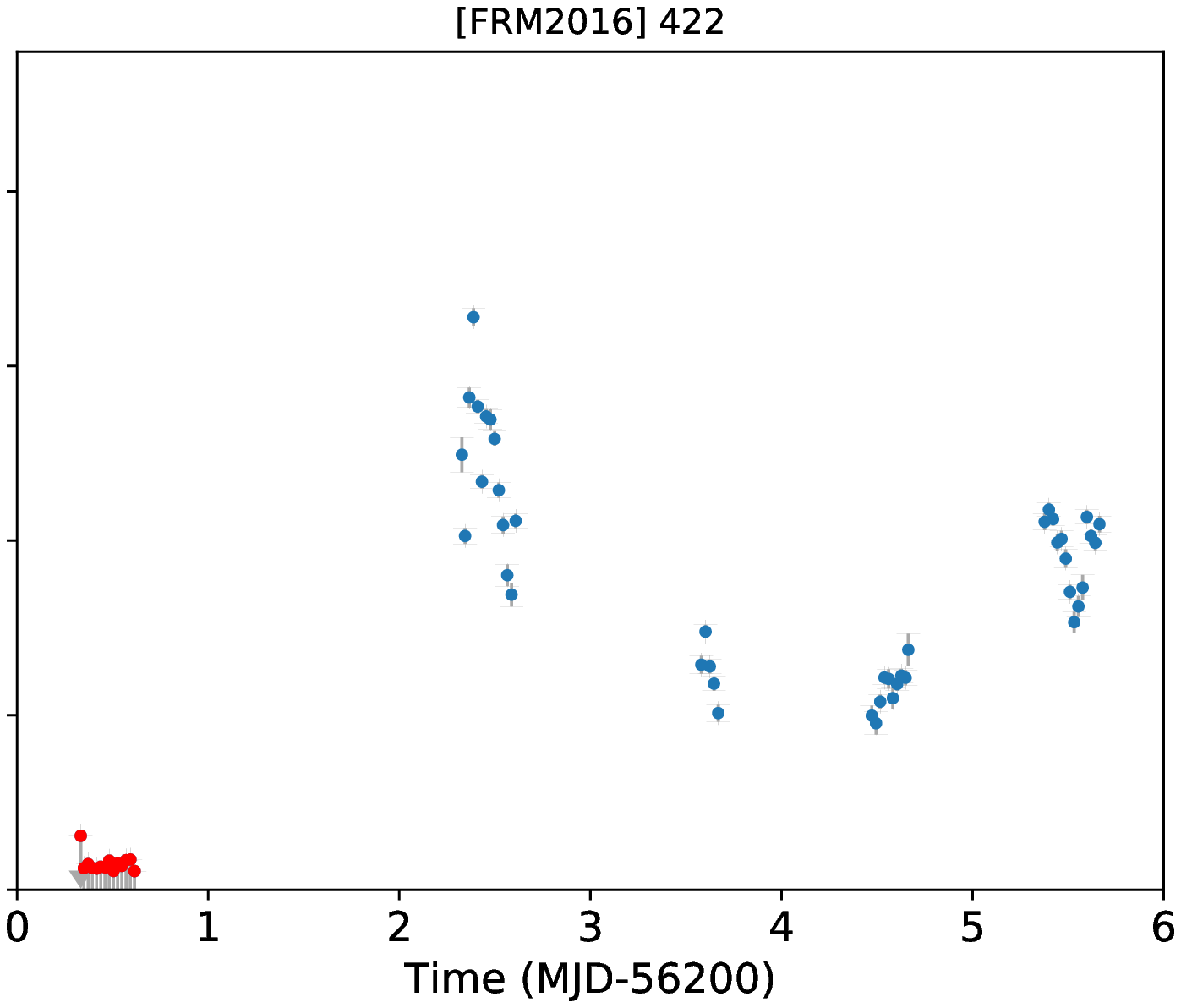}
\includegraphics*[bb=110 219 511 560, width=0.2330\linewidth]{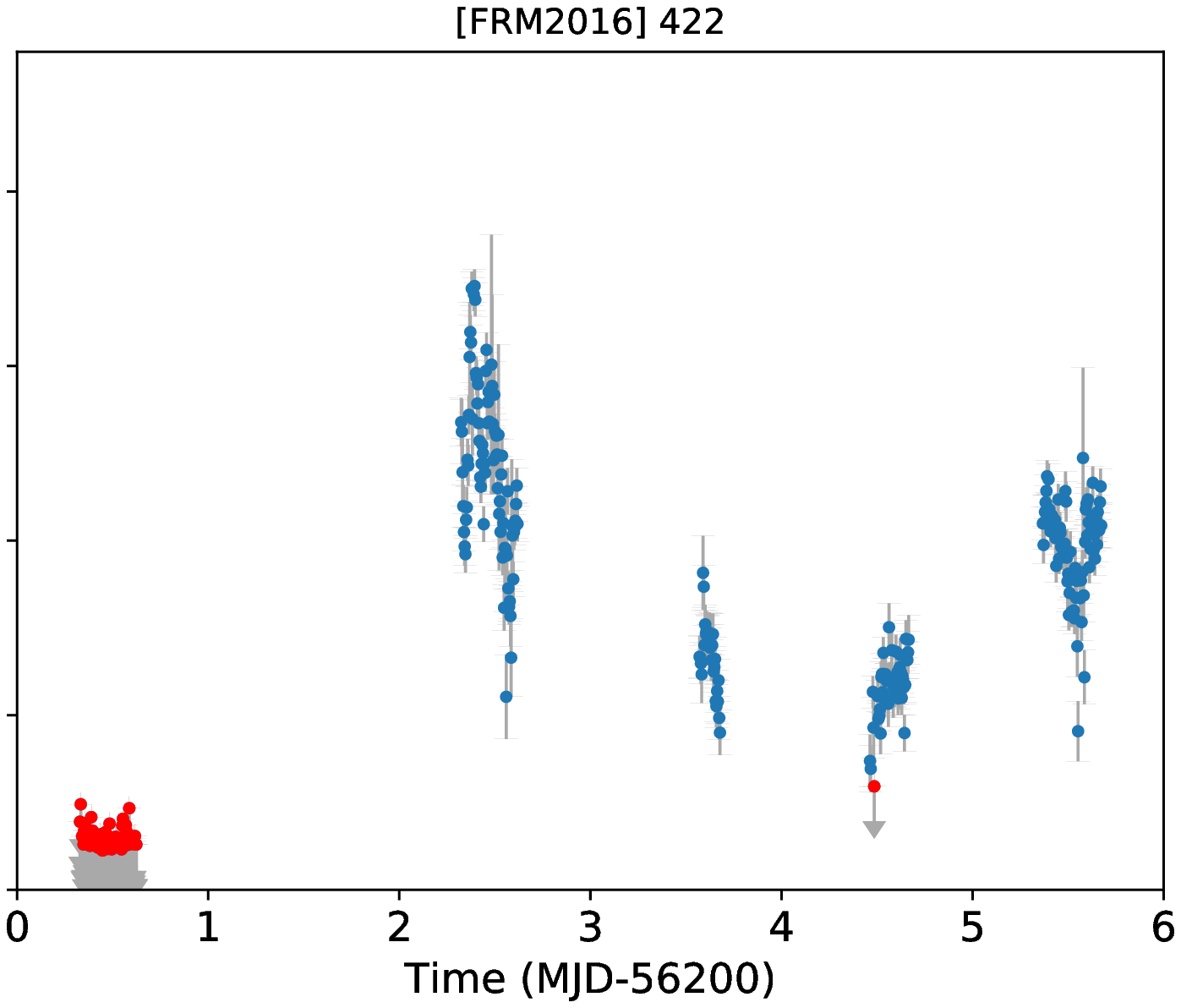}
\includegraphics*[bb=110 219 511 560, width=0.2330\linewidth]{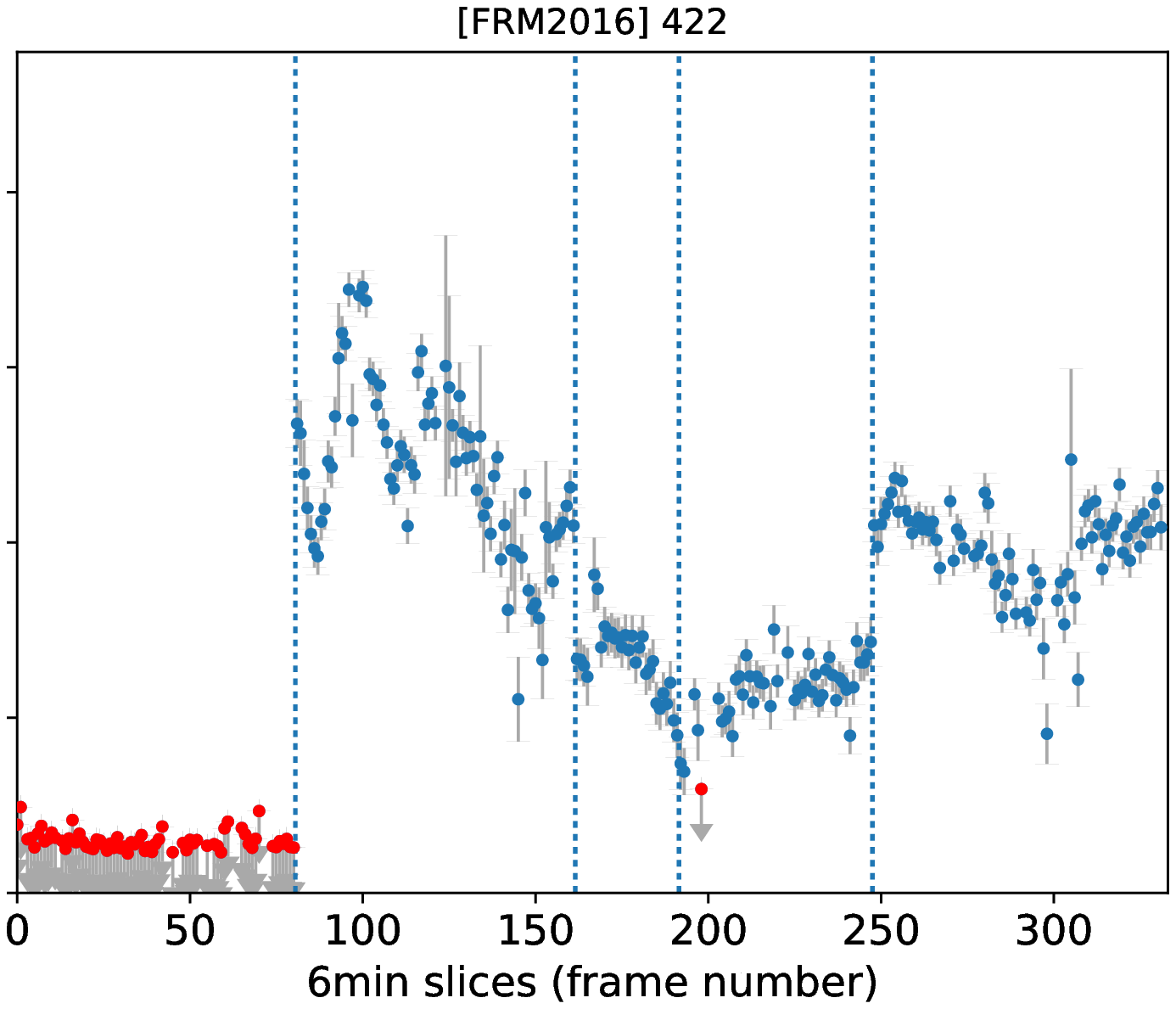}
\end{minipage} 
\vspace*{1mm} 
\begin{minipage}{\linewidth} 
\includegraphics*[bb= 60 219 511 560, width=0.2620\linewidth]{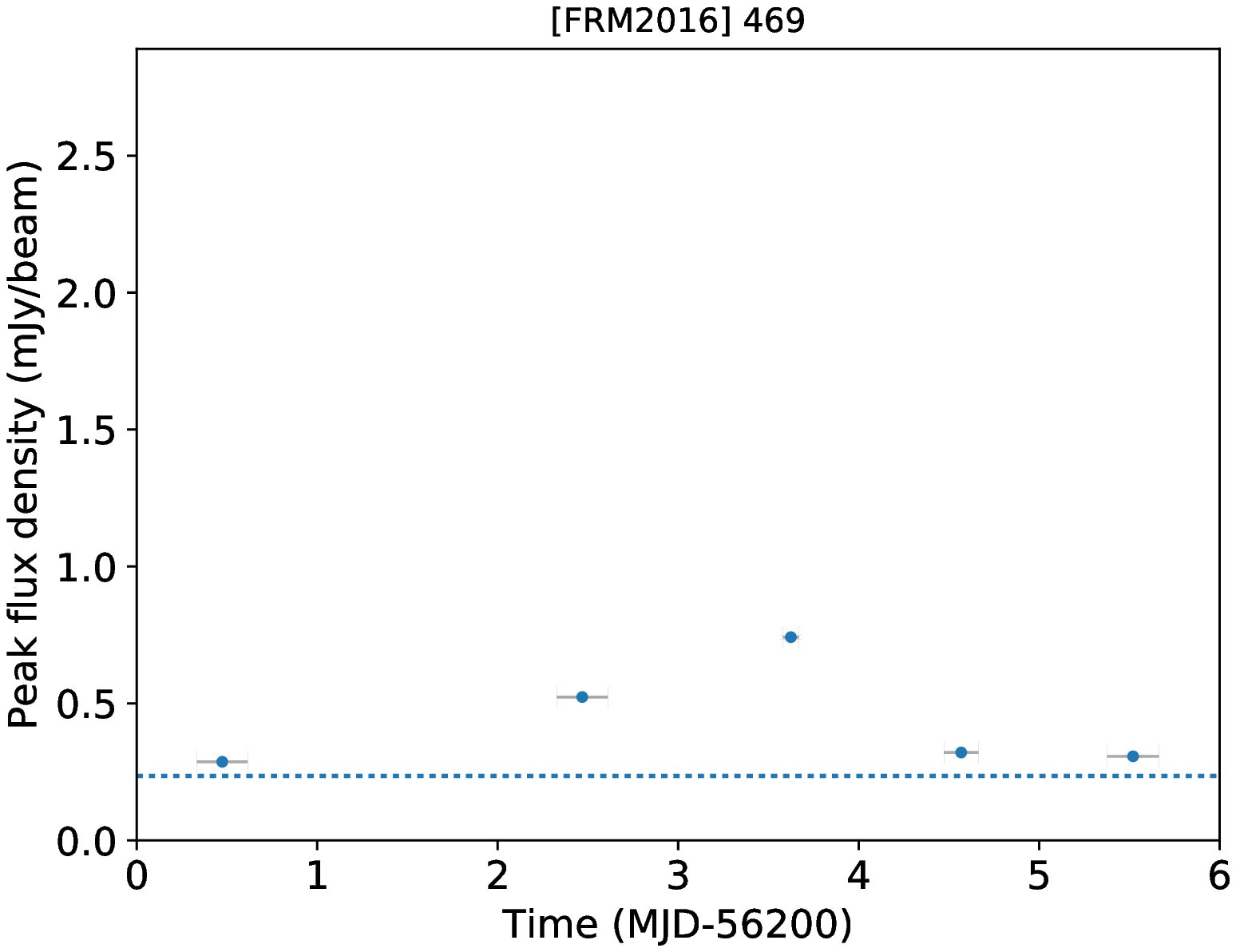}
\includegraphics*[bb=110 219 511 560, width=0.2330\linewidth]{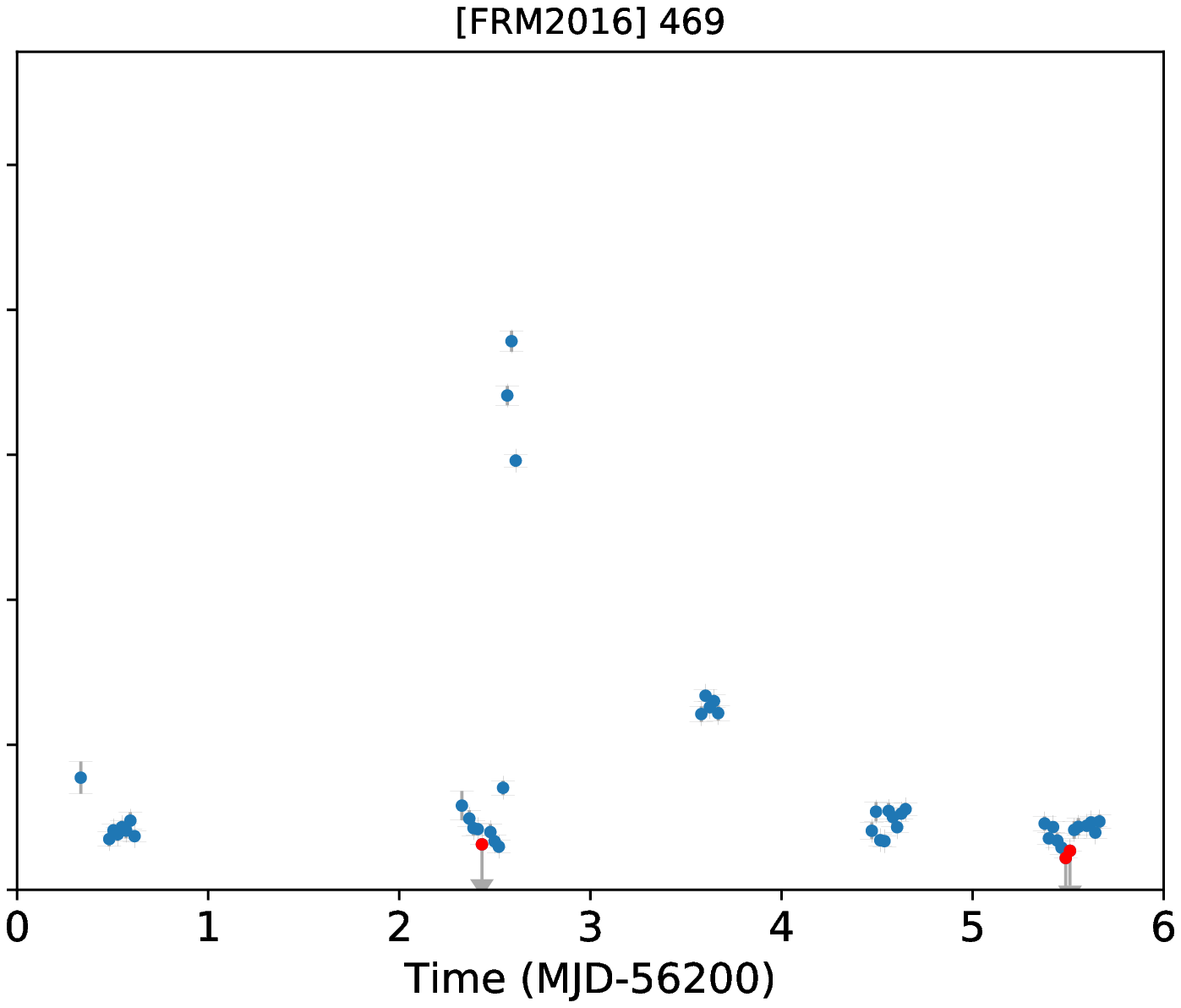}
\includegraphics*[bb=110 219 511 560, width=0.2330\linewidth]{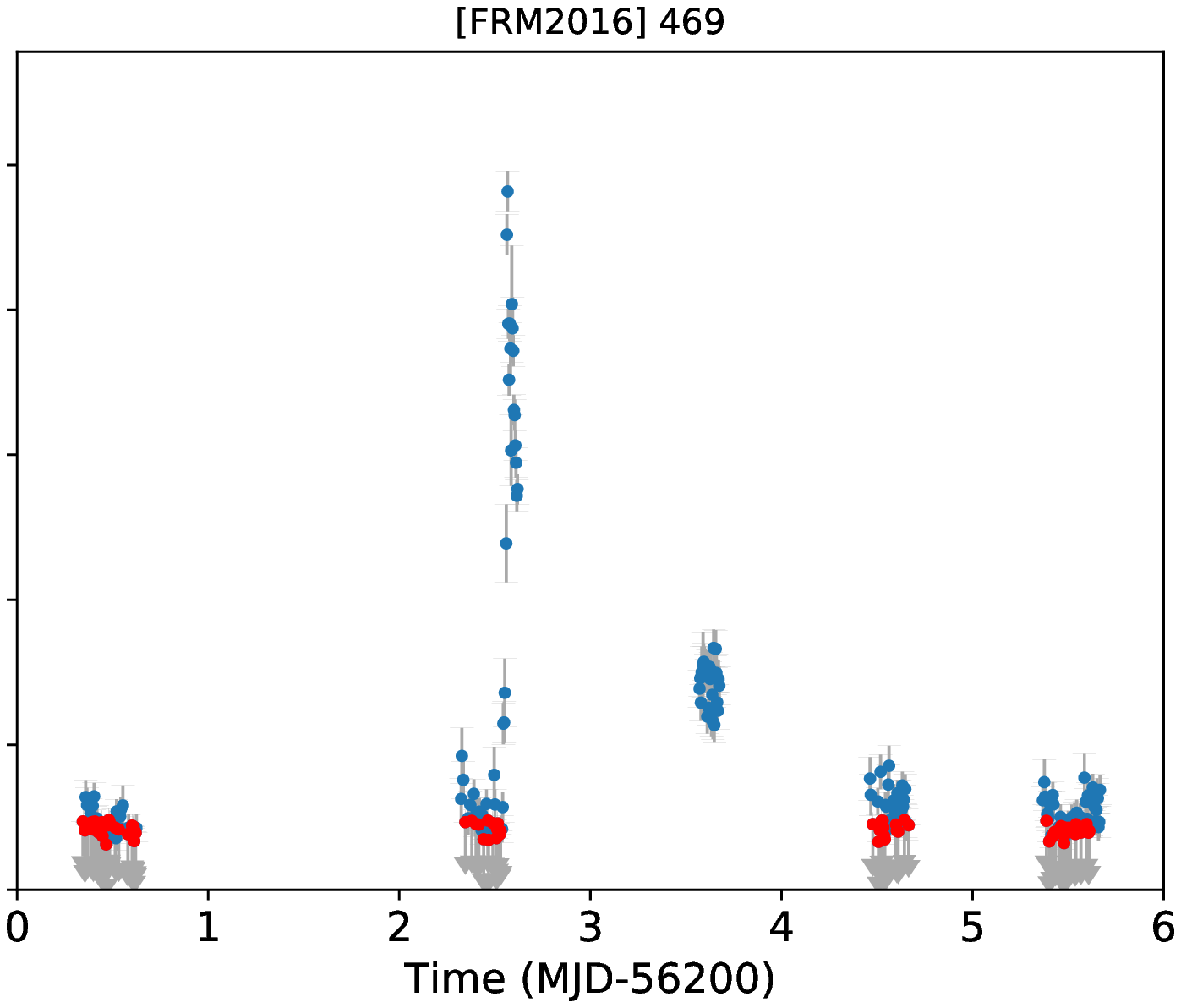}
\includegraphics*[bb=110 219 511 560, width=0.2330\linewidth]{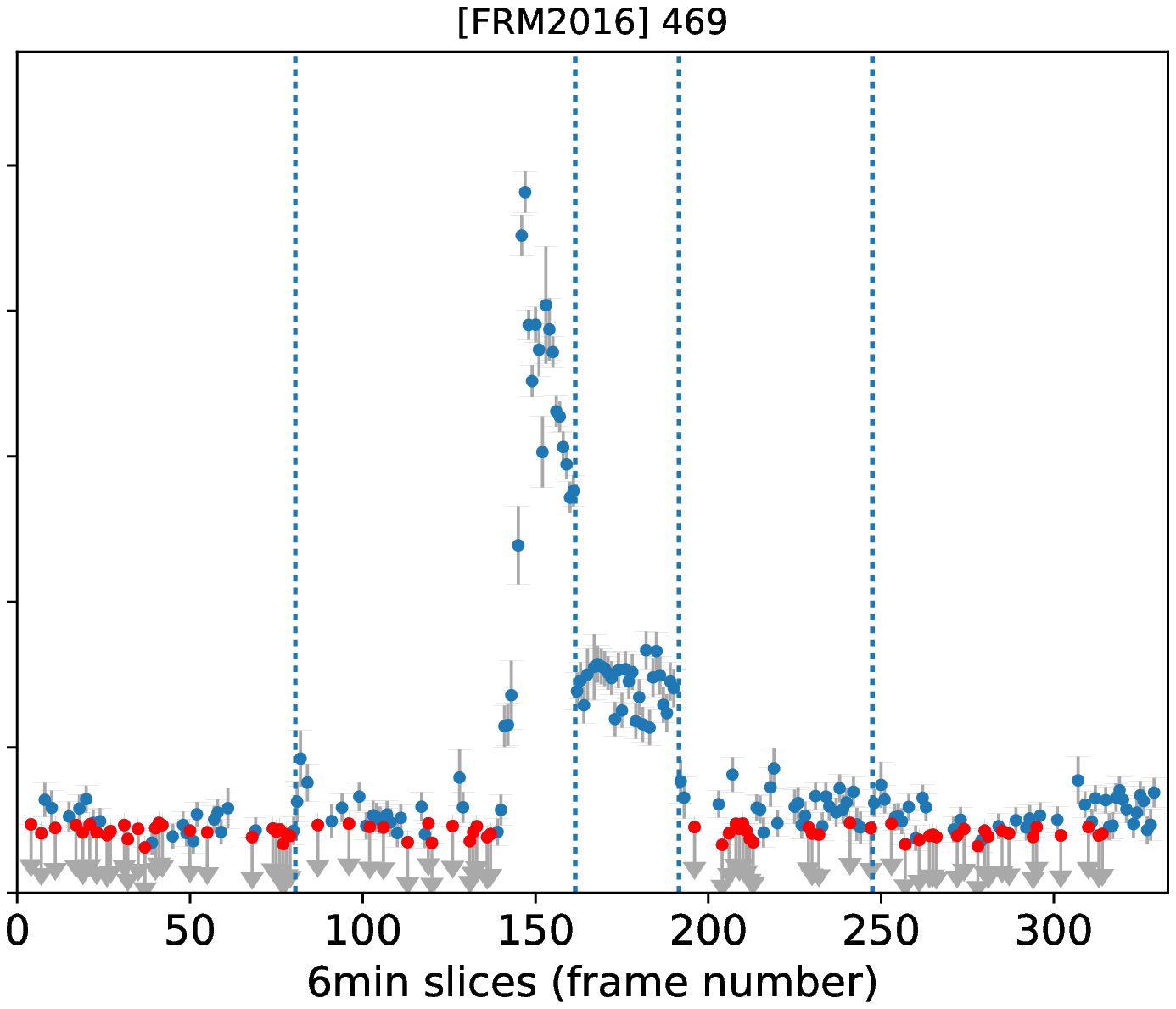}
\end{minipage} 
 
{\bf Figure~\ref{fig_var_all}} cntd. 
\end{figure*} 
 
\begin{figure*} 
\begin{minipage}{\linewidth} 
\includegraphics*[bb= 60 219 511 560, width=0.2620\linewidth]{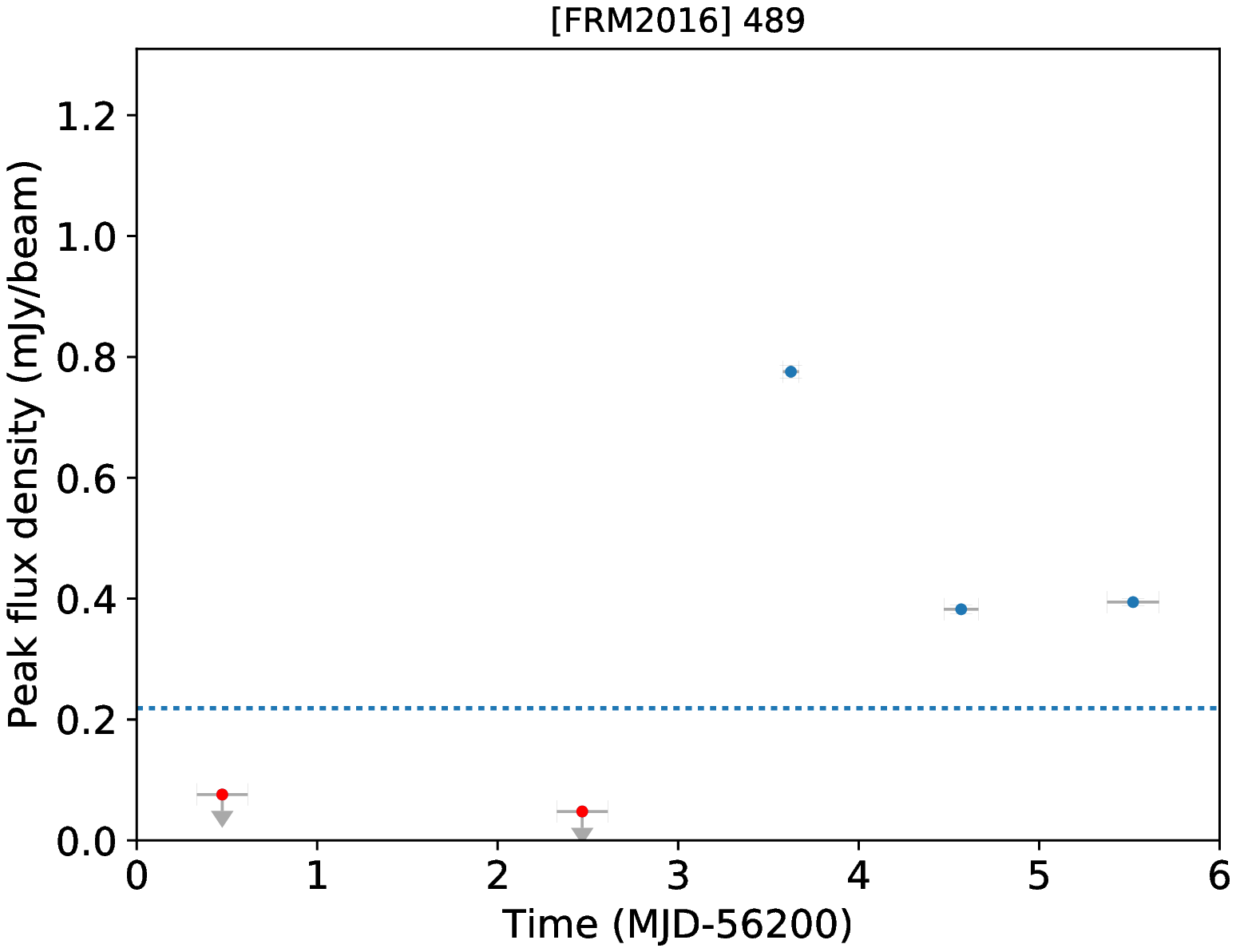}
\includegraphics*[bb=110 219 511 560, width=0.2330\linewidth]{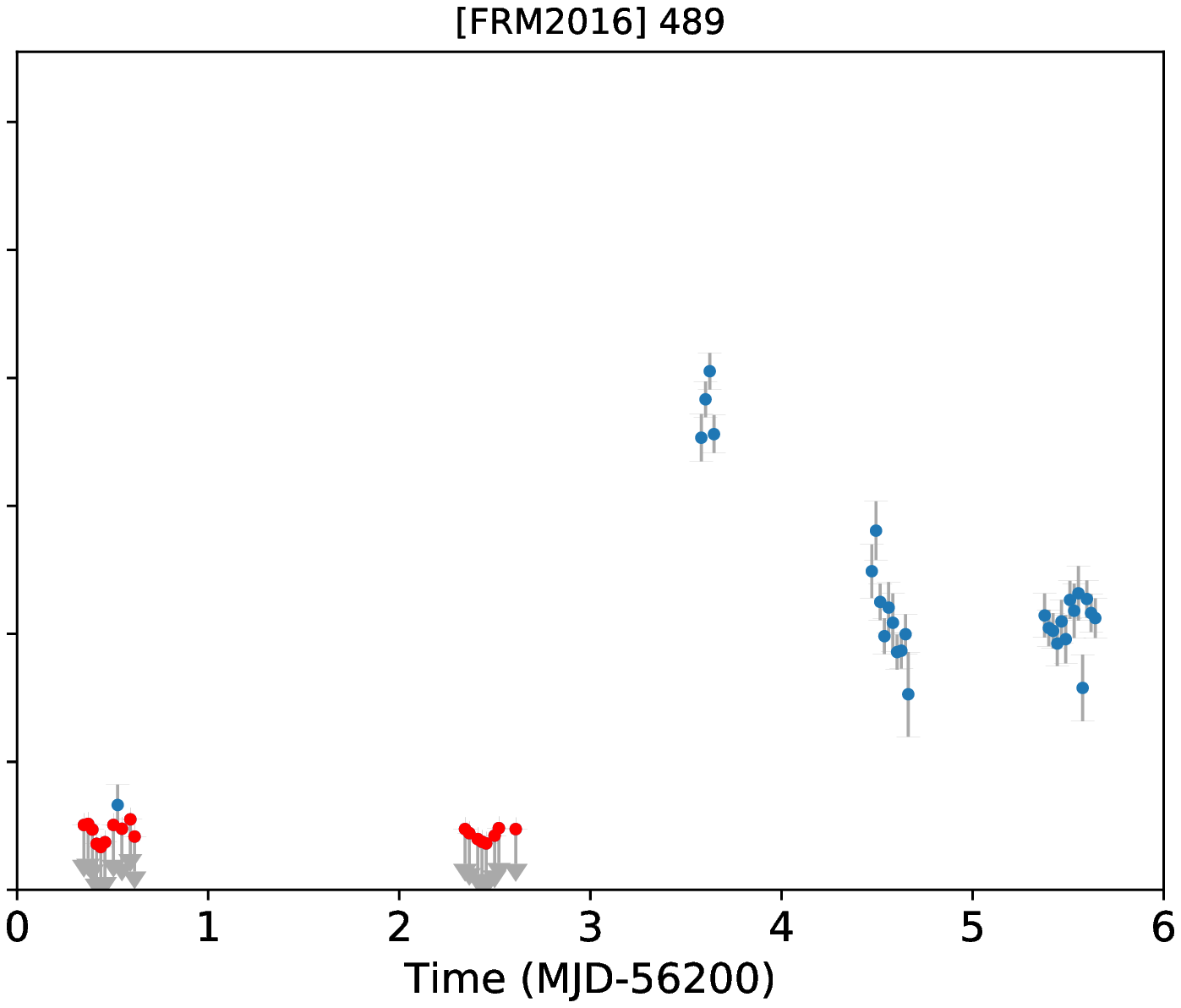}
\includegraphics*[bb=110 219 511 560, width=0.2330\linewidth]{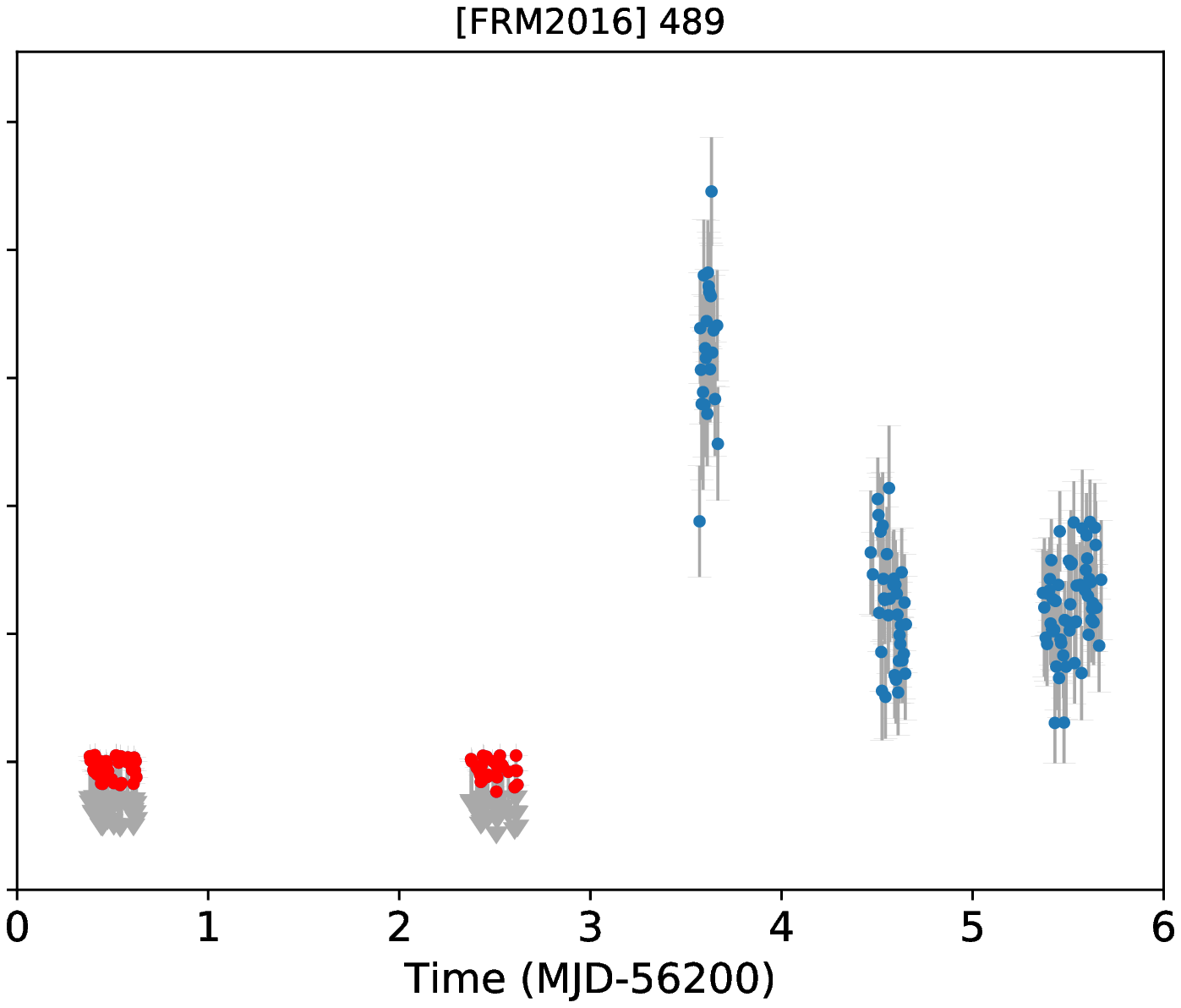}
\includegraphics*[bb=110 219 511 560, width=0.2330\linewidth]{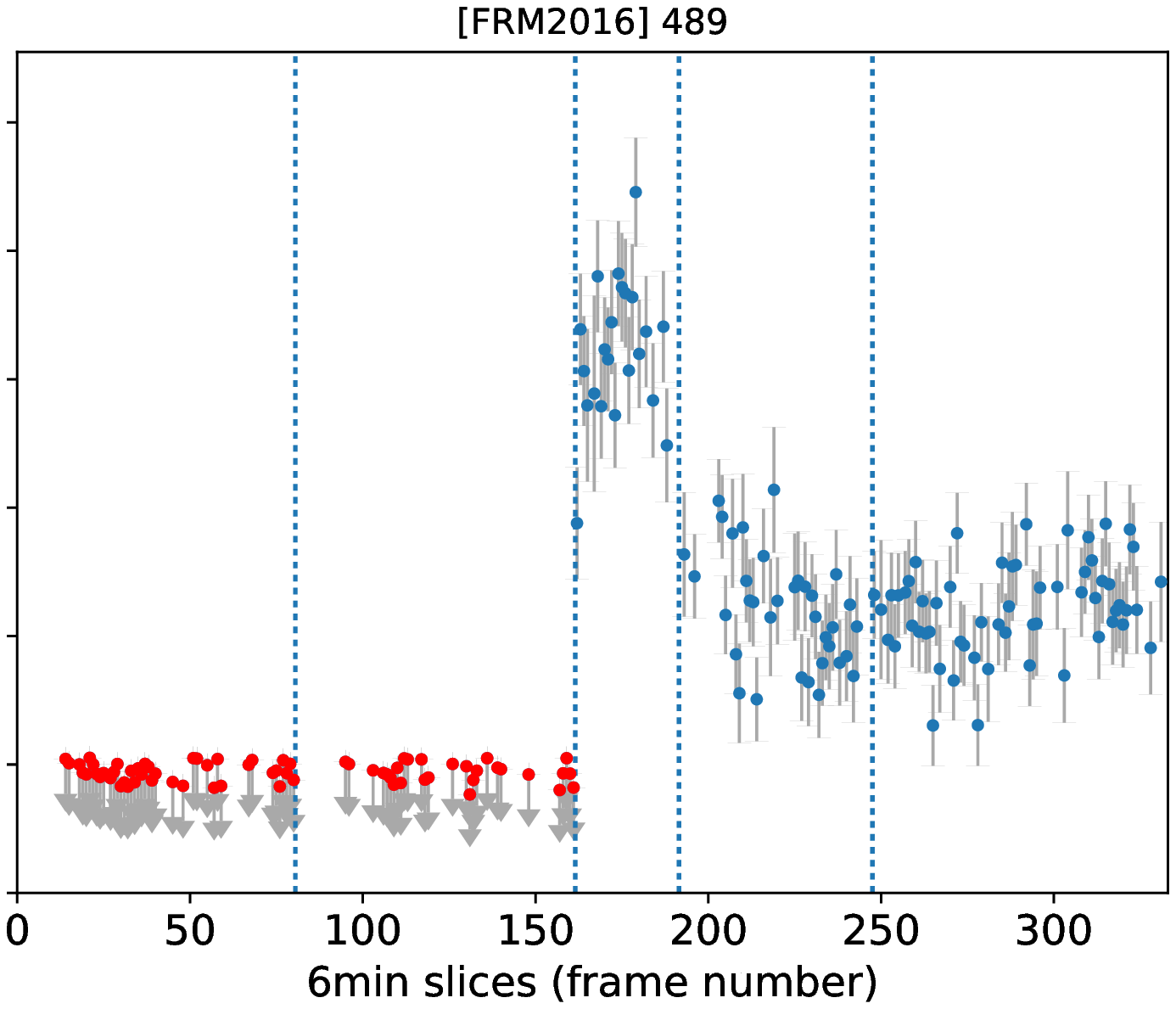}
\end{minipage} 
\vspace*{1mm} 
\begin{minipage}{\linewidth} 
\includegraphics*[bb= 60 219 511 560, width=0.2620\linewidth]{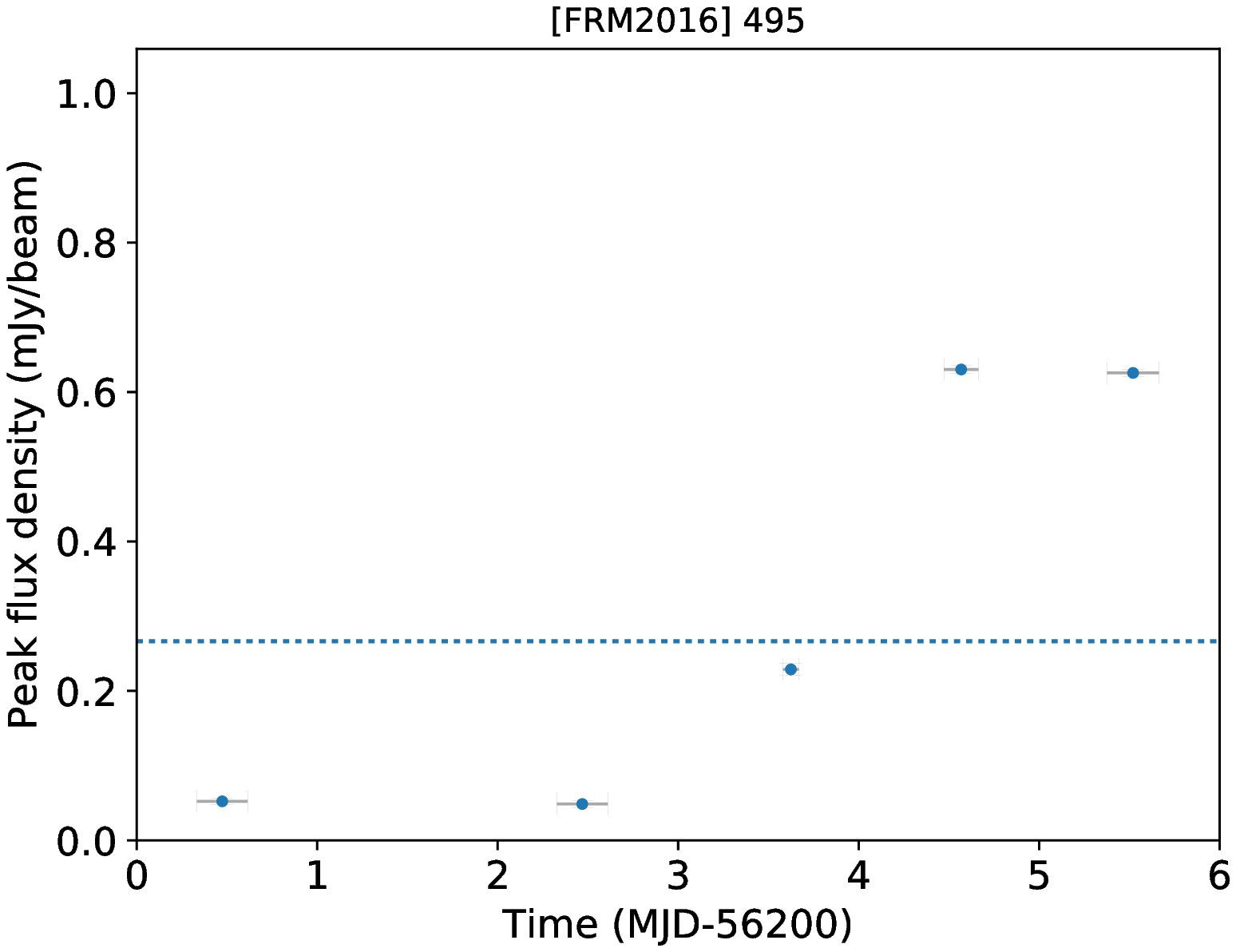}
\includegraphics*[bb=110 219 511 560, width=0.2330\linewidth]{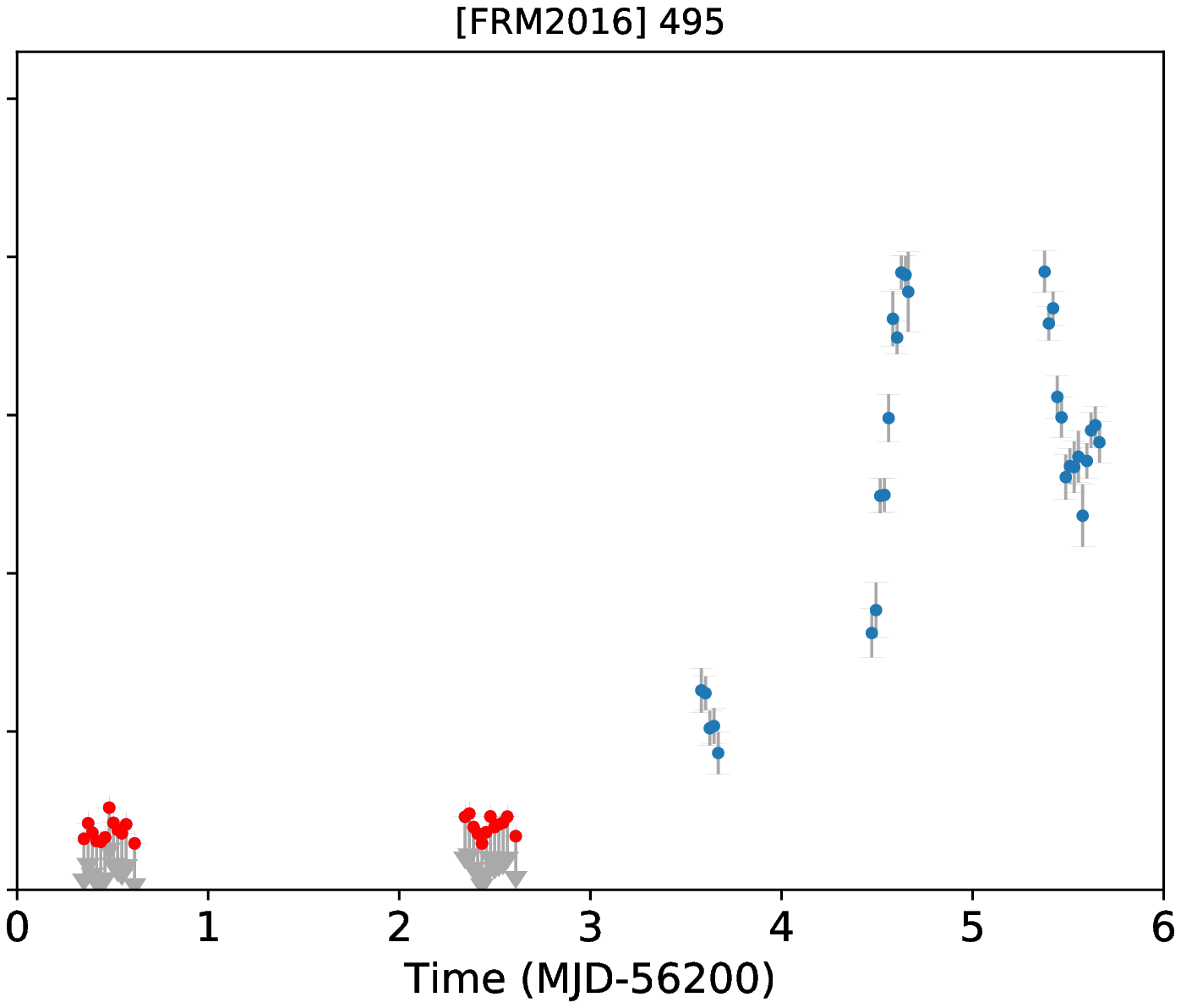}
\includegraphics*[bb=110 219 511 560, width=0.2330\linewidth]{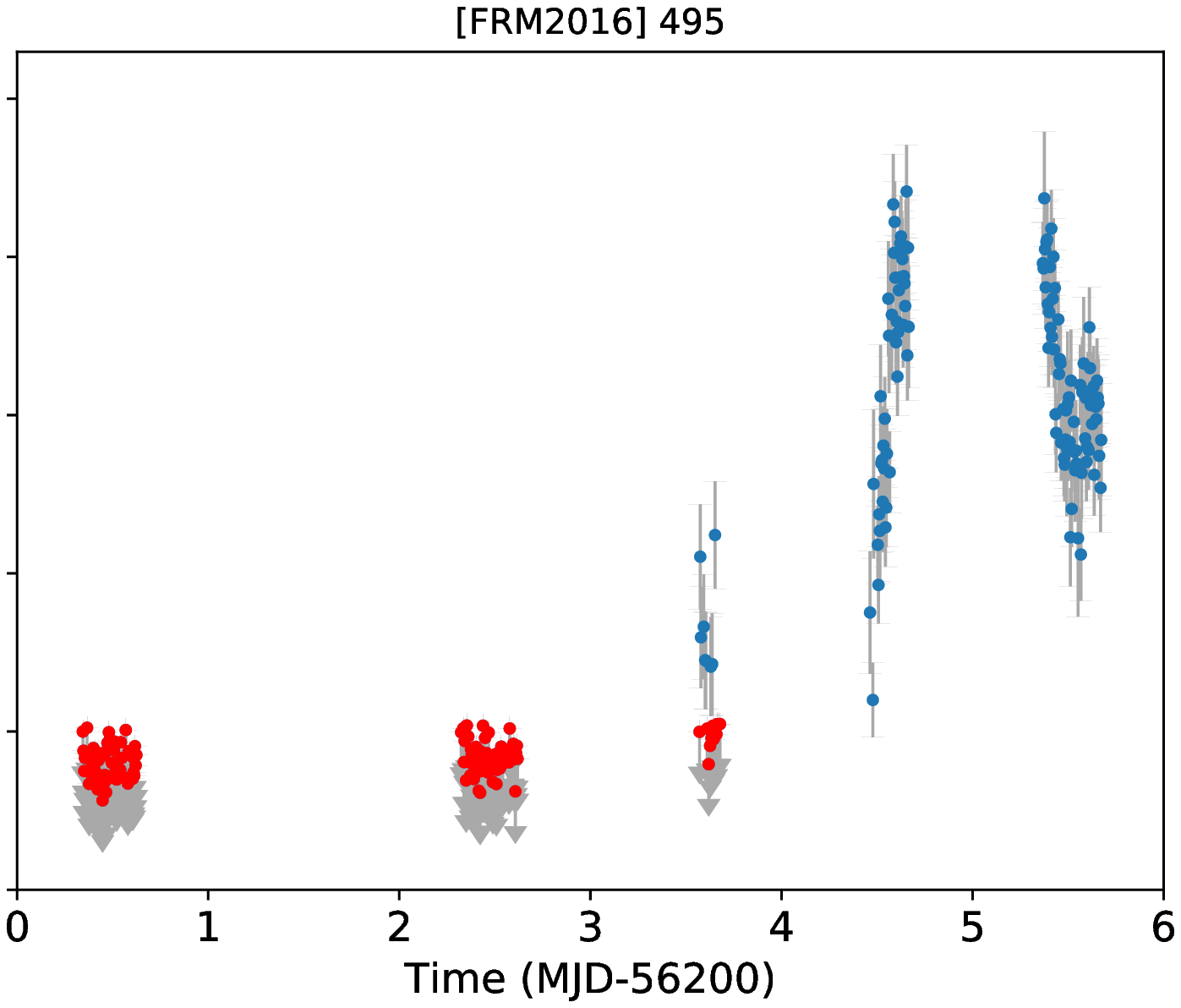}
\includegraphics*[bb=110 219 511 560, width=0.2330\linewidth]{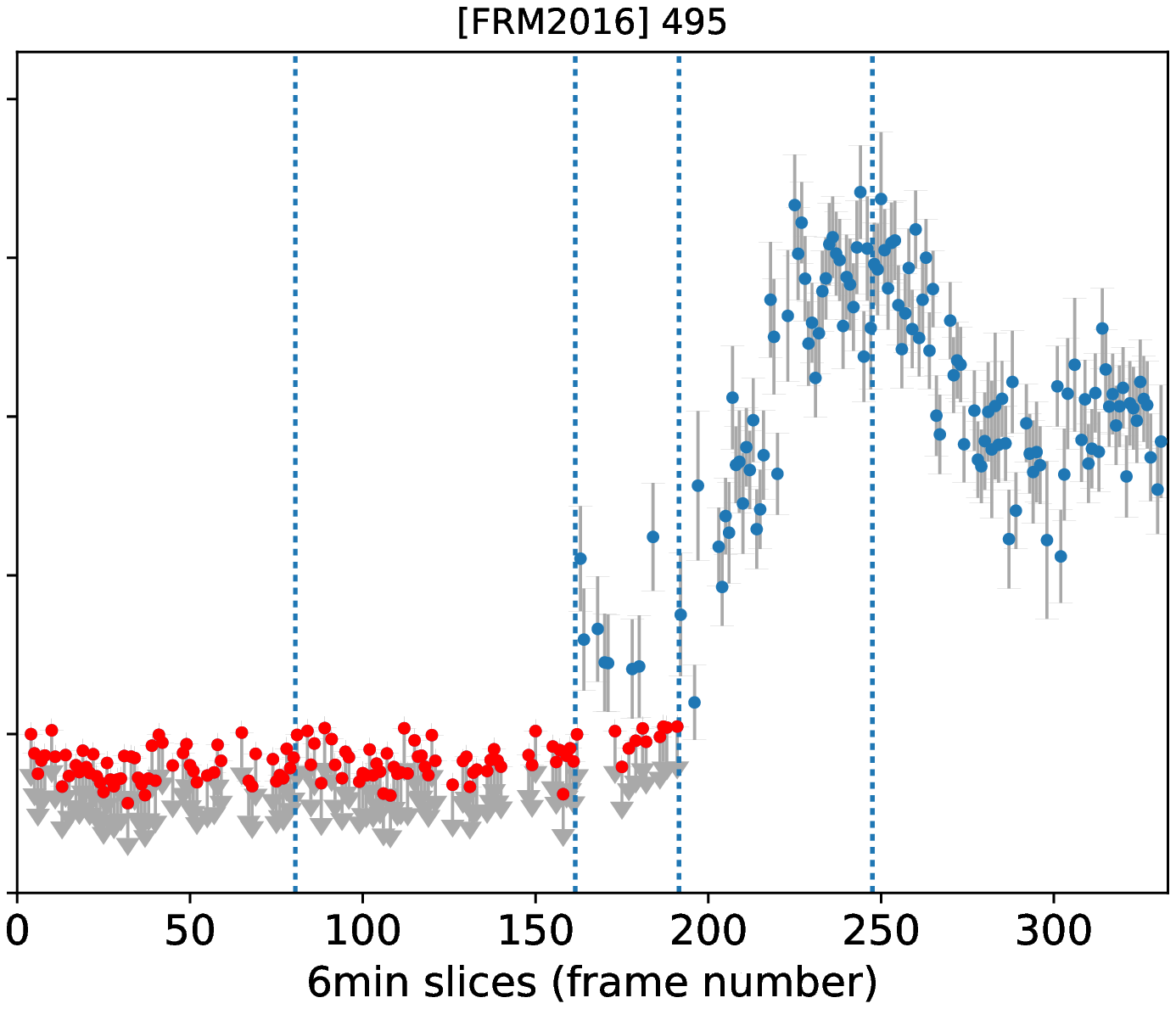}
\end{minipage} 
\vspace*{1mm} 
\begin{minipage}{\linewidth} 
\includegraphics*[bb= 60 219 511 560, width=0.2620\linewidth]{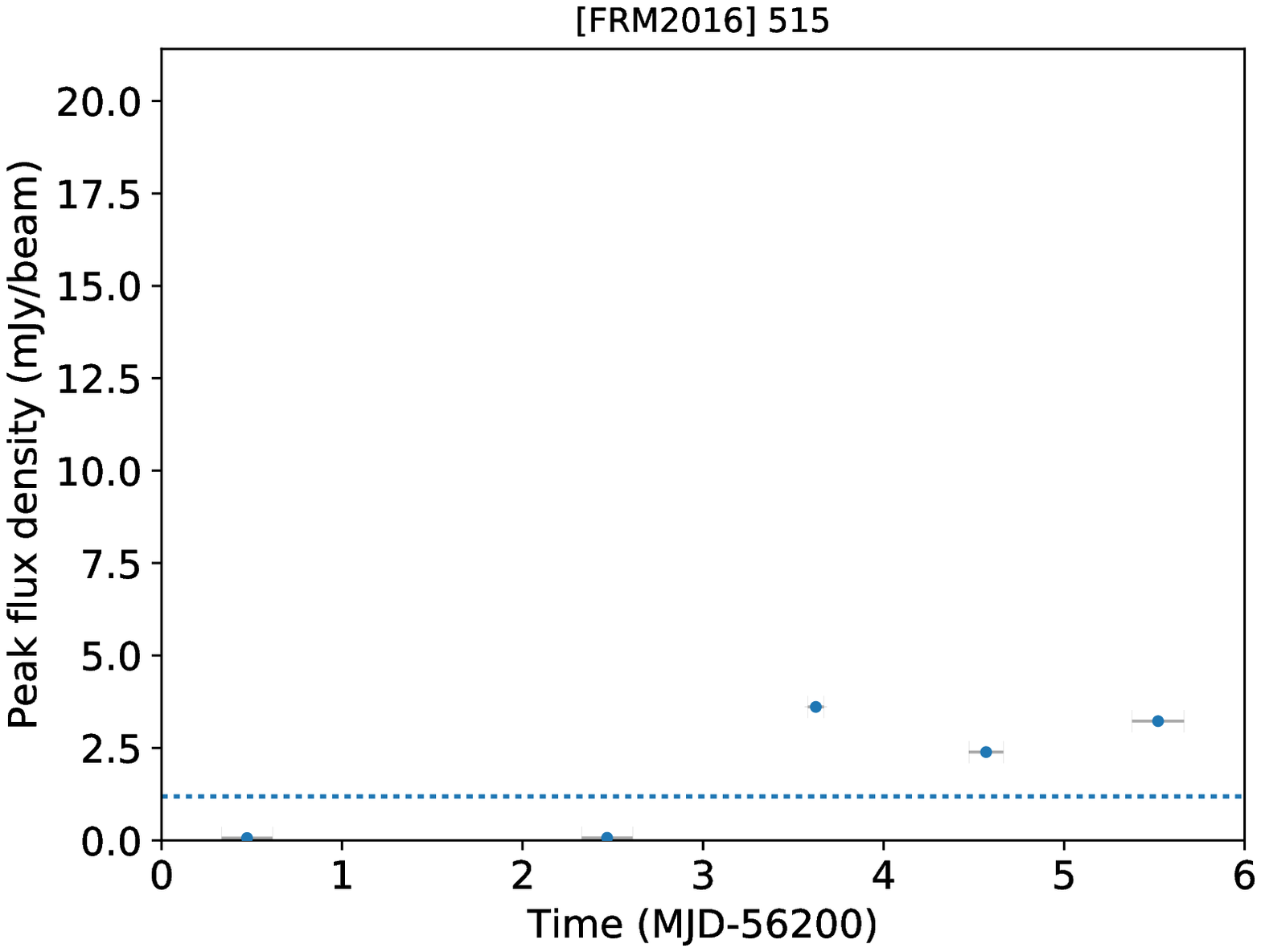}
\includegraphics*[bb=110 219 511 560, width=0.2330\linewidth]{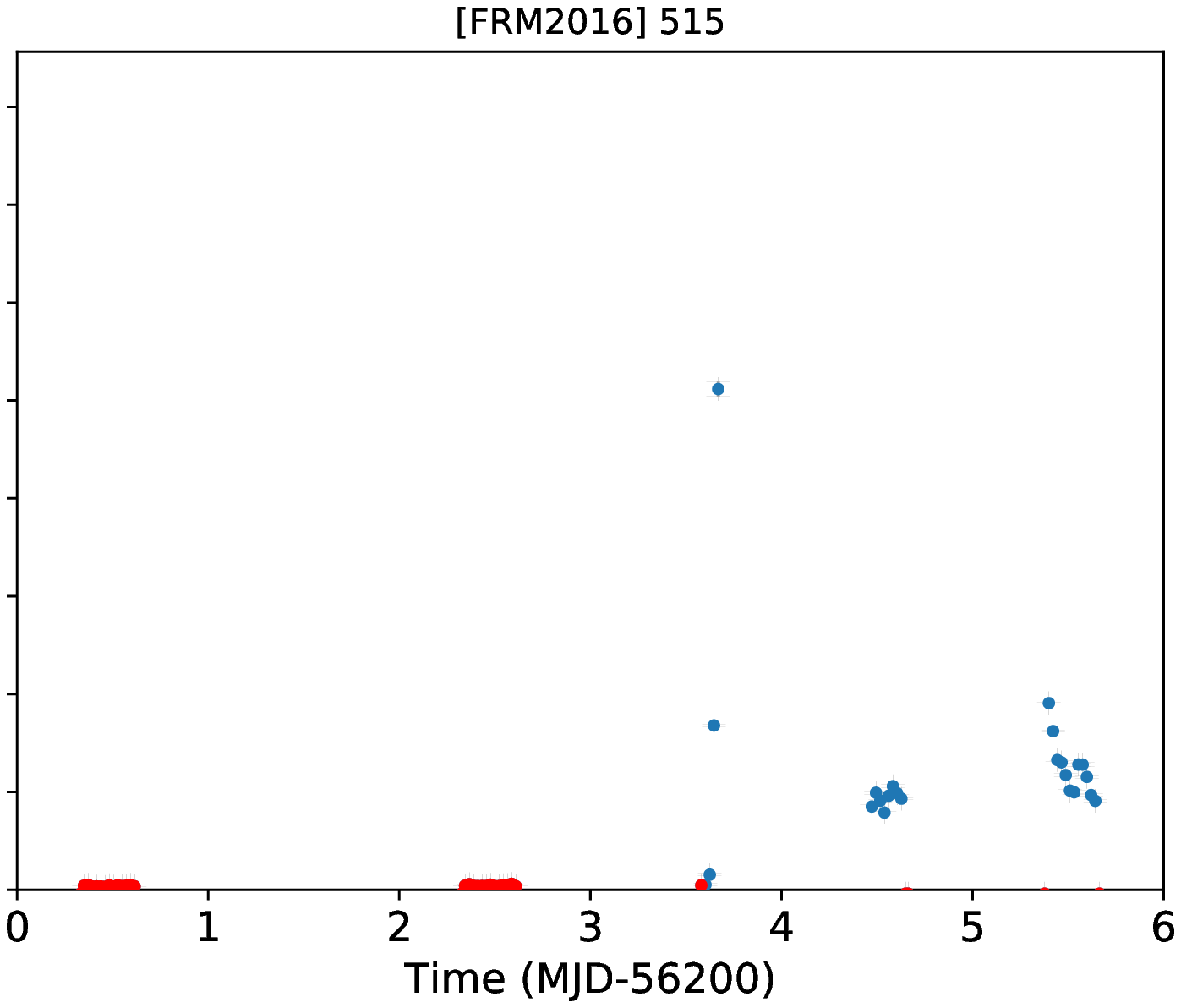}
\includegraphics*[bb=110 219 511 560, width=0.2330\linewidth]{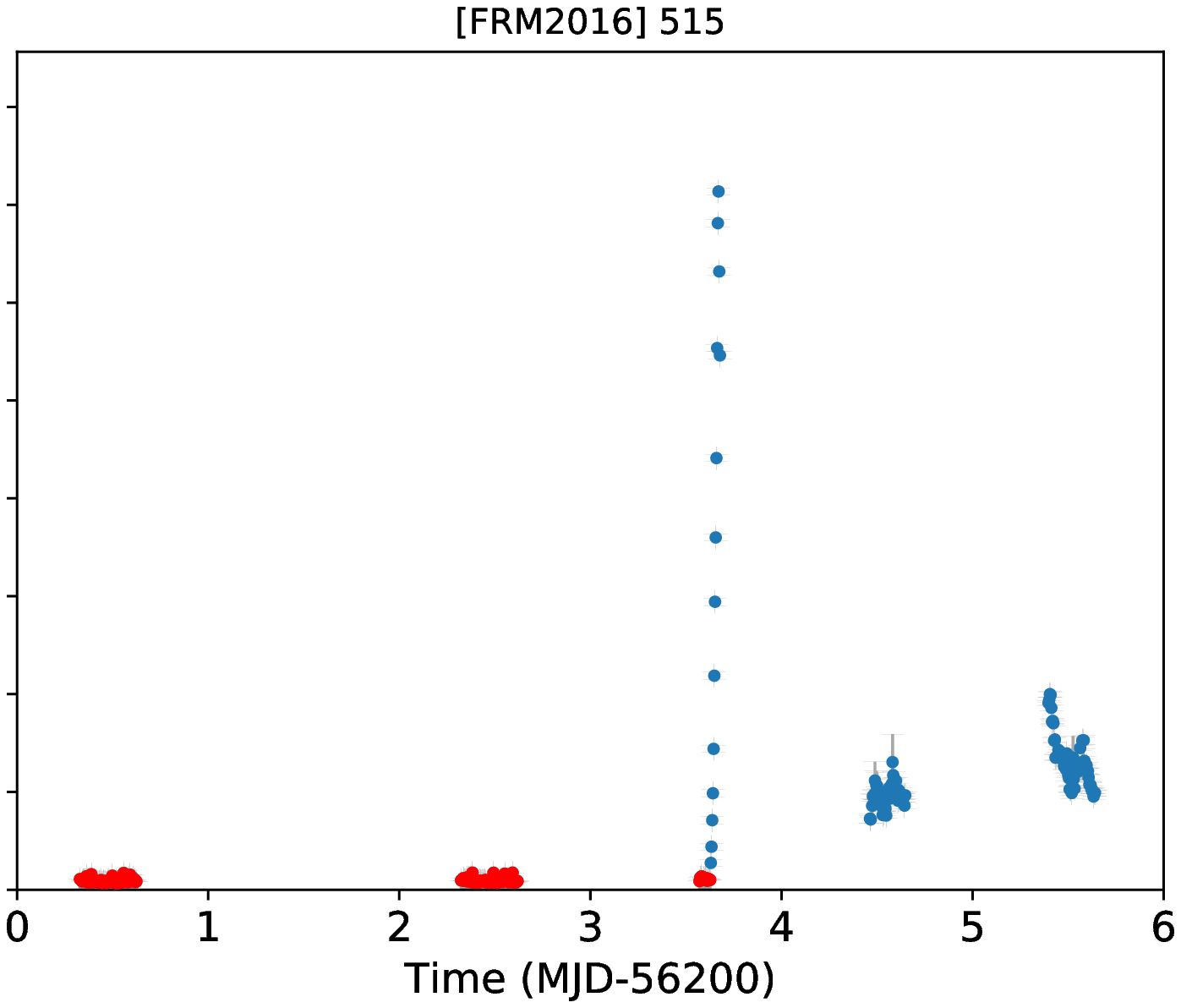}
\includegraphics*[bb=110 219 511 560, width=0.2330\linewidth]{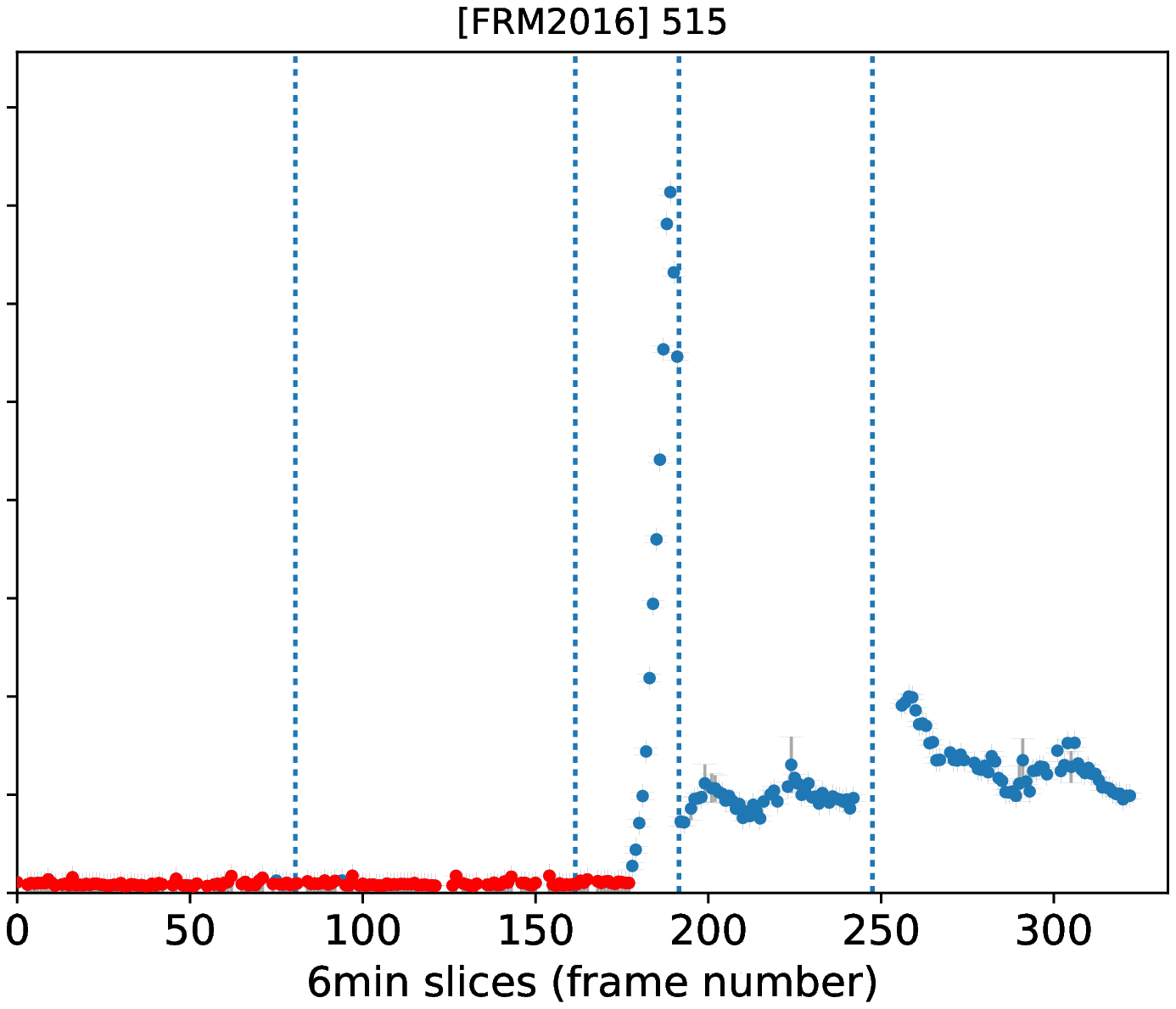}
\end{minipage} 
 
{\bf Figure~\ref{fig_var_all}} cntd. 
\end{figure*}

\subsection{X-ray and infrared identification} 
 
Interestingly, the 13 sources of extreme radio variability that we
identified all have X-ray counterparts in the deep \textit{Chandra}
observations reported by \citet{get05}, and also in the simultaneous
\textit{Chandra} observations reported here. All but three of them (sources
98, 110, and 189) also have near-infrared counterparts in
the VISION survey \citep{mei16}. Based on their X-ray and near-infrared
characteristics, all 13 sources have previously been identified as YSOs
\citep{bro13}.  Sources without
near-infrared counterparts could be missed if they are deeply embedded 
or are projected on bright nebulosity. Eight of the sources have reported 
spectral types in the literature, ranging from O to M, even if it remains 
unclear, in the case of the most massive stars, whether the radio emission
might instead come from an unresolved low-mass companion. 
Source 469 is a well-resolved M0+M3.5 binary\footnote{Both
components are detected in our radio observations, if marginally
in the case of the M3.5 star.} \citep{dae12}. The extreme variability originates
on the M0 main component.                                               
The cross-identifications are summarized in Table~\ref{tab_exvar}.

\section{Simultaneous X-ray observations}

In the analysis of our simultaneous X-ray observations, our focus
here is to determine if the extreme radio variability is associated
with unusual X-ray variability, since general studies of X-ray flares
of YSOs have appeared elsewhere.  We analyzed our
\textit{Chandra} observations with the acis\_extract software package \citep{bro10} 
in order to obtain X-ray lightcurves and time series of
median photon energy for all sources that had already been studied
by \citet{get05}.  Here we only discuss the X-ray lightcurves of the
extreme variable radio sample.                                        
 
The net X-ray counts for the present sample of 13 sources are listed
in Table~\ref{tab_exvar}; they range from a few to more than
8000 counts. The brightest X-ray sources show some photon pile-up,
where the count rate is high enough that two or more photons may
have been counted as a single event, affecting a more detailed analysis.
The acis\_extract program estimates that this could affect seven out of 13
sources, marked with asterisks in Table~\ref{tab_exvar}. The strongest
pile-up effect is expected toward the brightest source 254, with
an estimated count rate of 0.4~ct/frame. We have used the pile-up
correction tool described in \citet{bro11}, and we estimate a
pile-up correction of $\sim$45\% for this strong source. 
The pile-up correction, where applicable, thus ranges 
from $\sim10$\% to $\sim$45\%. This effect might alter the shape of 
extracted lightcurves and the assignment of photon energies, but not at
a level that would be a concern for the analysis presented here.
 
\begin{figure*} 
 
\begin{minipage}{0.32\linewidth} 
\includegraphics[width=\linewidth, bb=54 290 559 522]{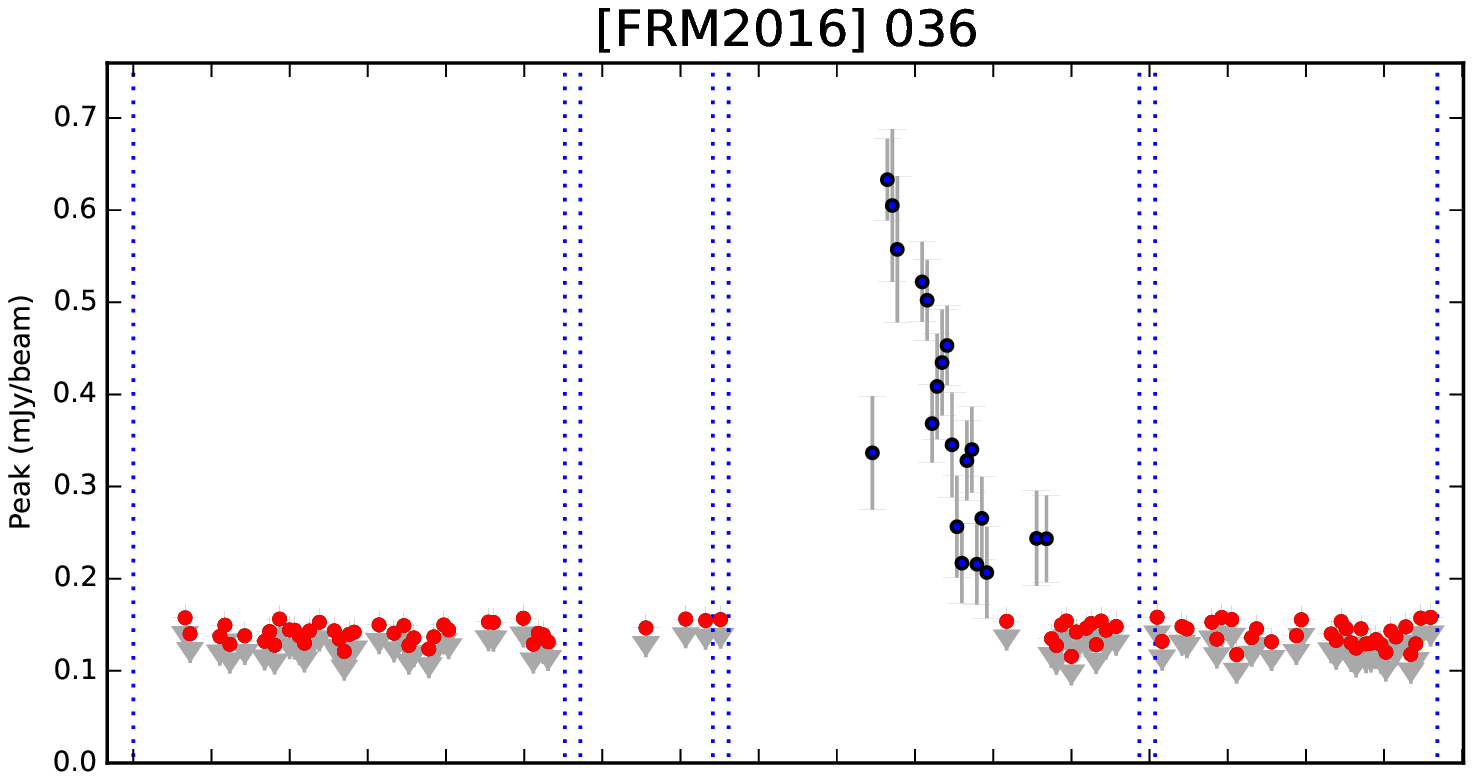}\\
 \hspace*{2.1mm}\includegraphics*[width=0.655\linewidth,angle=90,bb=35
35 557 757]{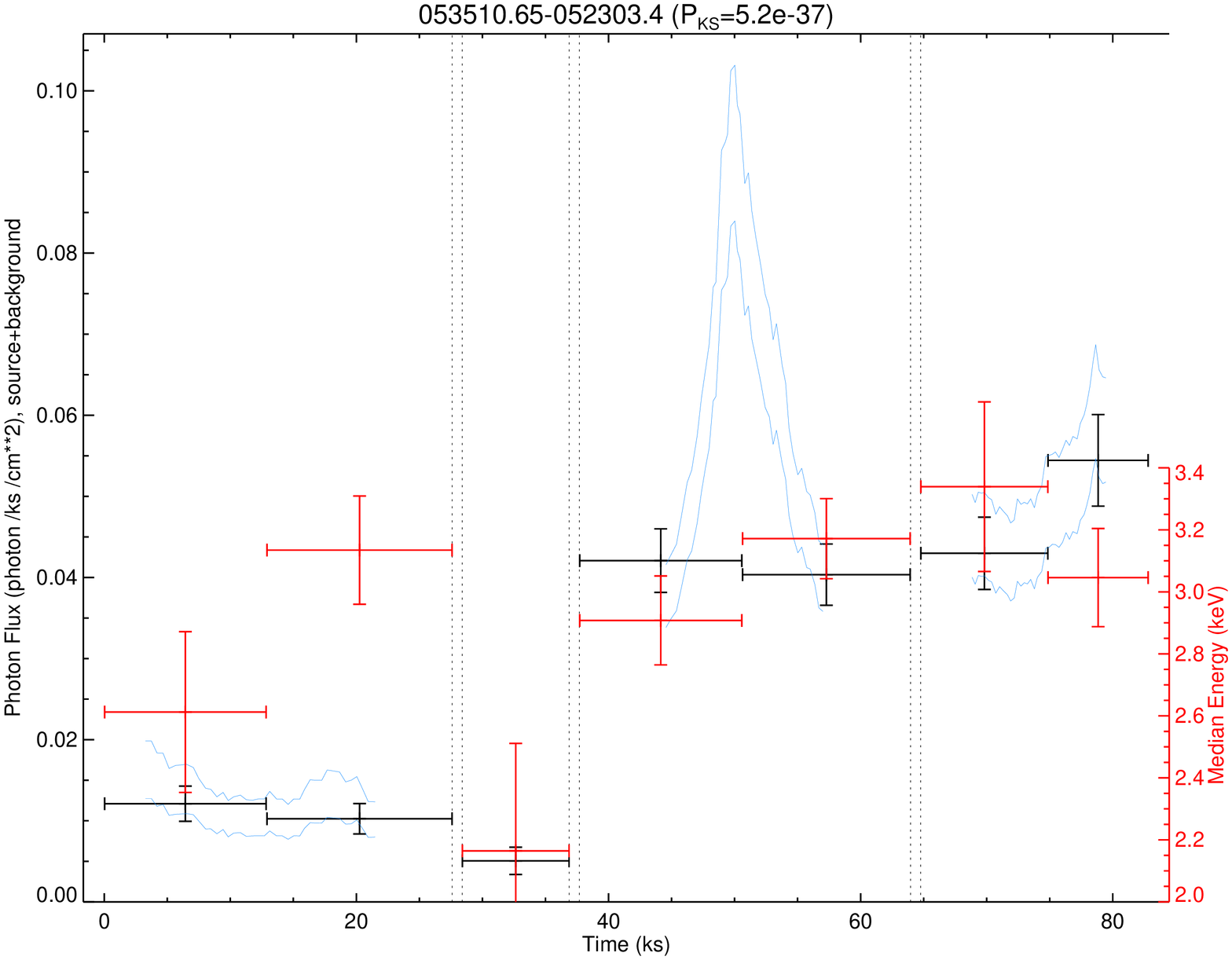}                                              
\end{minipage} 
\begin{minipage}{0.32\linewidth} 
\includegraphics[width=\linewidth, bb=54 290 559 522]{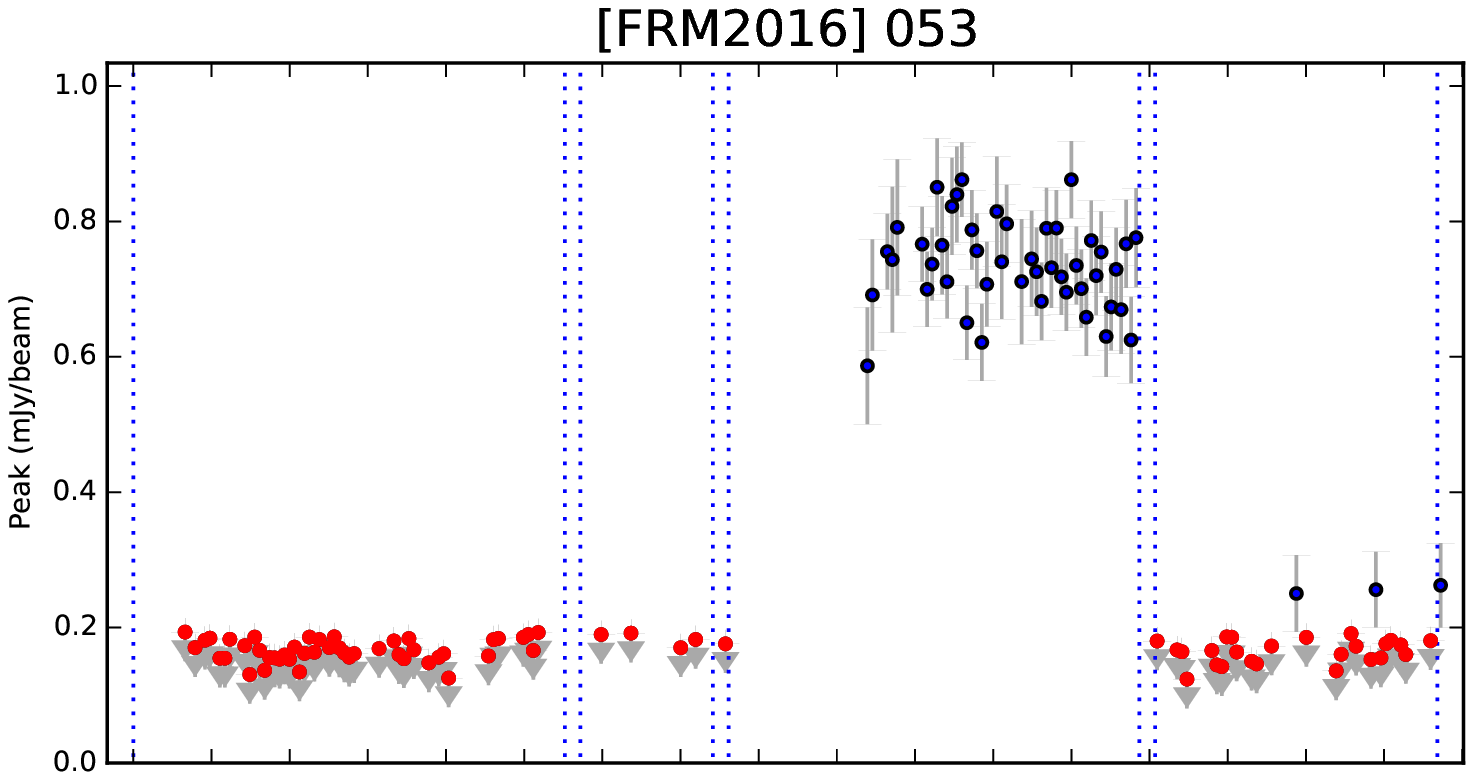}\\
 \hspace*{2.1mm}\includegraphics*[width=0.655\linewidth,angle=90,bb=35
35 557 757]{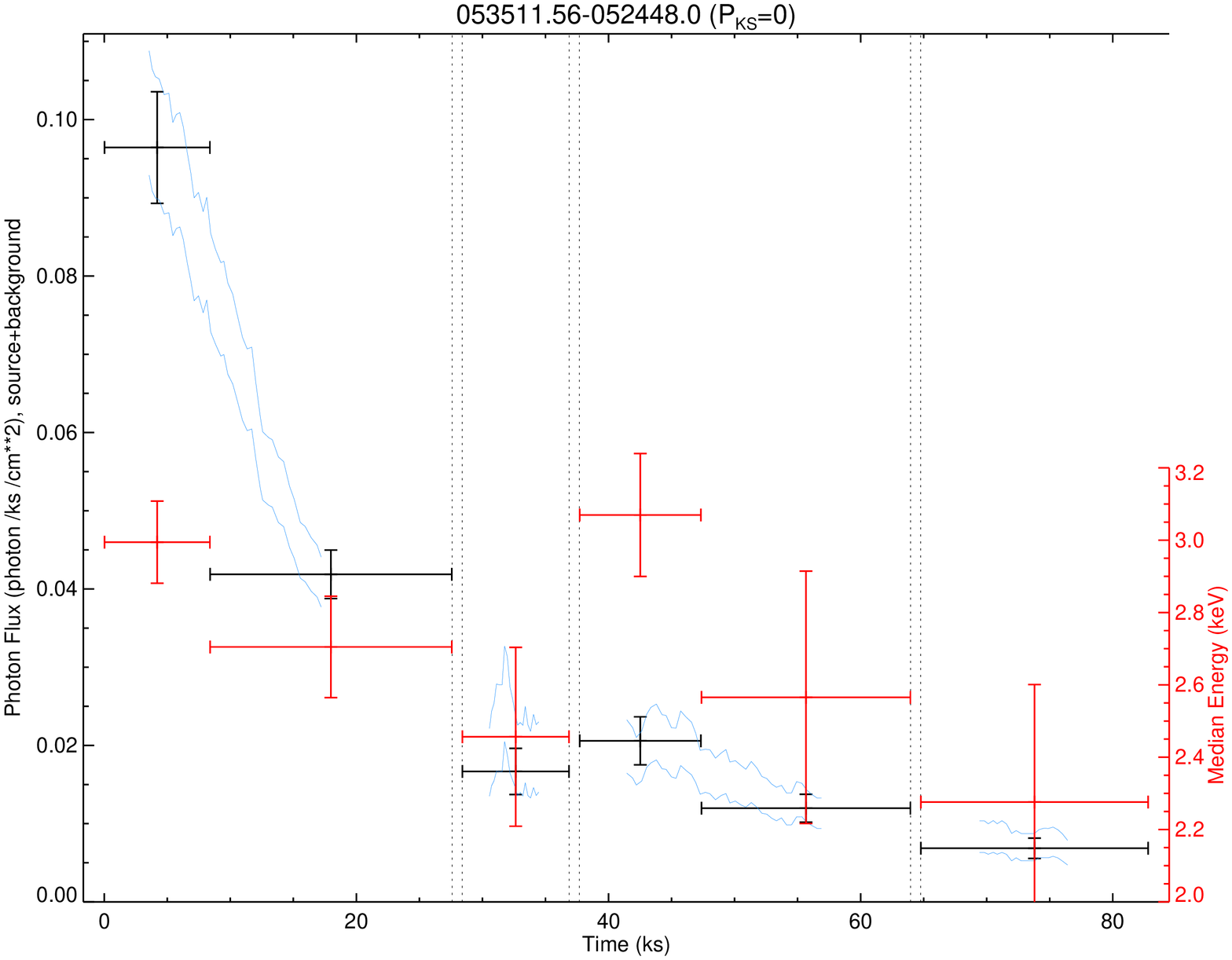}                                              
\end{minipage} 
\begin{minipage}{0.32\linewidth} 
\includegraphics[width=\linewidth, bb=54 290 559 522]{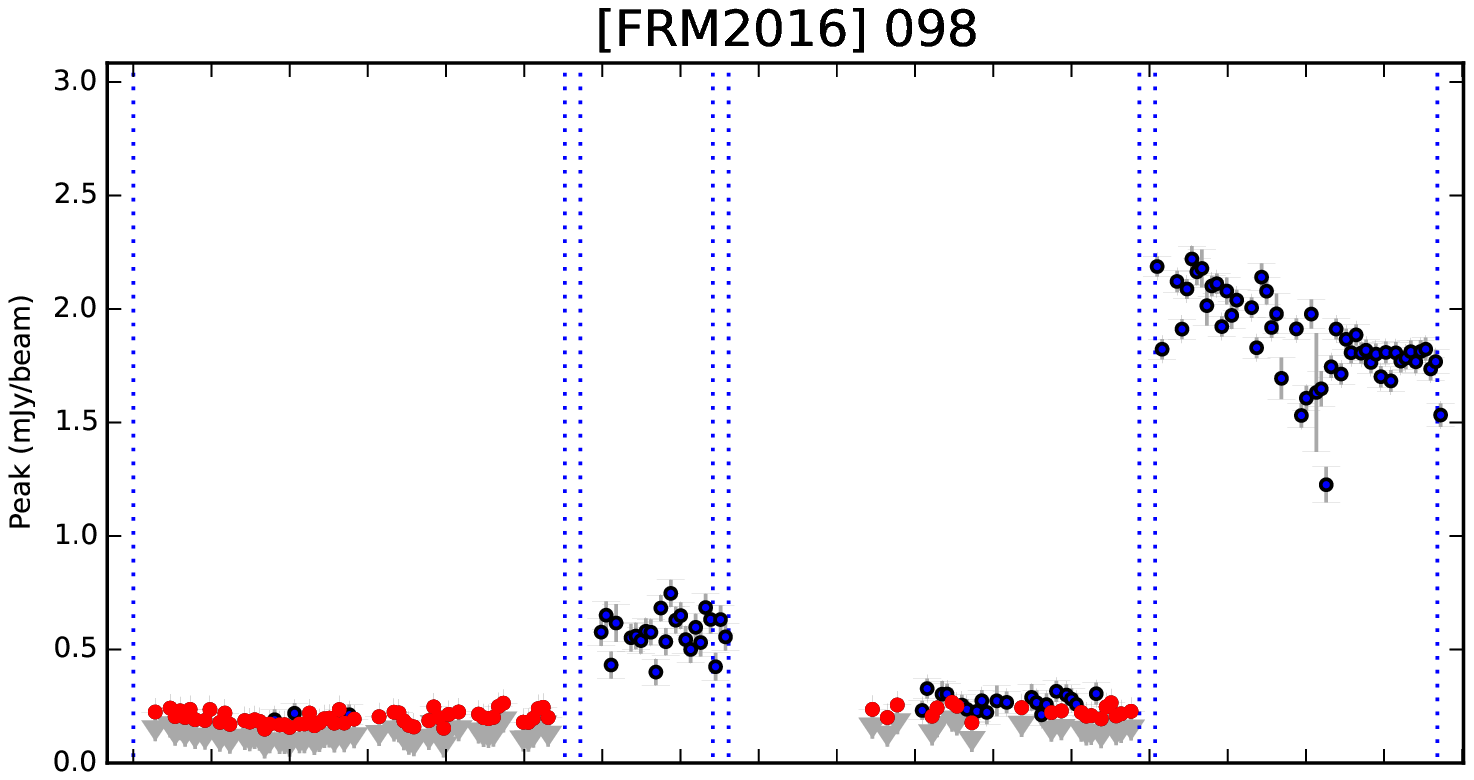}\\
 \hspace*{2.1mm}\includegraphics*[width=0.655\linewidth,angle=90,bb=35
35 557 757]{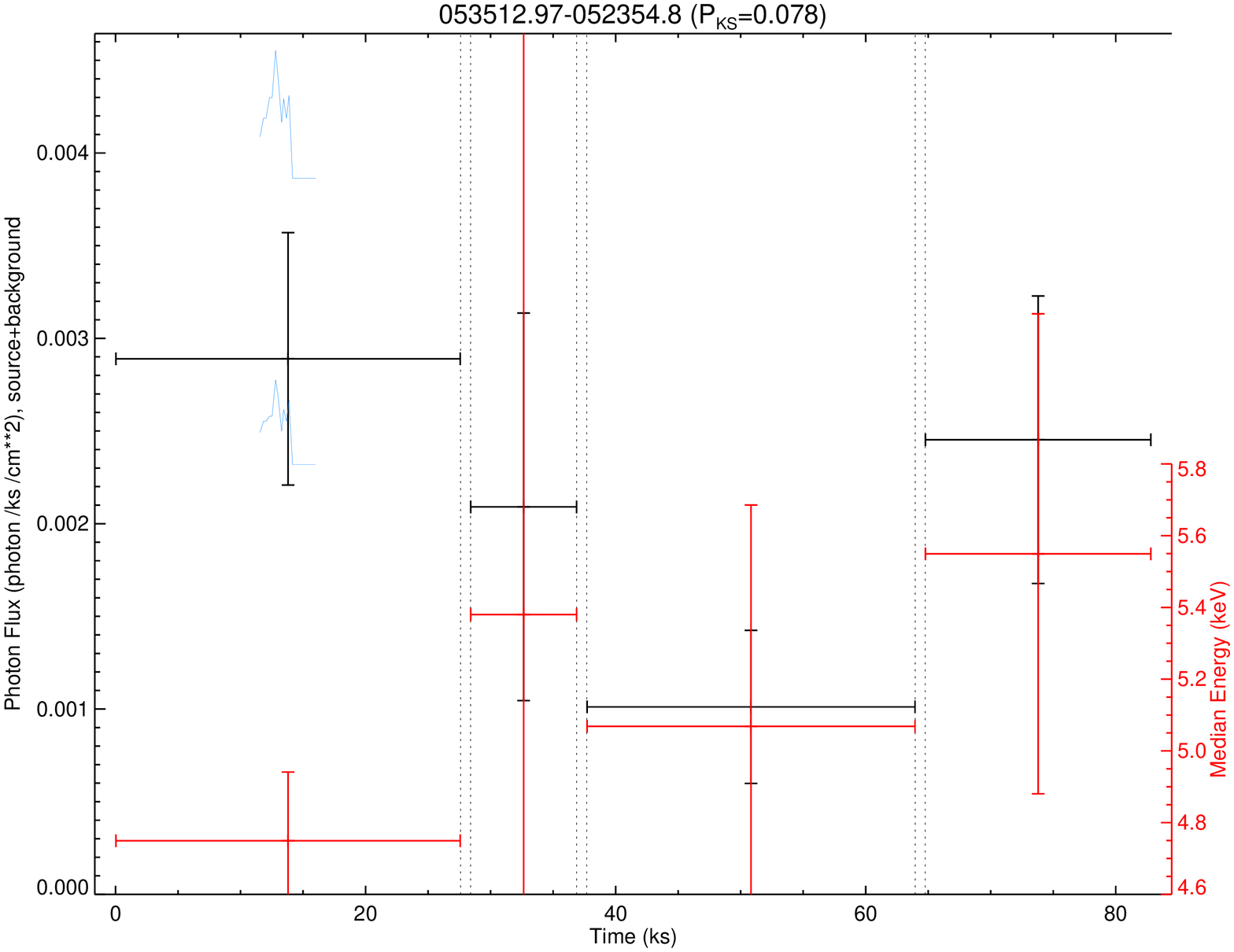}                                              
\end{minipage} 
 
\begin{minipage}{0.32\linewidth} 
\includegraphics[width=\linewidth, bb=54 290 559 522]{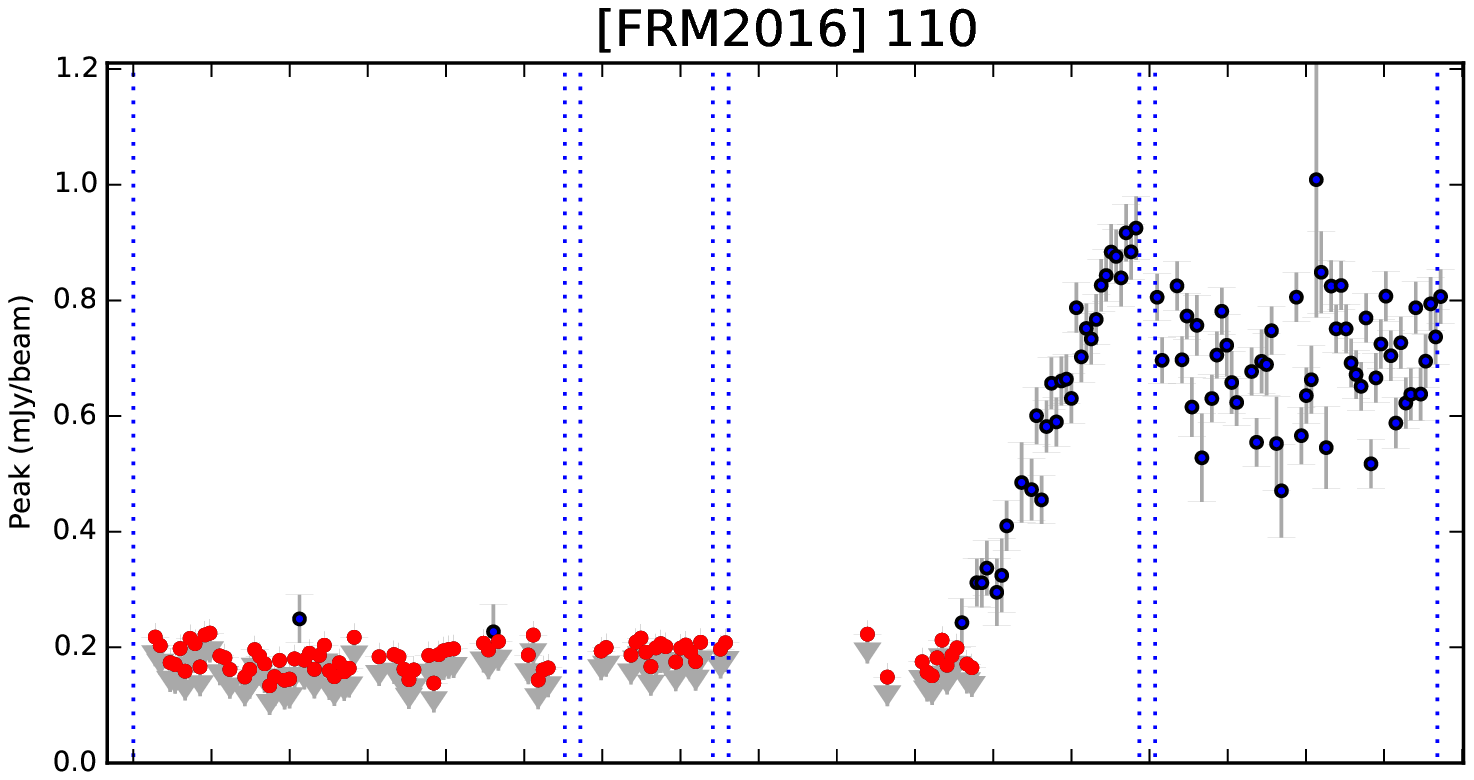}\\
 \hspace*{2.1mm}\includegraphics*[width=0.655\linewidth,angle=90,bb=35
35 557 757]{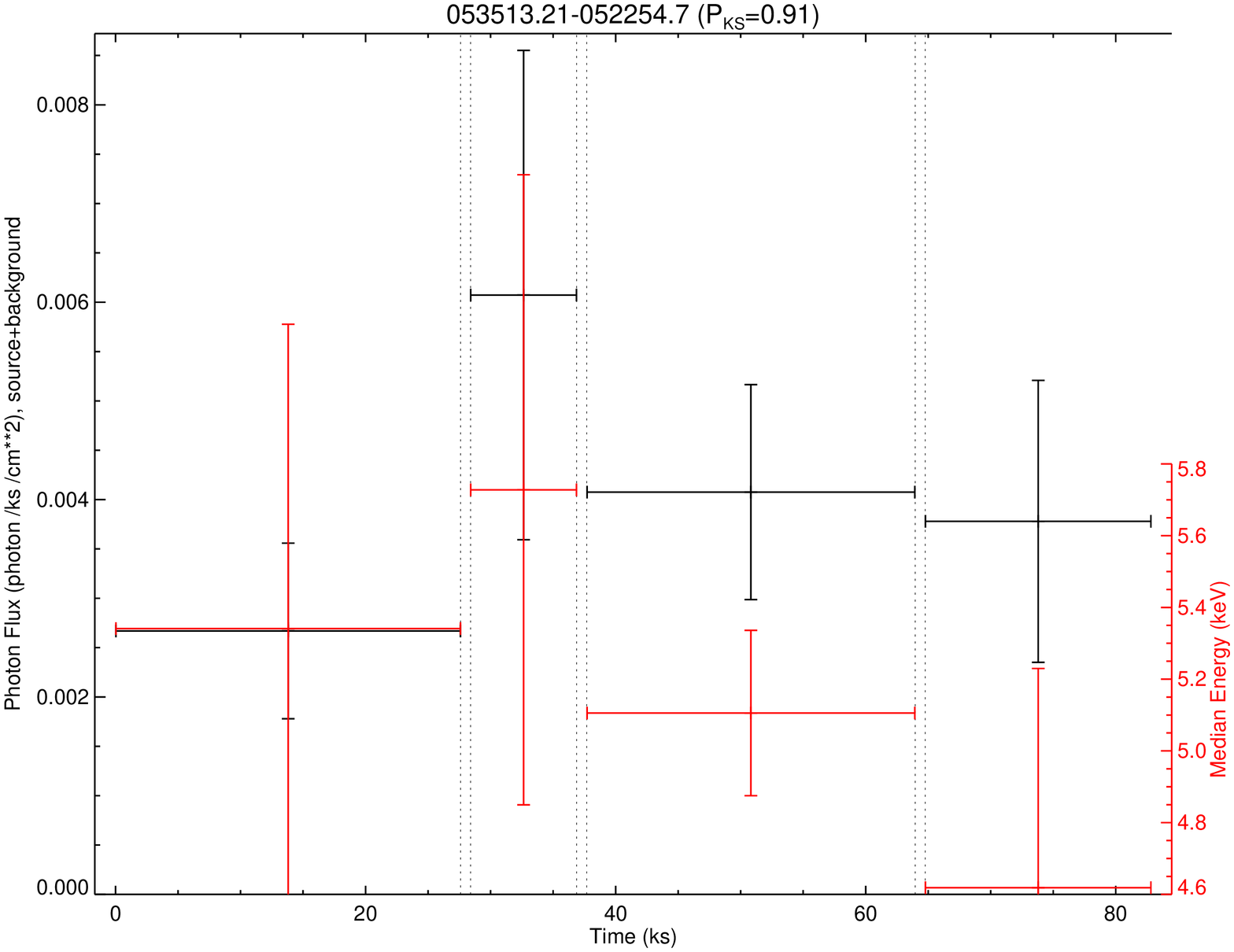}                                              
\end{minipage} 
\begin{minipage}{0.32\linewidth} 
\includegraphics[width=\linewidth, bb=54 290 559 522]{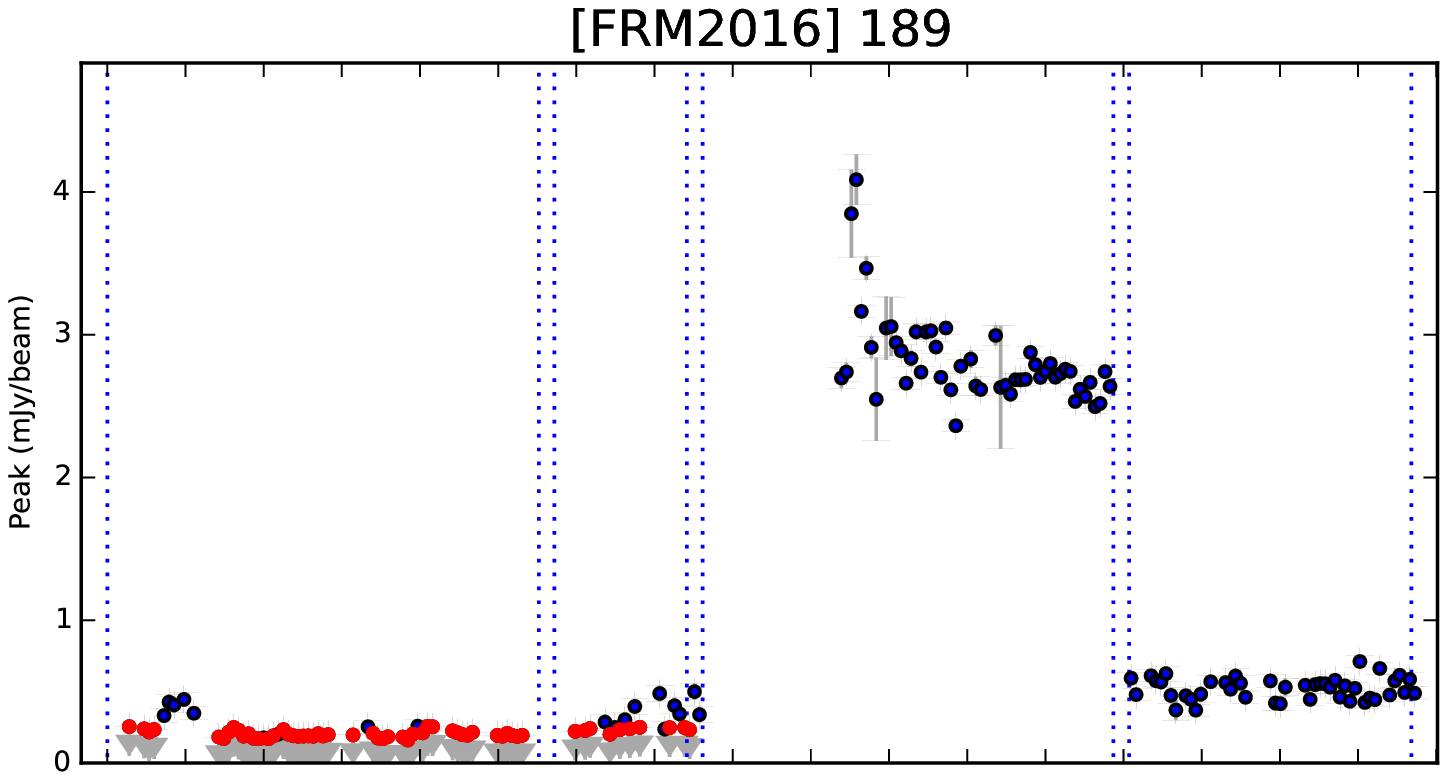}\\
 \hspace*{2.1mm}\includegraphics*[width=0.655\linewidth,angle=90,bb=35
35 557 757]{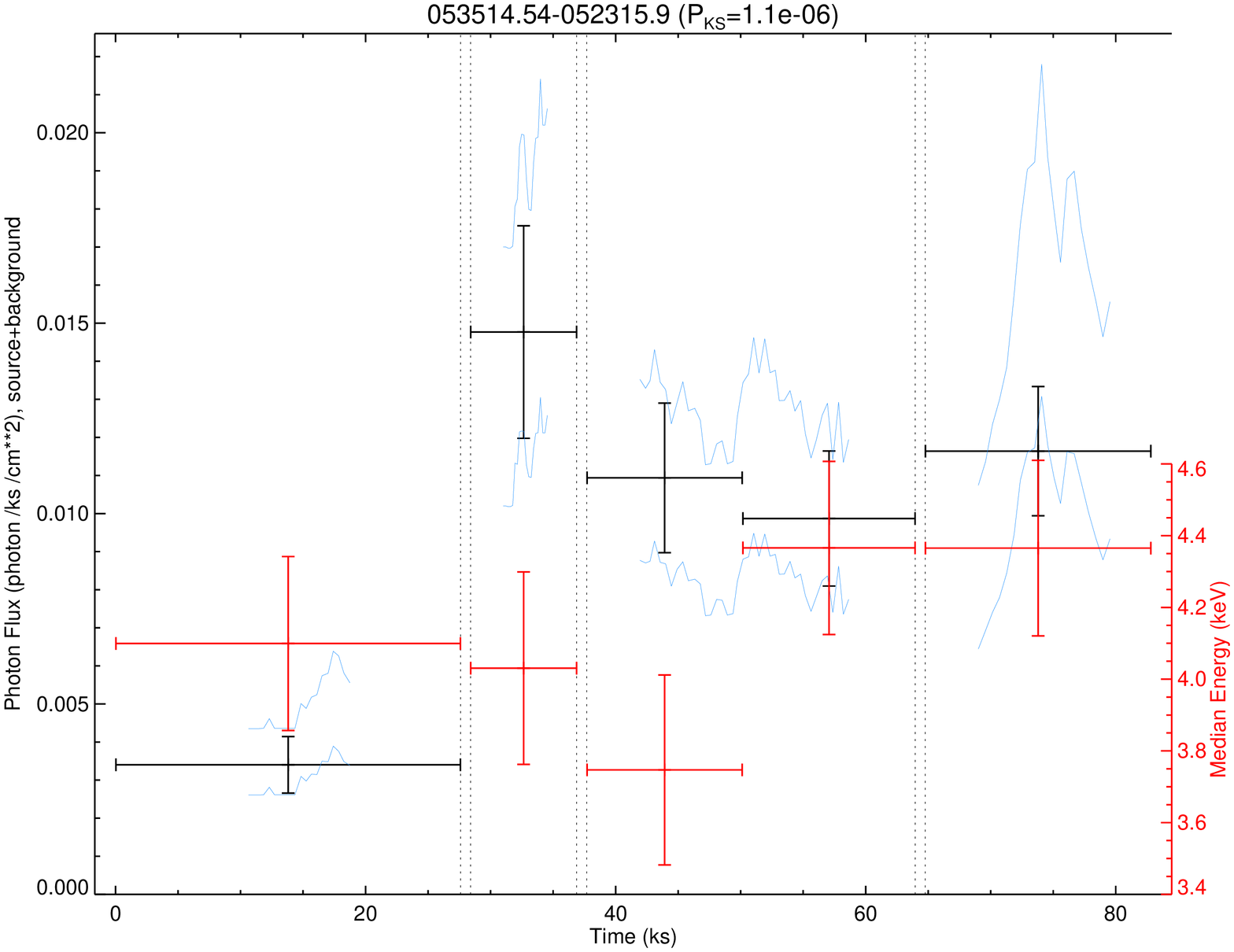}                                              
\end{minipage} 
\begin{minipage}{0.32\linewidth} 
\includegraphics[width=\linewidth, bb=54 290 559 522]{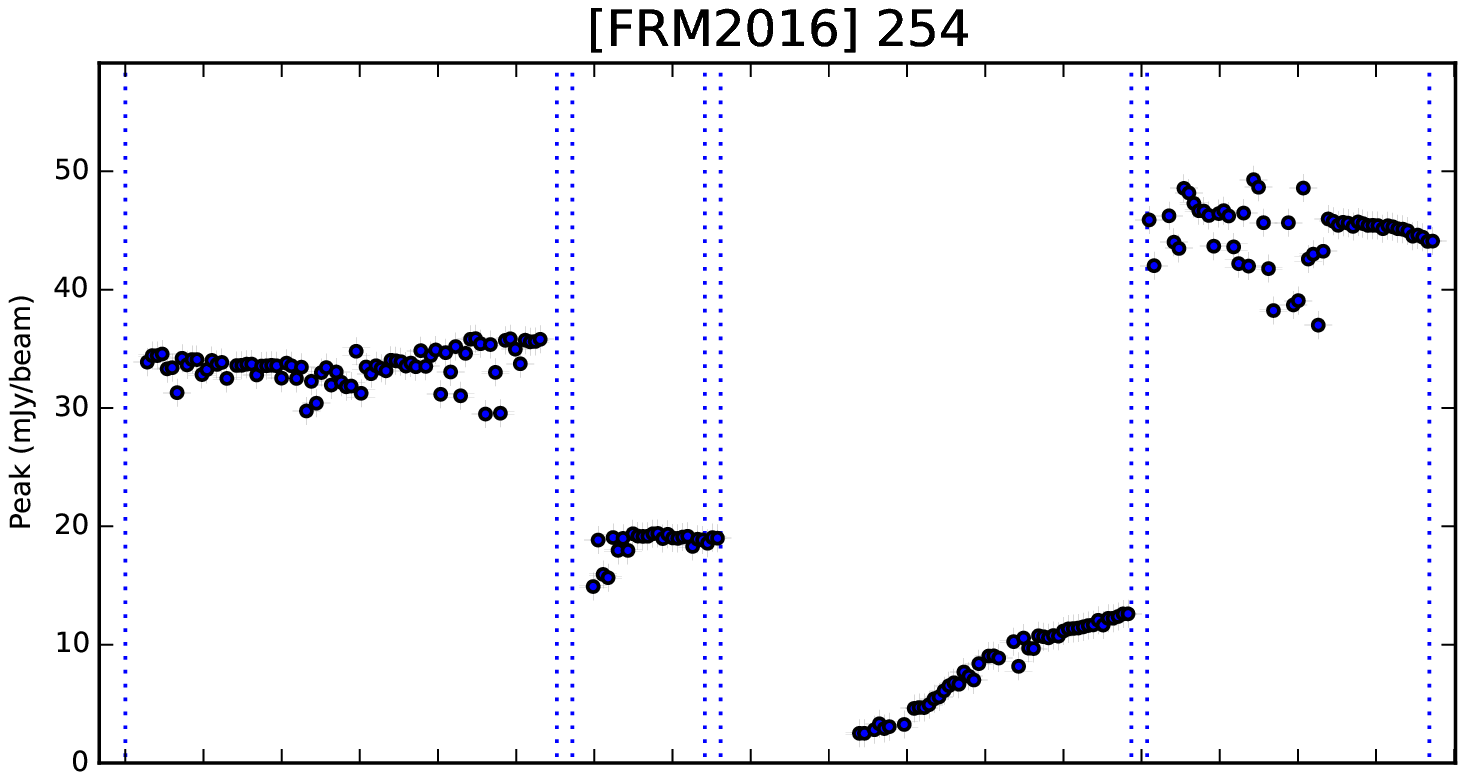}\\
 \hspace*{2.1mm}\includegraphics*[width=0.655\linewidth,angle=90,bb=35
35 557 757]{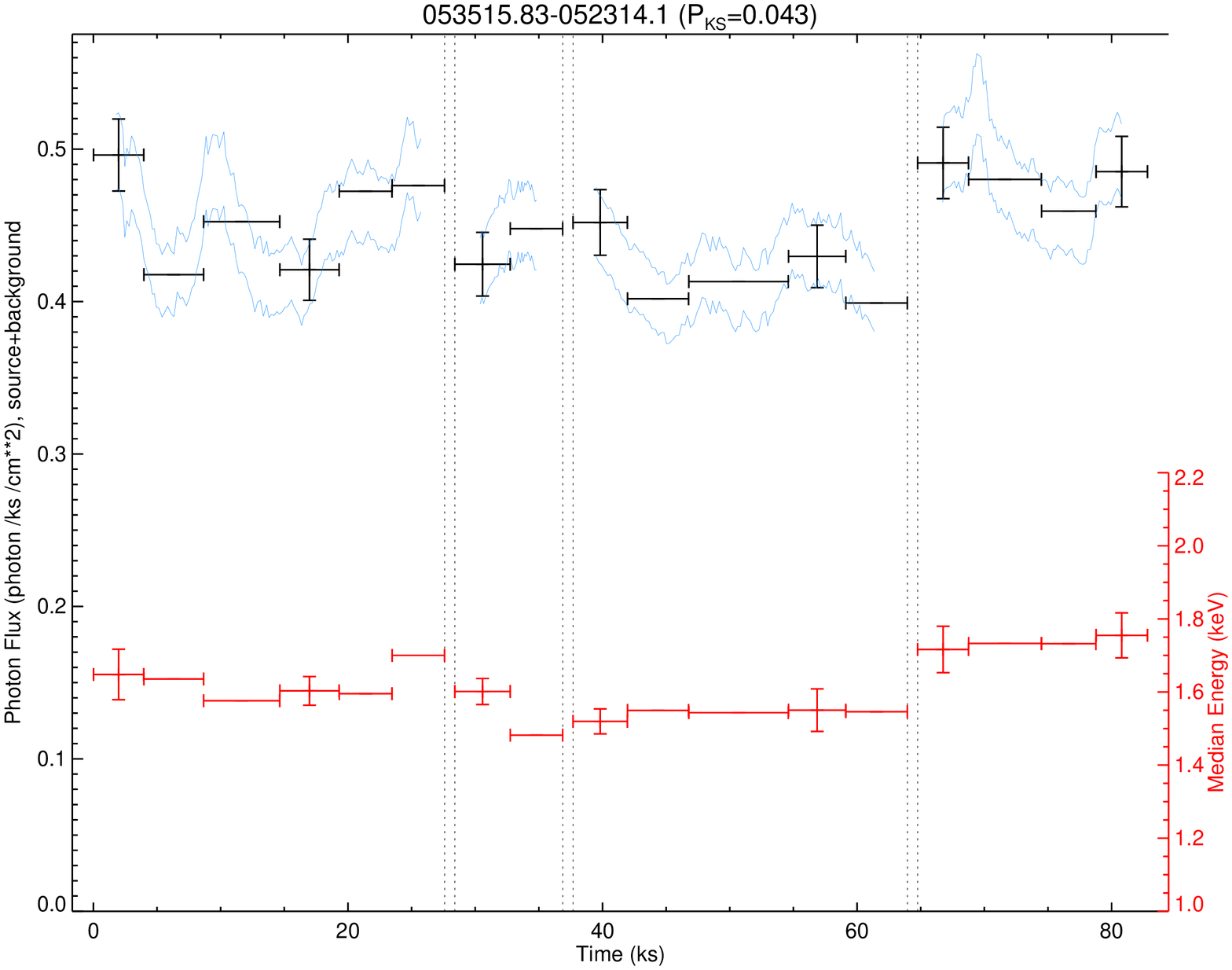}                                              
\end{minipage} 
 
\begin{minipage}{0.32\linewidth} 
\includegraphics[width=\linewidth, bb=54 290 559 522]{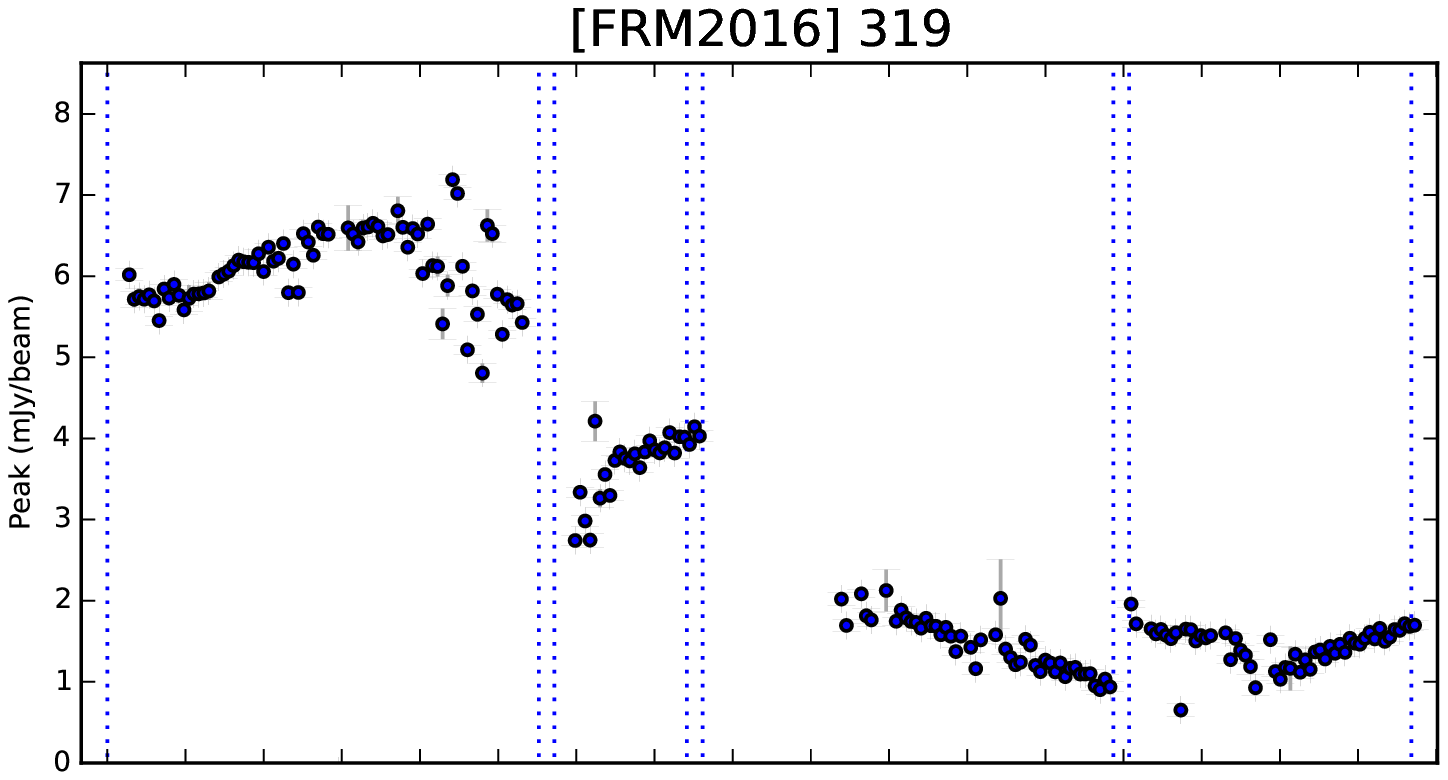}\\
 \hspace*{2.1mm}\includegraphics*[width=0.655\linewidth,angle=90,bb=35
35 557 757]{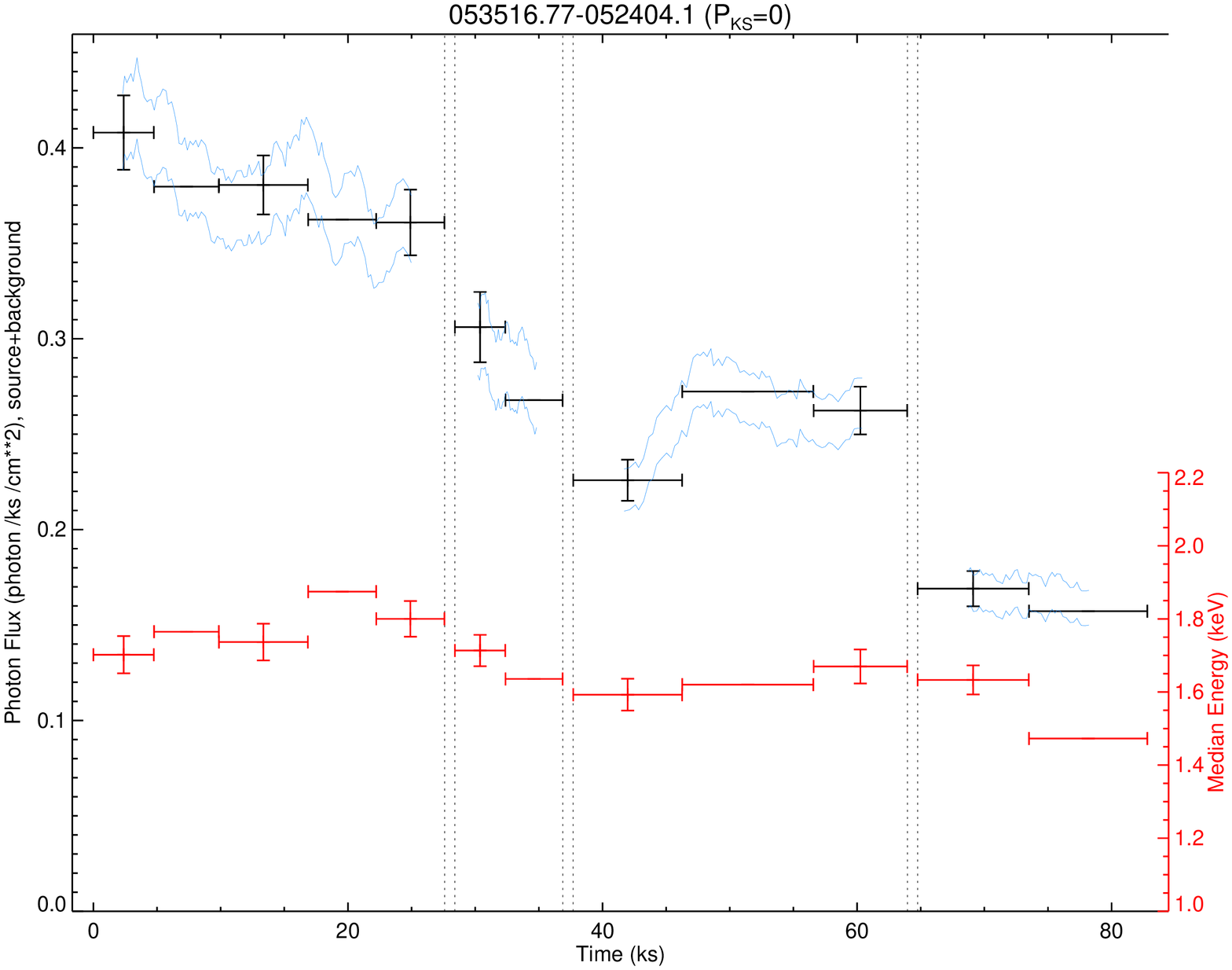}                                              
\end{minipage} 
\begin{minipage}{0.32\linewidth} 
\includegraphics[width=\linewidth, bb=54 290 559 522]{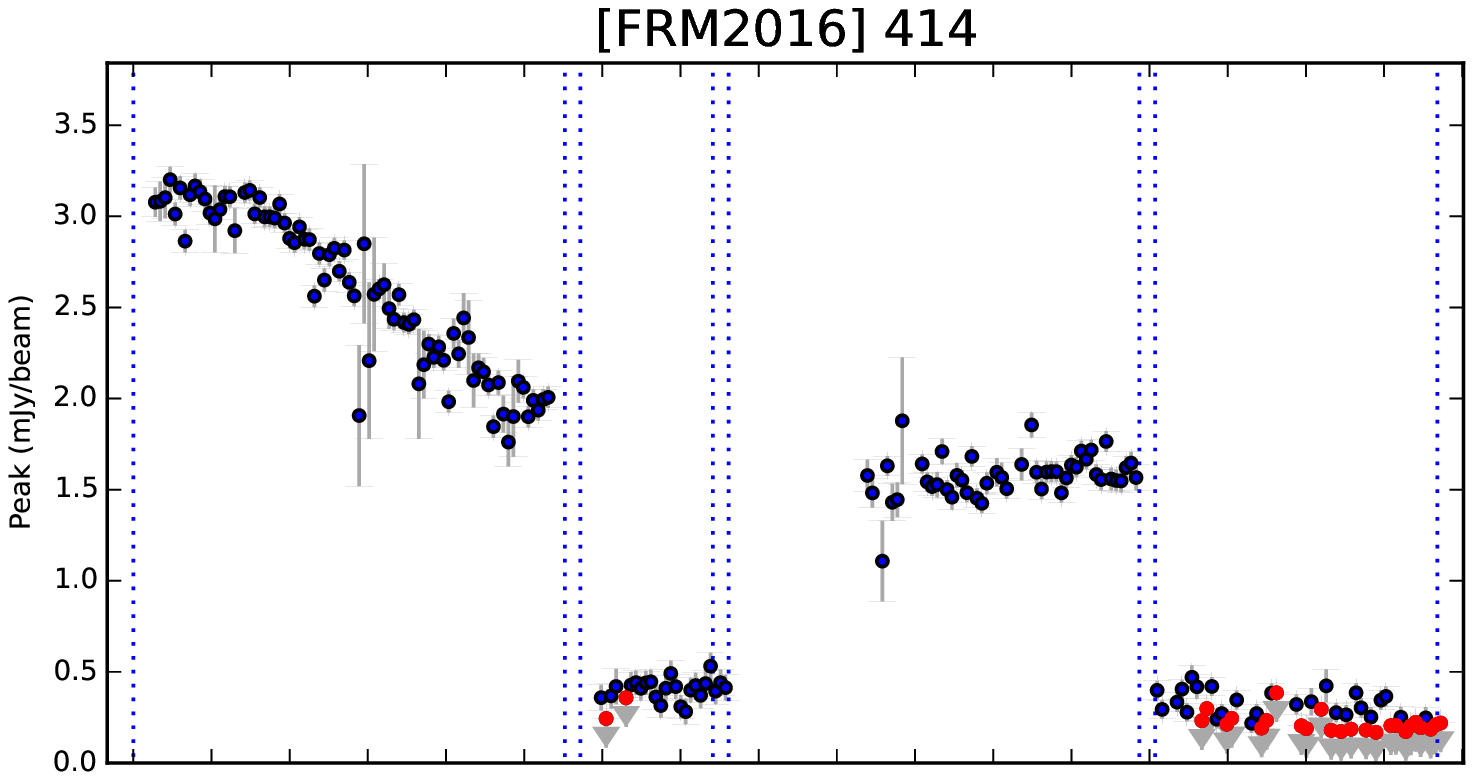}\\
 \hspace*{2.1mm}\includegraphics*[width=0.655\linewidth,angle=90,bb=35
35 557 757]{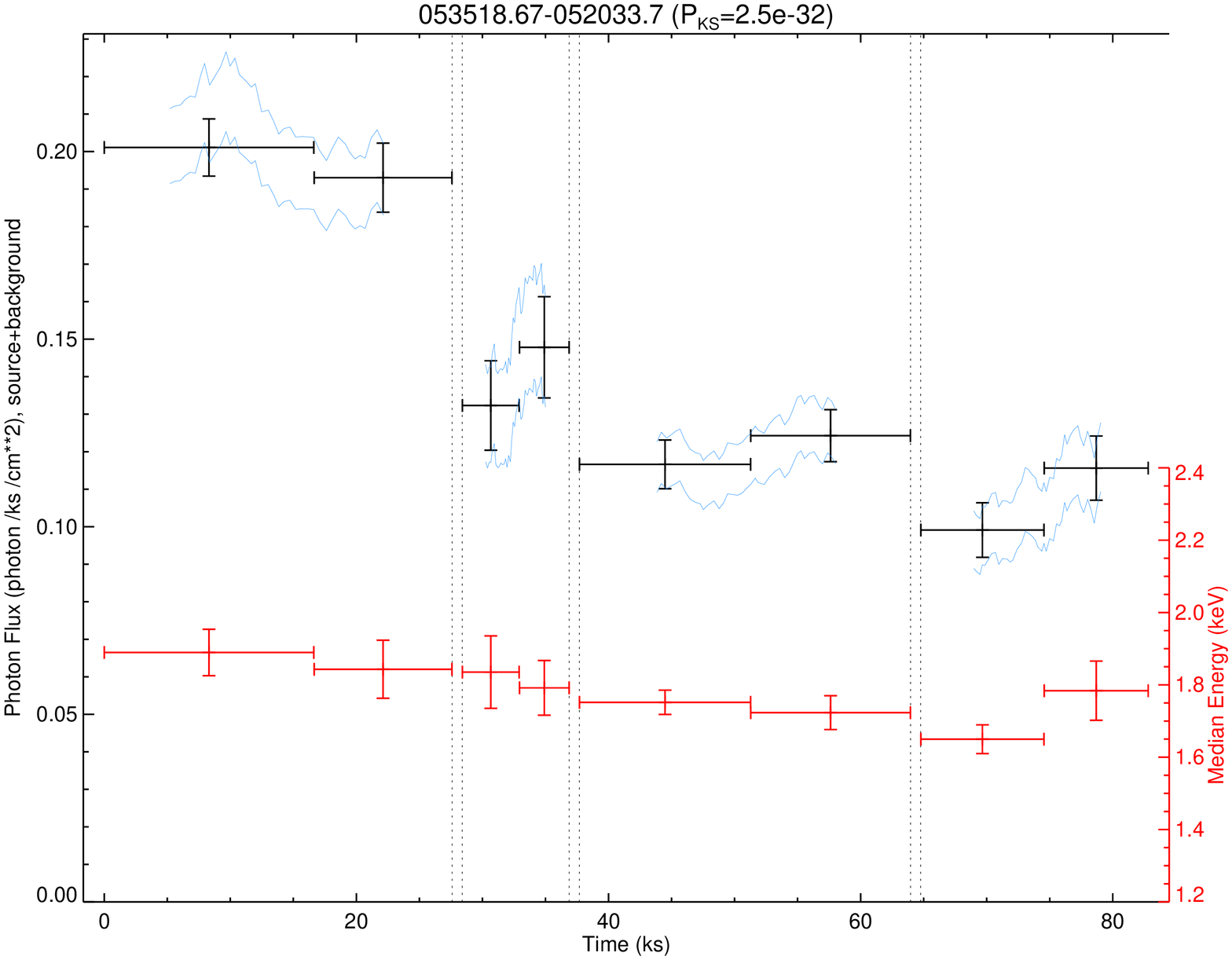}                                              
\end{minipage} 
\begin{minipage}{0.32\linewidth} 
\includegraphics[width=\linewidth, bb=54 290 559 522]{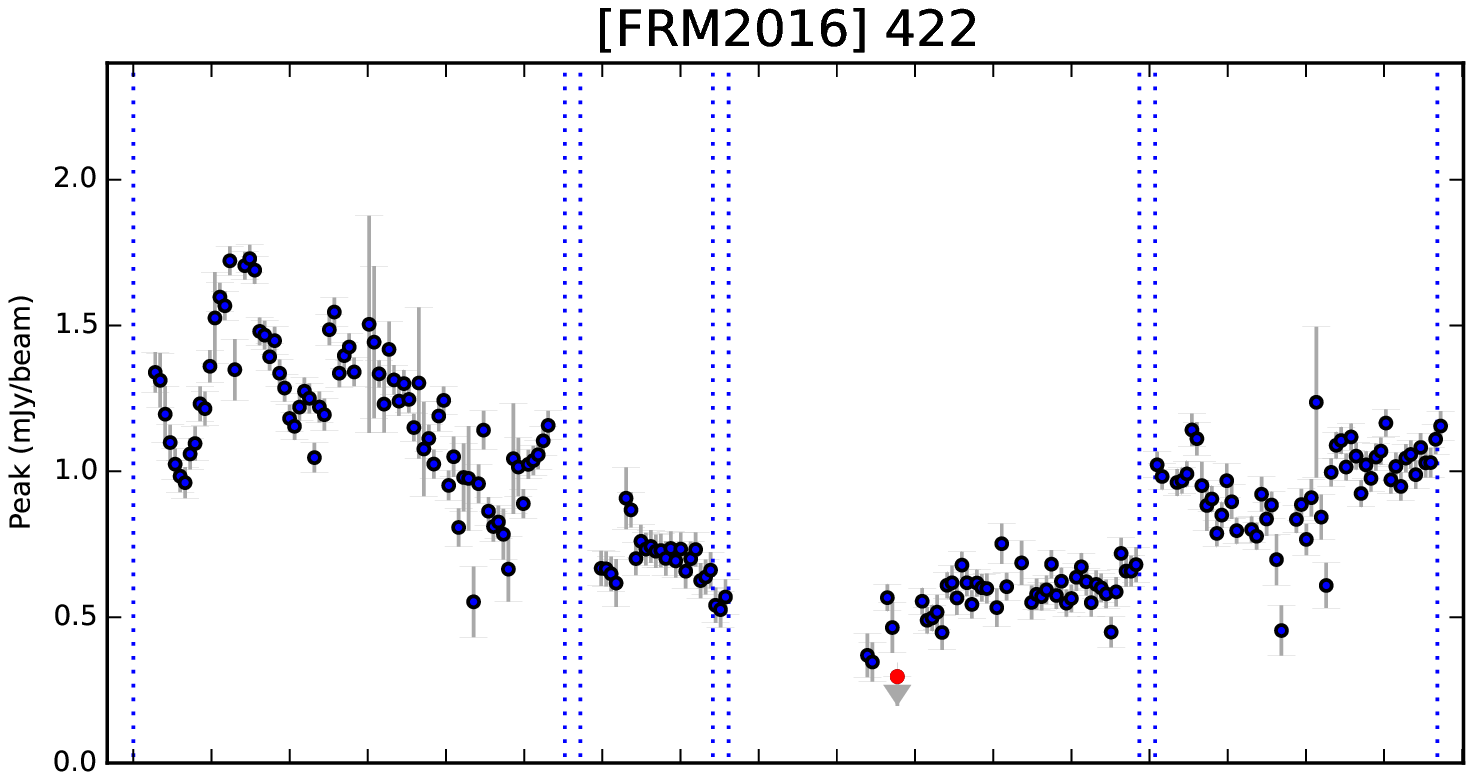}\\
 \hspace*{2.1mm}\includegraphics*[width=0.655\linewidth,angle=90,bb=35
35 557 757]{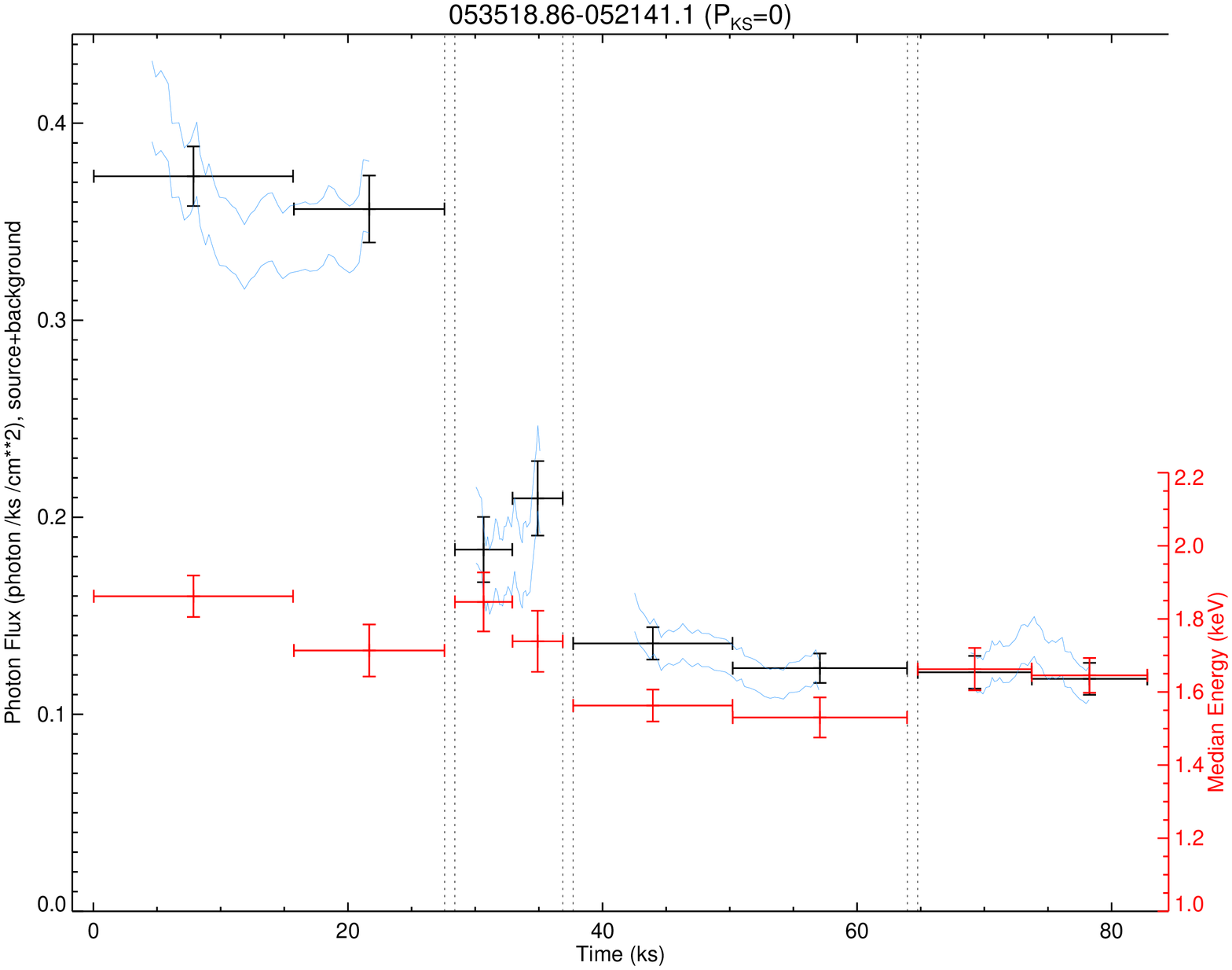}                                              
\end{minipage} 
 
\caption{Simultaneous radio and X-ray lightcurves. The upper
panel for each source shows the same radio data, at 6~min time resolution,
that is also shown in Figure~\ref{fig_var_all}, but only if the data
coincide with X-ray observations. The time axis of the radio lightcurves has been converted to match the cumulative {\it Chandra} elapsed time in ksec. The lower panel shows both a binned
X-ray lightcurve (black crosses) and a binned median photon energy
curve (red crosses). Additionally, it shows the upper and lower confidence
intervals of the adaptively smoothed photon counts (blue), see \citet{bro10}.\label{fig_xr}}
 
\end{figure*}

To assess the X-ray variability in the sample, we follow the acis\_extract
approach to analyze both a binned lightcurve and also a smoothed
unbinned lightcurve, built using the photon arrival times and adaptive
kernel smoothing \citep{bro10}. The latter is particularly useful
to identify substructure that changes on
timescales shorter than the bin size. The X-ray lightcurves for the
extremely variable radio sources are shown in Figure~\ref{fig_xr},
together with the simultaneous portions of the radio lightcurves. To match up
the X-ray and radio data, we have converted the time axes of the radio 
lightcurves to the time axes of the X-ray lightcurves, in cumulative 
{\it Chandra} elapsed time (in kiloseconds).
 
\begin{figure*} 
 
\begin{minipage}{0.32\linewidth} 
\includegraphics[width=\linewidth, bb=54 290 559 522]{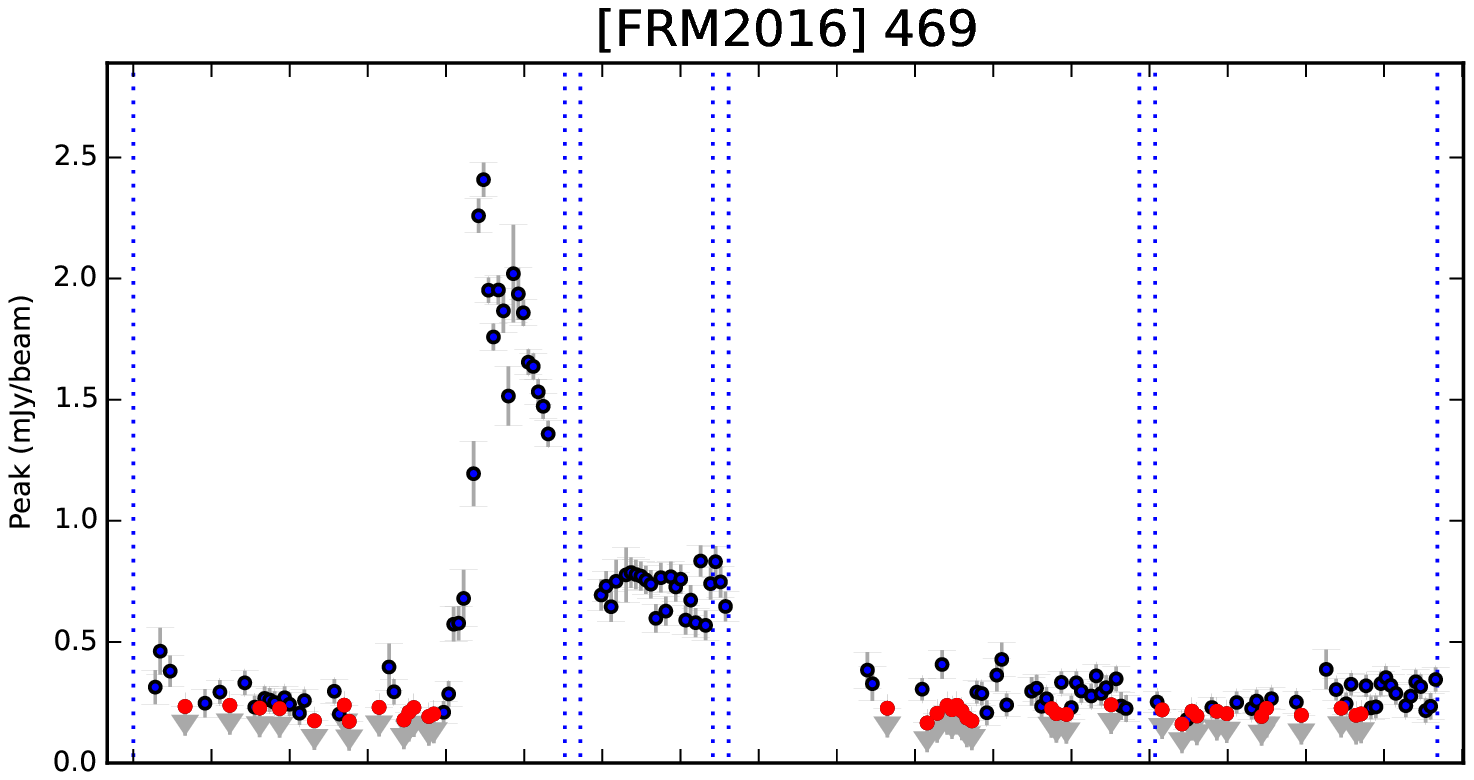}\\
 \hspace*{2.1mm}\includegraphics*[width=0.655\linewidth,angle=90,bb=35
35 557 757]{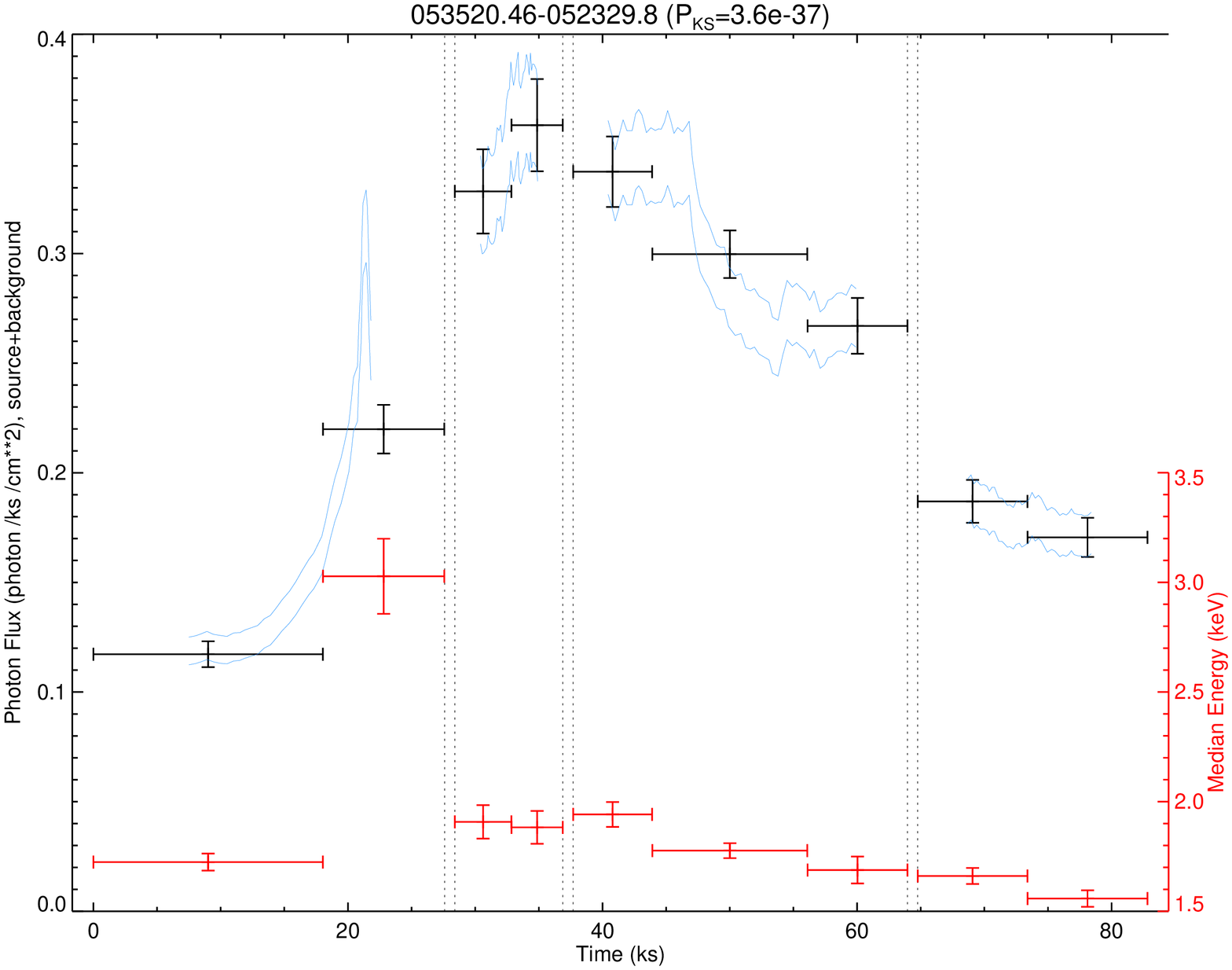}                                              
\end{minipage} 
\begin{minipage}{0.32\linewidth} 
\includegraphics[width=\linewidth, bb=54 290 559 522]{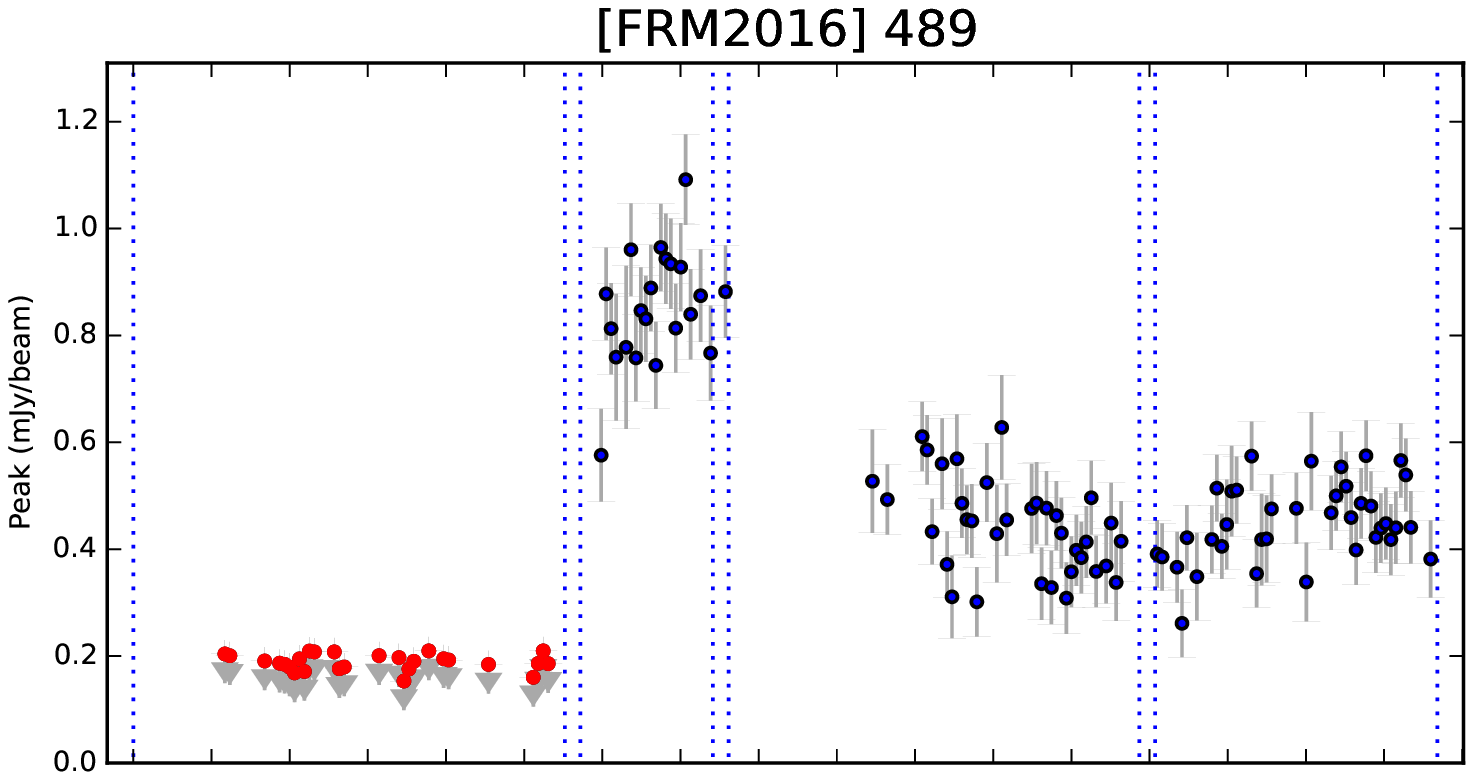}\\
 \hspace*{2.1mm}\includegraphics*[width=0.655\linewidth,angle=90,bb=35
35 557 757]{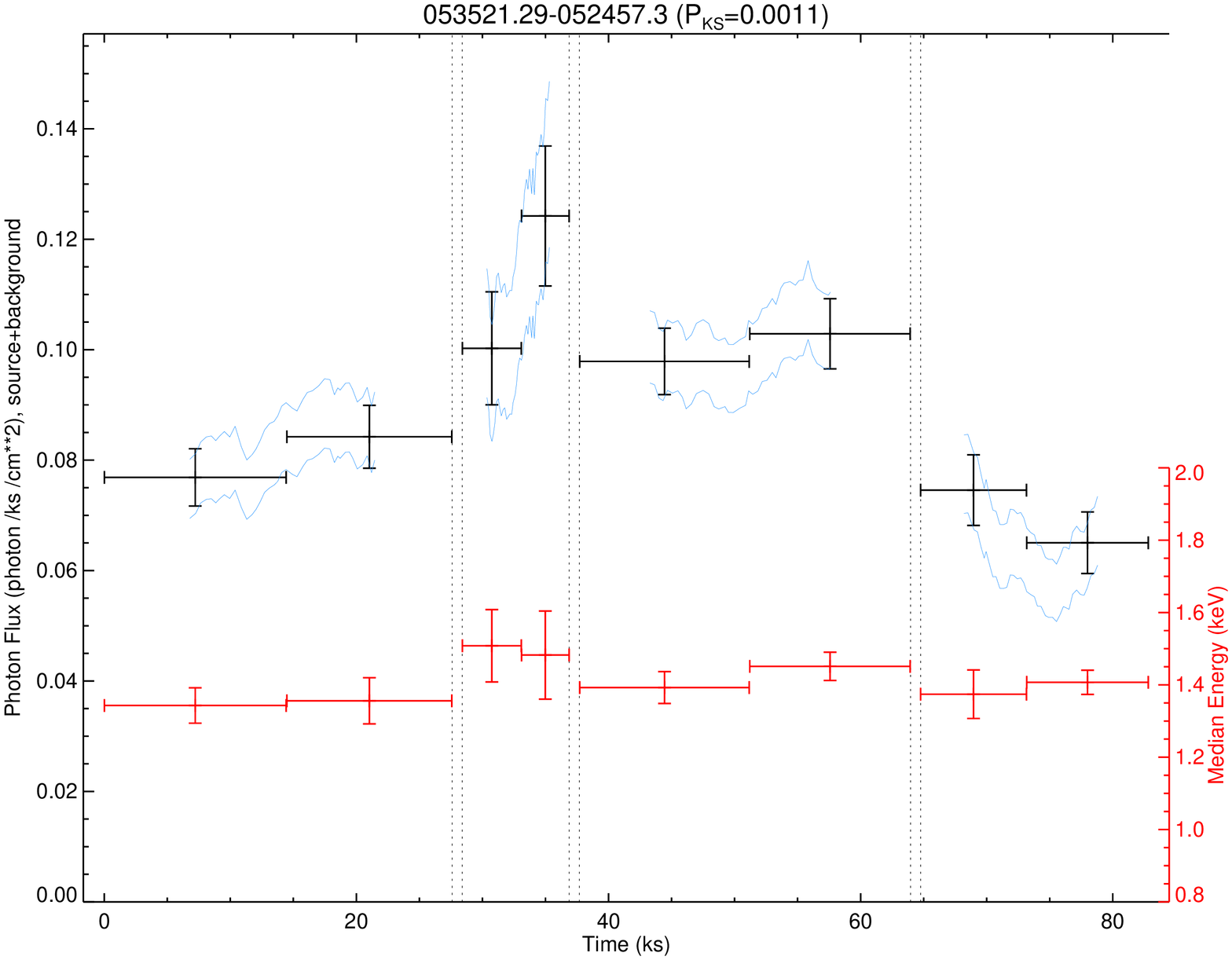}                                              
\end{minipage} 
\begin{minipage}{0.32\linewidth} 
\includegraphics[width=\linewidth, bb=54 290 559 522]{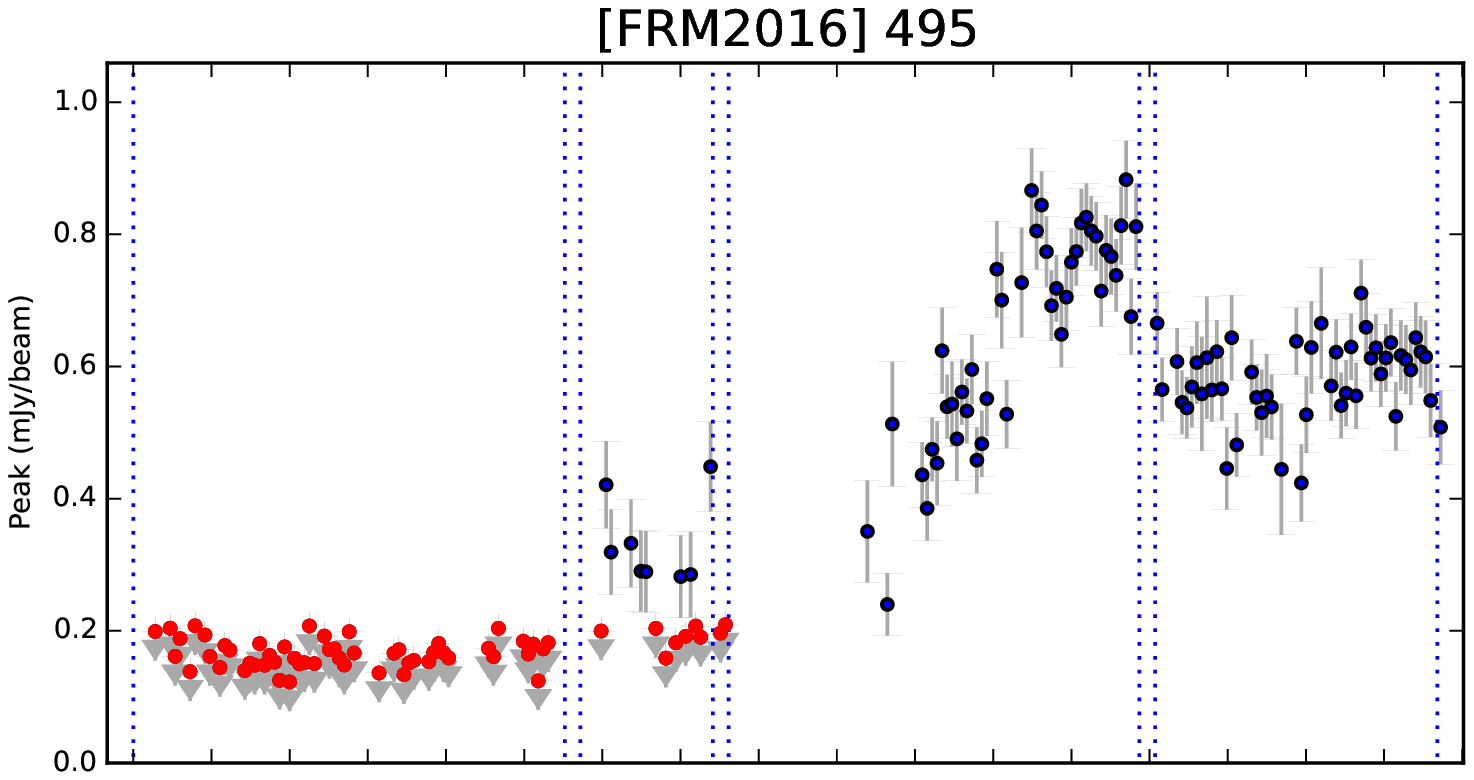}\\
 \hspace*{2.1mm}\includegraphics*[width=0.655\linewidth,angle=90,bb=35
35 557 757]{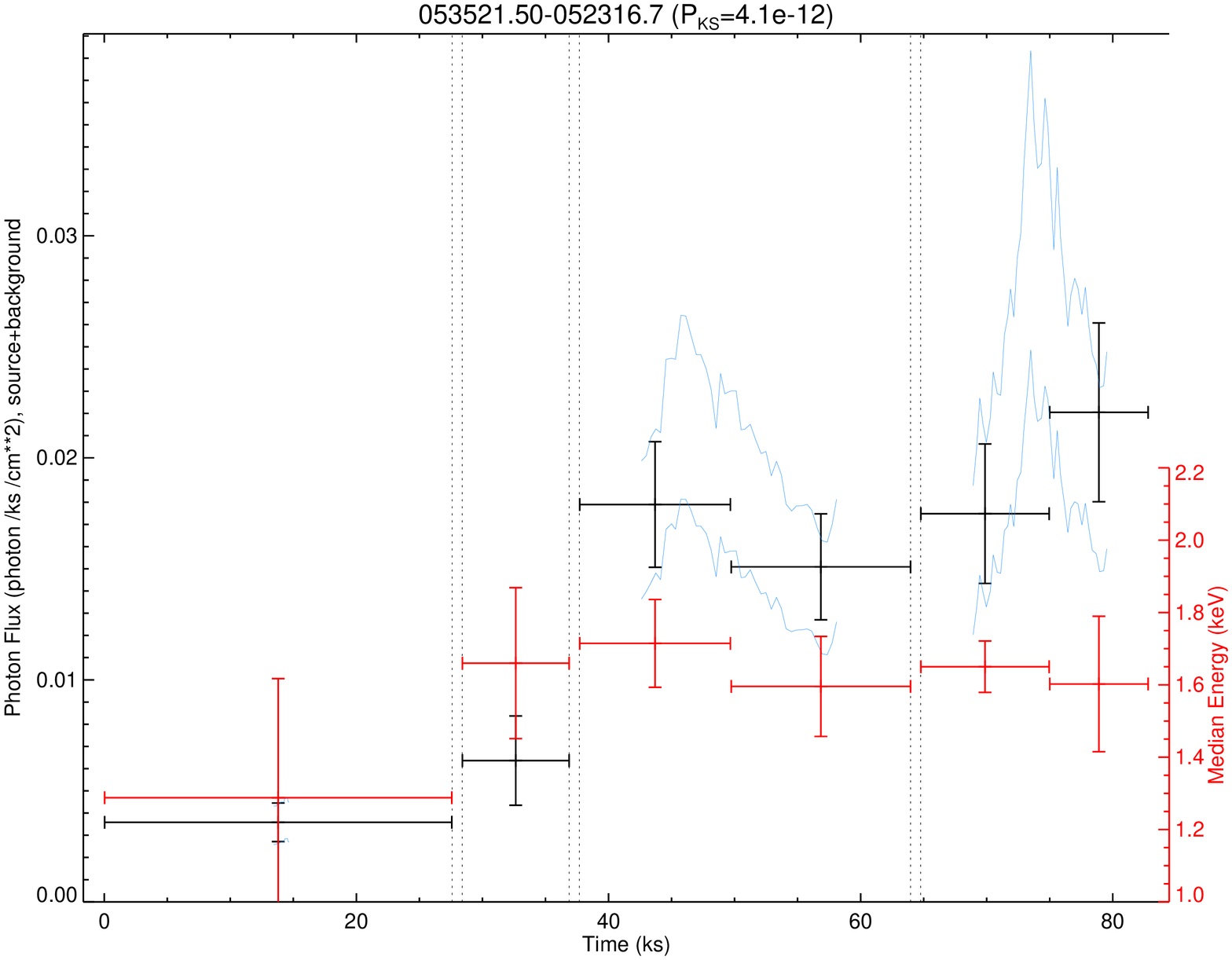}                                              
\end{minipage} 
 
\begin{minipage}{0.32\linewidth} 
\includegraphics[width=\linewidth, bb=54 290 559 522]{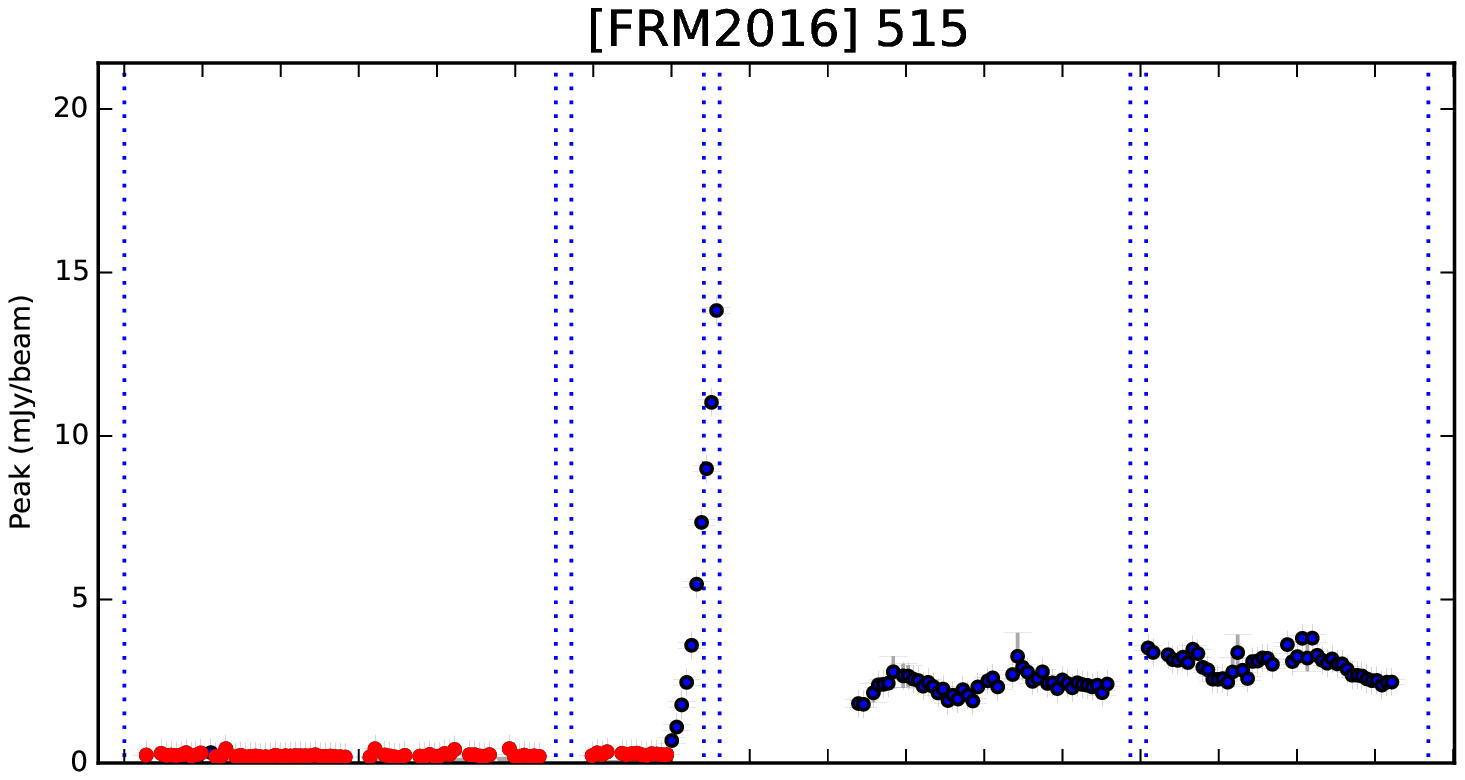}\\
 \hspace*{2.1mm}\includegraphics*[width=0.655\linewidth,angle=90,bb=35
35 557 757]{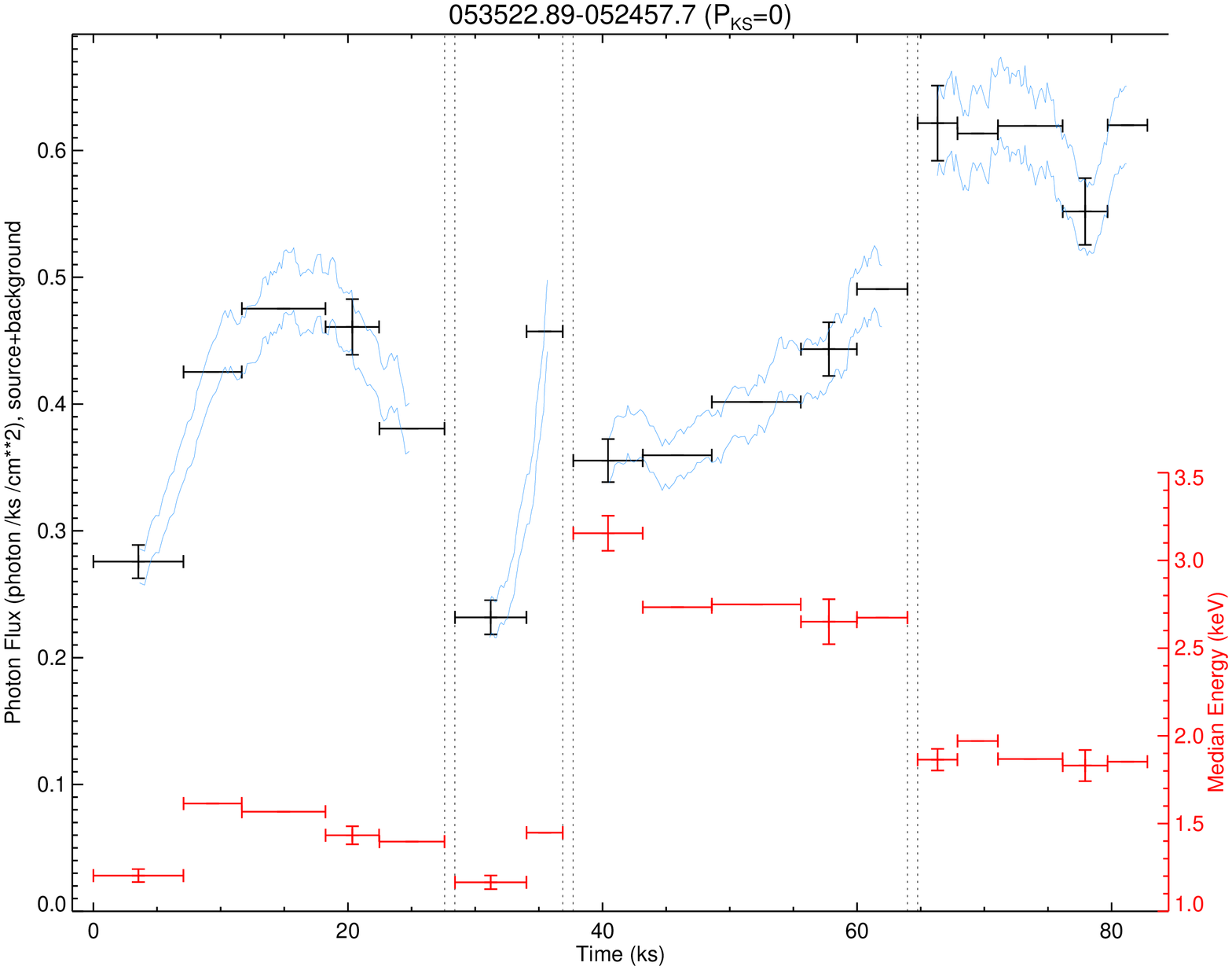}                                              
\end{minipage}

{\bf Figure~\ref{fig_xr}} cntd. 
 
\end{figure*}

While the focus of this paper is extreme radio variability, we have
also quantified extreme X-ray variability in the four epochs of {\it
Chandra} data. This allows us to quantify to what degree extreme
radio and X-ray variability are correlated.  To quantify extreme
X-ray variability, we use the ``glvary'' algorithm
of the Chandra Interactive Analysis of Observations software package (CIAO, \citealp{gre92,rot06}).
This program uses the Gregory-Loredo variability test algorithm
on the unbinned X-ray data, using multiple different time bins and
looking for significant deviations. As a result, it computes a variability
index (``varindex'') on a scale of 0 to 10, where 0 corresponds to a 
source that is definitely not variable and 6 for a source that
is definitely variable. A varindex of 10 indicates extreme variability akin to a flare. 
 
We determine the varindex for three different source samples on a
per-epoch basis. For the first sample, we take all detected X-ray sources, 
but we discard those sources with distances from the VLA phase center that 
are larger than the outermost radio detection of 6$\farcm$785, 
thereby ensuring an approximate comparability of the radio and X-ray 
samples. This step discards 164 out of 1129 X-ray
sources for a total sample size of 965 sources. The second sample
consists of those X-ray sources that have radio counterparts in our
observations, as listed in our catalog paper, resulting in a sample
of 228 sources. The third sample consists of the 13 sources
defined as extreme variable radio sources in the present paper.       
 
\begin{figure} 
\includegraphics[width=\linewidth]{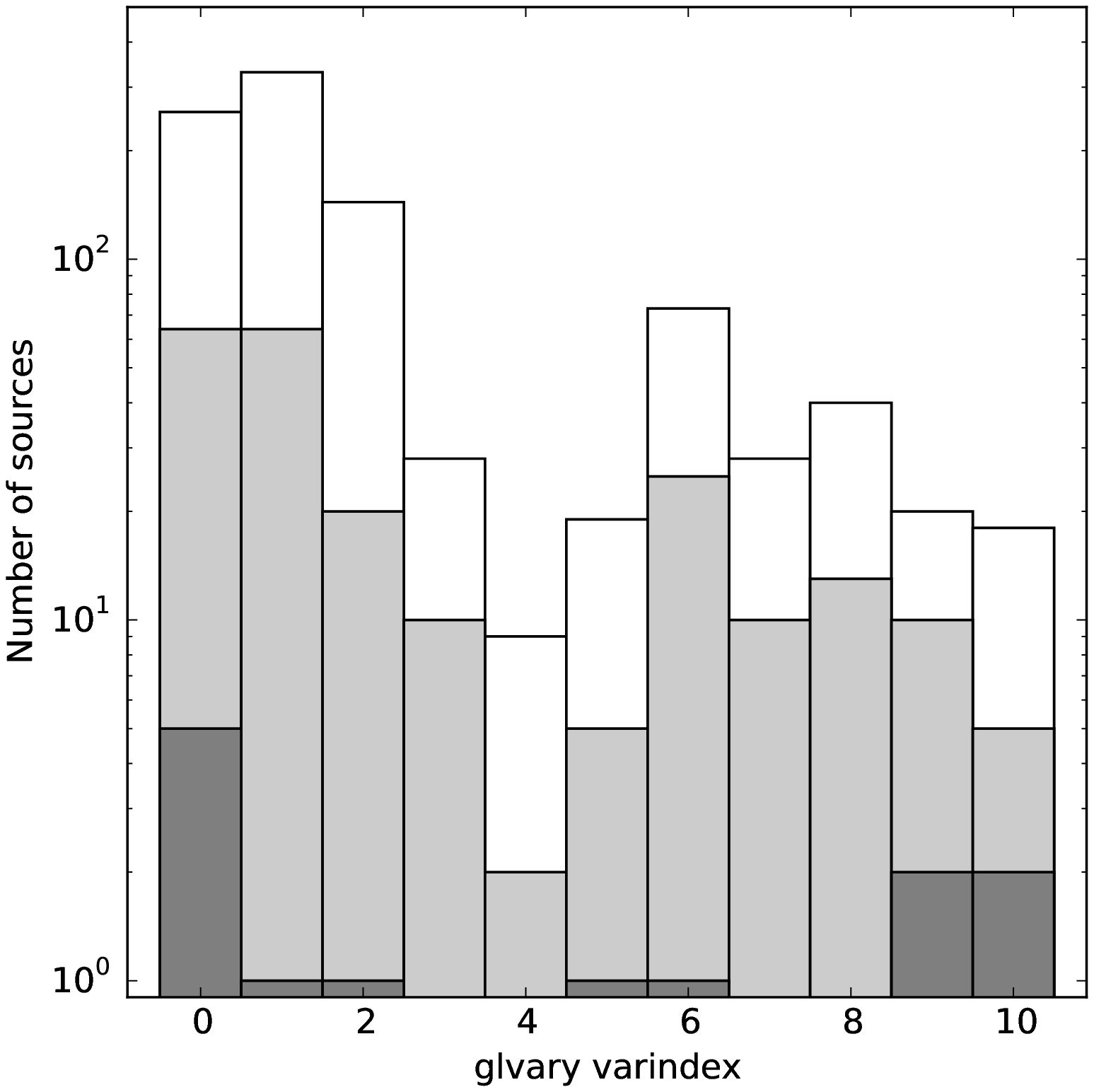} 
\caption{X-ray variability, as quantified with the glvary varindex
(see text), for the full sample of X-ray sources within the radio field 
of view (white bins), the
subsample with radio counterparts (light grey bins), and the subsample
of the 13 extremely variable radio sources identified in this paper
(dark grey bins). Note the logarithmic y axis.\label{fig_varindex}}                                 
\end{figure}

After determining the varindex values for these samples at all four
X-ray epochs, we pick the maximum varindex for each source to quantify
variability within the epochs. A source with varindex=10 in any one
epoch thus is listed with this varindex while a varindex of 0 means
that it has been zero in all four epochs. The resulting distributions
of the maximum varindex values per source for the three samples are
shown in Figure~\ref{fig_varindex}.                                     
 
The varindex distributions are spread across all possible values
from 0 to 10. Most sources have little or no variability,
but a total of 18 sources do show the highest possible varindex=10. 
Interestingly, only five of these sources have radio
counterparts in our sample, and only two of these are in the sample
of extremely variable radio sources.                                    
 
The extreme variable radio sources do not seem to have unique
X-ray varindex properties. Instead, they generally fall into two very
different groups. Most of them, 7 out of 13 sources, have
varindex=0 to 2, while 4 sources have varindex=9 to 10, with just
two sources in-between (at varindex=5 and 6). It is remarkable, though, that the three
radio sources that vary on hour timescales (sources 36, 469, 515) 
have a varindex of either 9 or 10.                                                                
 
\section{Simultaneous radio and X-ray variability} 
 
With the extreme radio variability and the corresponding X-ray variability
presented, we now assess the correlation between them.  The first step 
is to determine whether extreme radio variability within a single observing
epoch coincides with X-ray flaring. While there could well be a lag
between the two types of variability, we lose sensitivity to this
connection once the lag is longer than the duration of the observing
epochs, since a related but delayed flare could fall into one
of the gaps in-between our observing epochs.                            
 
We start with the three sources (98, 469, and 515) with extreme radio 
variability on less than hour timescales. The radio flare in source 98 occurs 
in the first epoch, which unfortunately does not have \textit{Chandra} coverage.
However, the radio flares of sources 469 and 515 both coincide with simultaneous
X-ray variability, although this only becomes apparent in the
adaptively smoothed lightcurves, as binning on timescales longer than
a flare can average away the detection. 
The X-ray flares are not very strong, and statistically the events coincide;
there is no evidence for a significant offset in time. 

Moving to other sources which exhibited extreme radio flaring on longer
timescales, source 36 also displayed near simultaneous radio and X-ray
flares.  Its X-ray flare is best detected in the adaptively
binned lightcurve and its radio variability stays just below an order of 
magnitude change in its flux density within a single epoch, but it reaches this 
threshold when considering neighboring epochs.
Interestingly, source 110 displays a radio flare with a steep slope, there
is no detected X-ray flare.  However, this source is generally a very weak 
X-ray source, and the non-detection may thus be due to insufficient sensitivity.
 
Overall, the results shown in Figure~\ref{fig_xr} qualitatively show
varying levels of radio--X-ray correlation across multiple epochs. 
In many cases, there is no correlation between overall flux levels or trends 
between epochs. In some cases, this is influenced by insufficient sensitivity
in either band, but even bright X-ray sources show phenomena as diverse
as no correlation whatsoever (source 254) or significant correlation
(source 319).                                                           
 
When comparing the radio and X-ray lightcurves, it thus emerges that
extreme radio variability is clearly correlated with X-ray variability
{\it only on very short timescales}.  However, on timescales of days, 
we find complexity in the lightcurves, but do not see clear trends.
associating X-ray and radio variability. Because of the unavoidable gaps in 
coverage for the ground-based radio observations, it is impossible to know 
whether or not an apparent overall correlation as seen toward source 319
traces a common event or is a coincidence of unrelated activity.
 
\subsection{Outlook} 
 
Our first results demonstrate how the upgraded VLA, with greatly increased 
sensitivity, is providing us with a new perspective on high-energy processes in
YSOs, by allowing us to systematically study YSO radio emission on timescales
of minutes.  This is complementary to the X-ray view, which by now is
much better documented.   Understanding connections between emissions in
these wavebands may be particularly important for a full assessment of the 
high-energy irradiation of protoplanetary disks. 
 
Based on this first estimate of extreme centimeter-wavelength variability,
it will now also be interesting to systematically probe extreme variability
involving even higher energy electrons in the millimeter-wavelength range,
using the Atacama Large Millimeter Array (ALMA), where synchrotron may 
dominate over gyro-synchrotron emission (e.g., \citealp{mas06}).
 
A general assessment of YSO radio variability will be presented after
complex corrections for time- and position-dependent effects of the
wideband primary beam, impacting variability, polarization, and spectral
indices \citep{bha13}.                                                  
Finally, our results highlight the inadequacy of a constant-sky assumption
in interferometric imaging in this scenario. For the extreme
radio variability reported here, the single integrated flux density
and spectral index, as for example reported in our deep catalog, is of 
questionable value. It will soon be possible, however, to incorporate a model
of a variable sky in the imaging process (e.g., \citealp{rau12}).
The upgrade to wideband continuum receivers thus has resulted in
fantastic new opportunities while also posing new calibration and
imaging challenges, mostly involving the abandonment of the assumptions
of monochromatic imaging of a constant sky.

\section{Summary} 
 
This study presents a first systematic assessment of extreme radio
variability in YSOs and the connection with X-ray variability from
simultaneous observations. Our main results can be summarized as
follows:                                                                
 
\begin{itemize} 
 
\item Thirteen of 507 radio sources show extreme variability, defined
as a factor of $>10$ and on timescales shorter than two days. 
All of these extremely variable sources are X-ray sources,
so 13 out of 222 radio sources with X-ray counterparts fall into this 
category, or about 6\%. The most extreme variability on timescales of 
less than two days is by a factor of $>138$.     
        
\item Three sources show such extreme radio variability on timescales 
of just 0.4 to 0.7 hours. We estimate an average occurrence rate of one 
such extreme short-timescale flare per star in about three months. 
 
\item For two of the short-timescale radio events (in sources 469 and 515) we had
simultaneous X-ray observations, and these radio flares were accompanied by 
quasi-simultaneous, equally short-duration, X-ray flares. A third source (36) shows similar behavior, even if it stays short of a variability factor of 10 at the highest time resolution. 
Beyond these short-duration radio flares, the correlation of radio and 
X-ray variability is less clear, and we find cases both of apparent 
correlation as well as seemingly uncorrelated behaviour.                                                               

\item The sample of variable radio sources contains stars with spectral
types ranging from O to M stars, indicating that extreme radio variability
occurs across the stellar mass range, unless the emission
from higher mass stars comes from unresolved low-mass companions.        
 
\item In X-rays, we find that only five out of 18 of
the most variable X-ray sources (varindex=10) in the common field of view 
have radio counterparts, and only two of these are in the sample
of extremely variable radio sources.  However, five of the 13 extremely
variable radio sources show no X-ray variability (varindex=0),
indicating a complex time domain connection, or lack thereof,
between YSO radio and X-ray emission.                                   
 
\end{itemize}

\acknowledgments 
Support for this work was provided by the National Aeronautics and
Space Administration through Chandra Award Number GO2-13019X issued
by the Chandra X-ray Observatory Center, which is operated by the
Smithsonian Astrophysical Observatory for and on behalf of the National
Aeronautics Space Administration under contract NAS8-03060. The National
Radio Astronomy Observatory is a facility of the National Science
Foundation operated under cooperative agreement by Associated Universities,
Inc. JF would like to thank Pat Broos, for insightful help with and
advice on acis\_extract, and acknowledge \citet{lie74} as a continuing
source of inspiration for time domain science. V.M.R. is funded by the Italian Ministero dell'Istruzione, Universit\`a e Ricerca through the grant Progetti Premiali 2012 - iALMA.

\facilities{VLA, CXO} 
 

\software{CASA, \citealp{mcm07},
          AIPS, \url{http://www.aips.nrao.edu},
          CIAO, \citealp{gre92,rot06},
          acis\_extract, \citealp{bro10}
	  }



 
\bibliography{orion} 

\begin{rotatetable} 
\begin{deluxetable*}{rrrrrrrrrrrrrr}
\tabletypesize{\scriptsize}
\tablecaption{Extreme radio variability in the ONC and X-ray properties\label{tab_exvar}}
\tablehead{
\colhead{Src\tablenotemark{a}}    & \colhead{COUP} & \colhead{prev. ID\tablenotemark{b}} & \colhead{$d$} & \multicolumn3c{$\Delta S_{\rm max}$ (0--2d)} & \multicolumn3c{$\Delta t_{\rm min}$ for x10 var.} & \colhead{$S_{\rm C}$\tablenotemark{c}} & \colhead{CXO cnts\tablenotemark{d}} & \colhead{varindex\tablenotemark{e}} & \colhead{SpT\tablenotemark{f}} \\
\colhead{} & \colhead{}                          & \colhead{(radio)}                   & \colhead{($'$)} & \colhead{(ep.)} & \colhead{(30 min)}  & \colhead{(6 min)}   & \colhead{(ep.)} & \colhead{(30 min)}  & \colhead{(6 min)}   & \colhead{(mJy\,bm$^{-1}$)} & \colhead{(0.5-8.0 keV)} & \colhead{(max)}
}
\startdata               
36  & 391  &	    --  & 1.10 & $>11$ & $>10$ &   $>5$ &   1d & 27.0h &   --  & 0.056$\pm$0.003 &  484.2	&  9 &  	 \\ 
53  & 427  &	    --  & 2.40 &   13  & $>12$ &   $>7$ &   1d & 17.7h &   --  & 0.165$\pm$0.004 &  514.4	&  9 &  	 \\ 
98  & 510  & Zap 7, S	& 1.45 &    8  & $>17$ &  $>17$ &   -- &  1.1h &  0.7h & 0.754$\pm$0.003 &     36.8	&  2 &  	 \\ 
110 & 530  & GMR Q, S	& 0.51 &    6  &   11  &   $>6$ &   -- & 24.9h &   --  & 0.303$\pm$0.003 &     35.3	&  5 &  	 \\ 
189 & 640  &	    S	& 0.76 &   10  &   34  &  $>25$ &   1d & 20.4h & 19.3h & 0.690$\pm$0.003 &  155.7	&  1 &  	 \\ 
254 & 745  & GMR 12, KS & 0.80 &    5  &   14  &    19  &   -- & 21.2h & 20.5h &23.208$\pm$0.003 & 8048.8$^*$   &  0 &  O9--B0.5 \\ 
319 & 828  &	     S  & 1.66 &   15  & $>17$ &  $>34$ &   2d & 41.1h & 40.7h & 2.358$\pm$0.003 & 5095.5$^*$   &  0 &  K2--K6   \\ 
414 & 985  &	     K  & 2.21 &    8  &    9  &  $>13$ &   -- &   --  & 22.9h & 1.324$\pm$0.003 & 2391.7$^*$   &  0 &  F8--K4   \\ 
422 & 997  &	     -- & 1.37 & $>54$ & $>30$ &  $>15$ &   2d & 41.1h & 40.7h & 0.676$\pm$0.003 & 2265.9$^*$   &  6 &  K8	 \\ 
469 & 1101 &	     S  & 1.79 &    2  &   12  &  $>14$ &   -- &  1.1h &  0.6h & 0.235$\pm$0.003 & 3740.3$^*$   & 10 &  M0+M3.5  \\ 
489 & 1143 &	     K  & 2.98 & $>16$ & $>11$ &   $>7$ &   1d & 27.6h &   --  & 0.219$\pm$0.004 & 1430.8$^*$   &  0 &  K1--K4   \\ 
495 & 1155 &	     -- & 1.91 &    5  & $>10$ &   $>7$ &   -- & 47.4h &   --  & 0.267$\pm$0.003 &  166.9	&  0 &  M3.5	 \\ 
515 & 1232 &	     K  & 3.22 &   52  &$>138$ & $>101$ &   1d &  0.5h &  0.4h & 1.188$\pm$0.004 & 7916.2$^*$   & 10 &  O9.5--B2 \\ 
\enddata
\tablenotetext{a}{[FRM2016]}
\tablenotetext{b}{In (Zap) \citet{zap04}, (K) \citet{kou14} and (S) \citet{she16}.}
\tablenotetext{c}{Peak flux density, derived from the concatenated data in \citet{for16}.}
\tablenotetext{d}{Net counts, all four 2012 X-ray epochs combined; an asterisk denotes possible photon pile-up (see text).}
\tablenotetext{e}{Maximum varindex from the four CXO epochs (see text).}
\tablenotetext{f}{Spectral types as reported by \citet{hil13}. Note that particularly in the case of the high-mass stars, the radio emission may be from unresolved lower-mass companions.}
\end{deluxetable*}
\end{rotatetable}

\end{document}